%% file: main.tex
\documentclass{aa}  
\usepackage{comment}
\usepackage{graphicx}
\usepackage{txfonts}
\usepackage{multirow}
\usepackage{subcaption}
\usepackage{lscape}          
\usepackage{adjustbox}
\usepackage{amssymb}
\usepackage{placeins}   
\usepackage{xcolor}
\usepackage{colortbl}
\usepackage{linenoaa}
\usepackage{makecell} 
\usepackage[colorlinks=true,allcolors=blue]{hyperref}

\makeatletter
\renewcommand*\aa@pageof{, page \thepage{} of \pageref*{LastPage}}
\makeatother

\begin{document}

\title{HYPERION. The cold ISM of rapidly growing $z>6$ quasars: diverse gas reservoirs, dust enrichment, and feedback signatures}

\titlerunning{The cold ISM of HYPERION QSOs}
\authorrunning{C. M. Pierro, R. Tripodi et al.}

\input{authors}

\abstract
{Luminous QSOs at $z>6$ host some of the most rapidly assembled SMBHs in the early Universe. Characterizing their cold ISM is essential to determine whether extreme SMBH growth is accompanied by similarly rapid host-galaxy assembly. We investigate the molecular gas, cold dust, star formation, gas-to-dust ratio, and ionized ISM of ten HYPERION QSOs using new ALMA Band 3 observations targeting CO(6--5) and the underlying $\sim100$ GHz continuum, complemented by archival and literature ALMA/NOEMA data. We detect $\sim100$ GHz continuum emission in eight targets and CO(6--5) emission in four QSO hosts, J025--33, J083+11, J231--20, and J0252--0503, as well as in the companion of J231--20. The inferred molecular gas masses are of order $10^{10}~M_\odot$, while the non-detections imply upper limits of a few $10^9~M_\odot$, indicating a broad range of molecular reservoirs within the HYPERION population. For J025--33 and J083+11, the FIR SEDs are well sampled and yield low dust temperatures, $T_{\rm dust}=36^{+13}_{-7}$ K and $32^{+4}_{-3}$ K, respectively, well below the average value for $z>6$ QSOs. Combining gas and dust masses, we find a gas-to-dust ratio for J083+11, ${\rm GDR}=16^{+5}_{-4}$, among the lowest measured in a high-redshift QSO host. We also detect [NII]$\lambda 205,\mu$m emission in J025--33 and tentatively in J083+11, suggesting dense or highly structured ionized gas. Finally, we identify a tentative connection among $T_{\rm dust}$, the X-ray photon index $\Gamma$, and the C\,{\sc iv} velocity shift. These trends may indicate that more powerful winds redistribute dust away from the central AGN heating source, lowering its temperature and weakening the connection between the large-scale dust reservoir and the X-ray corona. Overall, HYPERION QSOs emerge as a heterogeneous population in which SMBH growth, star formation, gas consumption, enrichment, and feedback are not necessarily synchronized.
}

   \keywords{}

   \maketitle

\section{Introduction}
\label{sec:intro}

\nolinenumbers

\begin{figure*}
    \centering
    \includegraphics[width=0.24\linewidth]{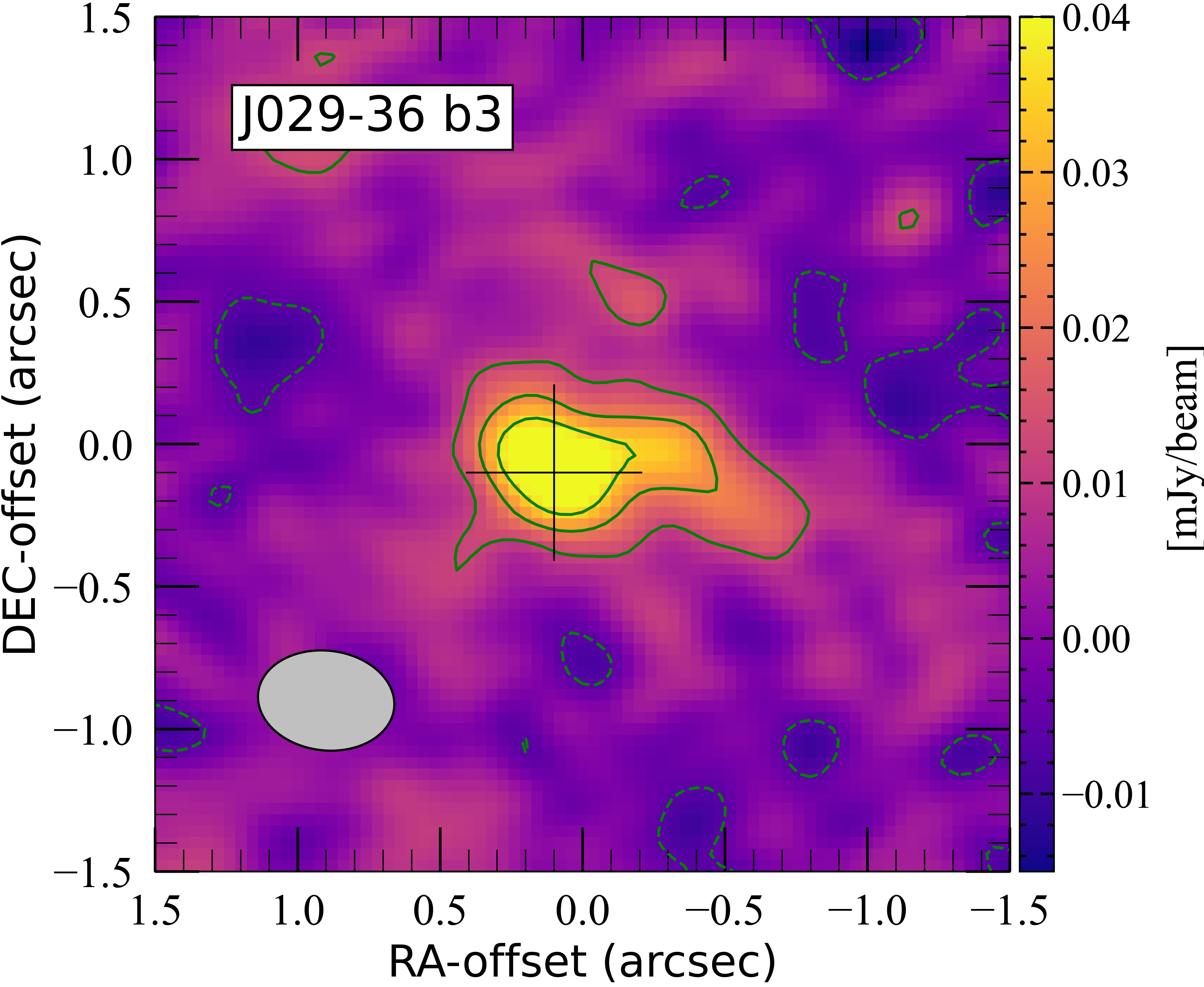}
    \includegraphics[width=0.24\linewidth]{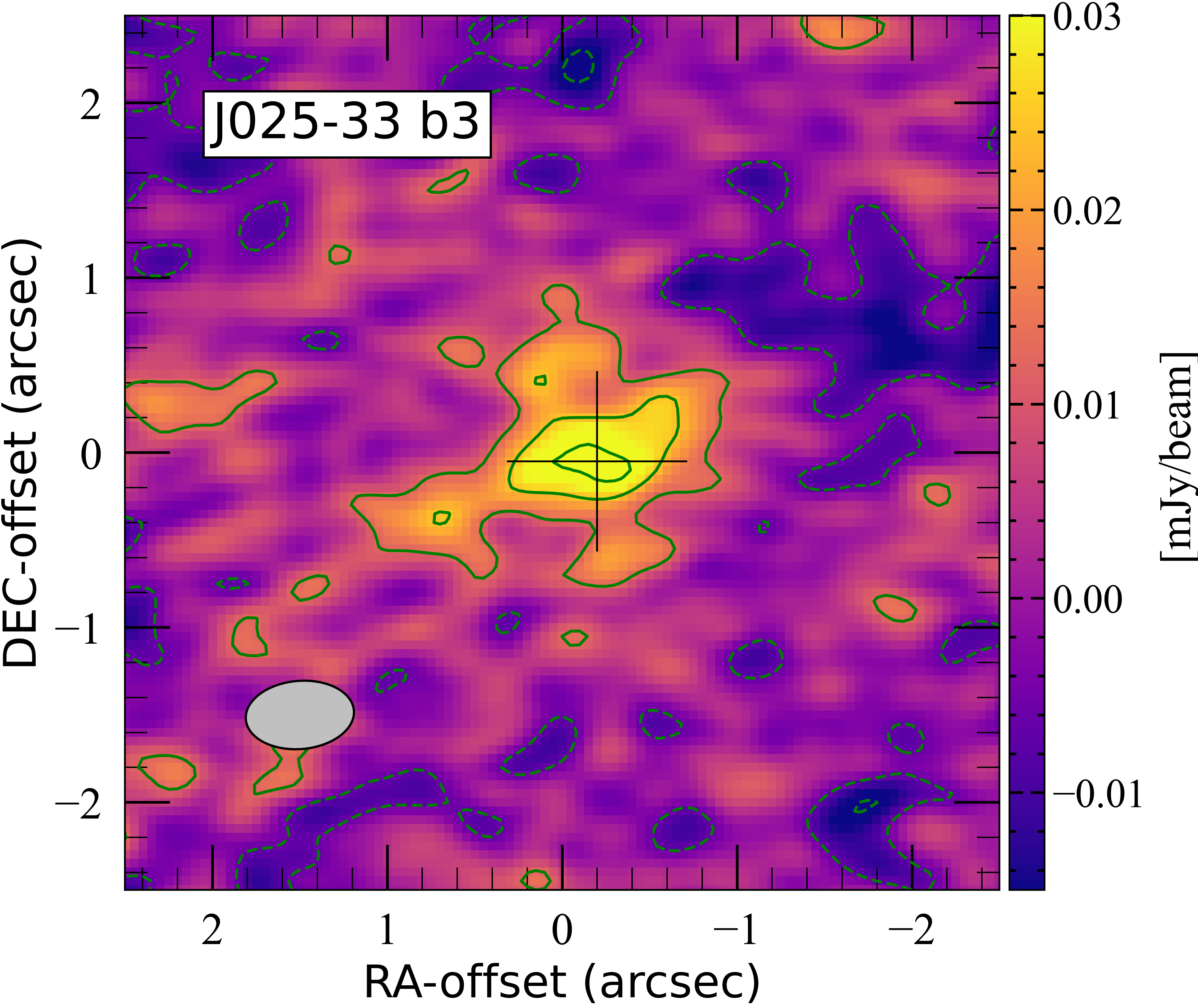}
    \includegraphics[width=0.24\linewidth]{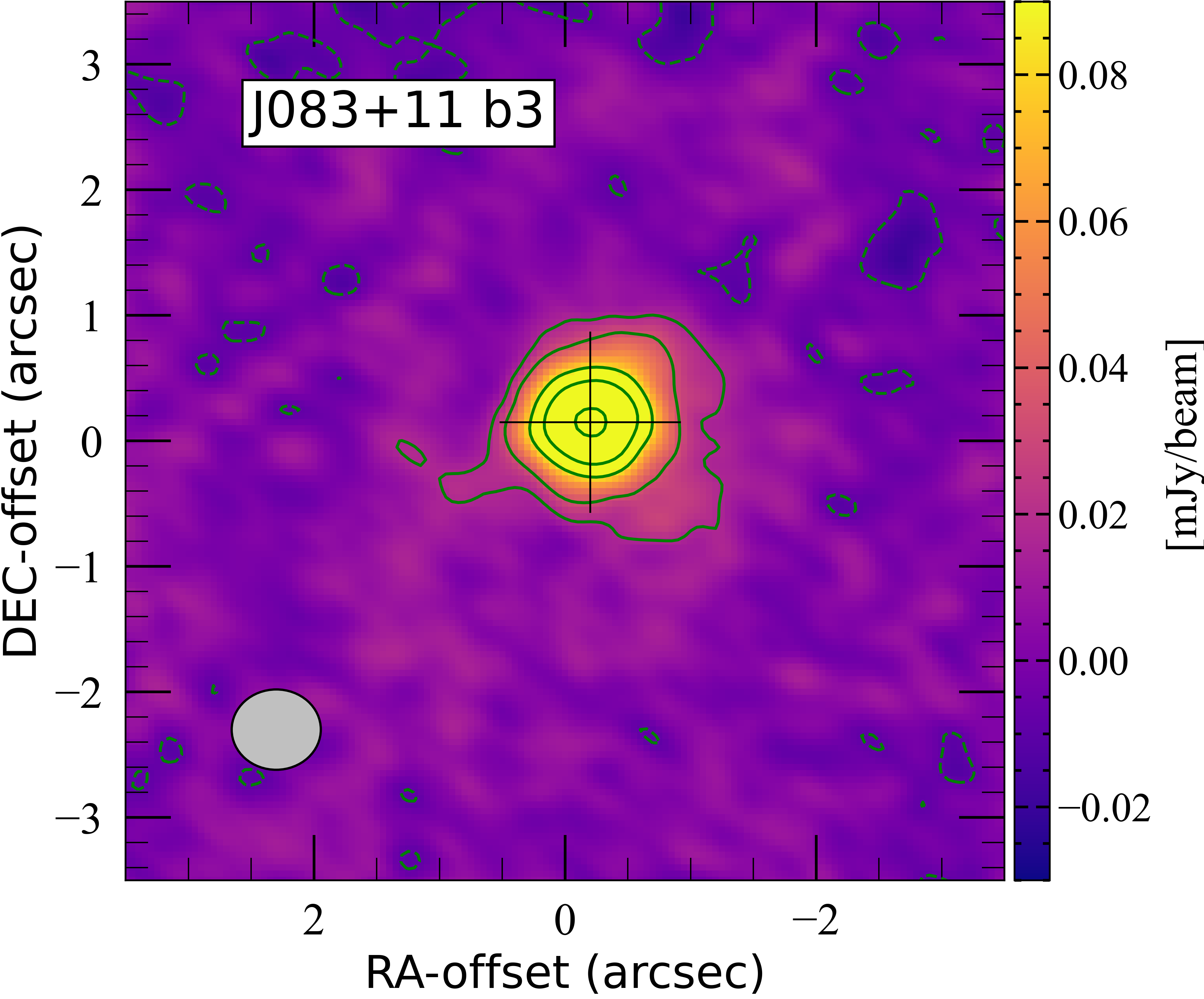}
    \includegraphics[width=0.24\linewidth]{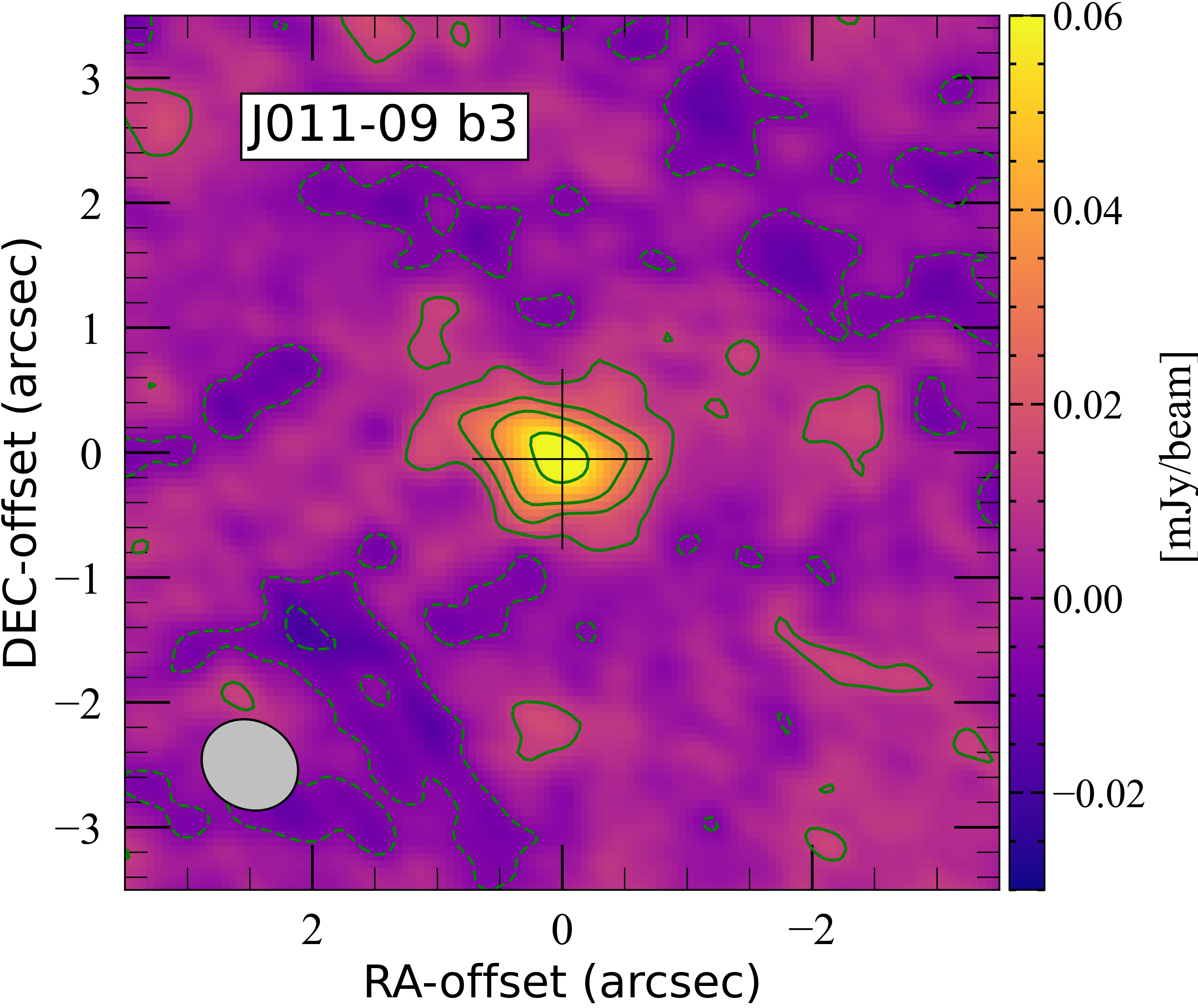}\\
    
    \includegraphics[width=0.24\linewidth]{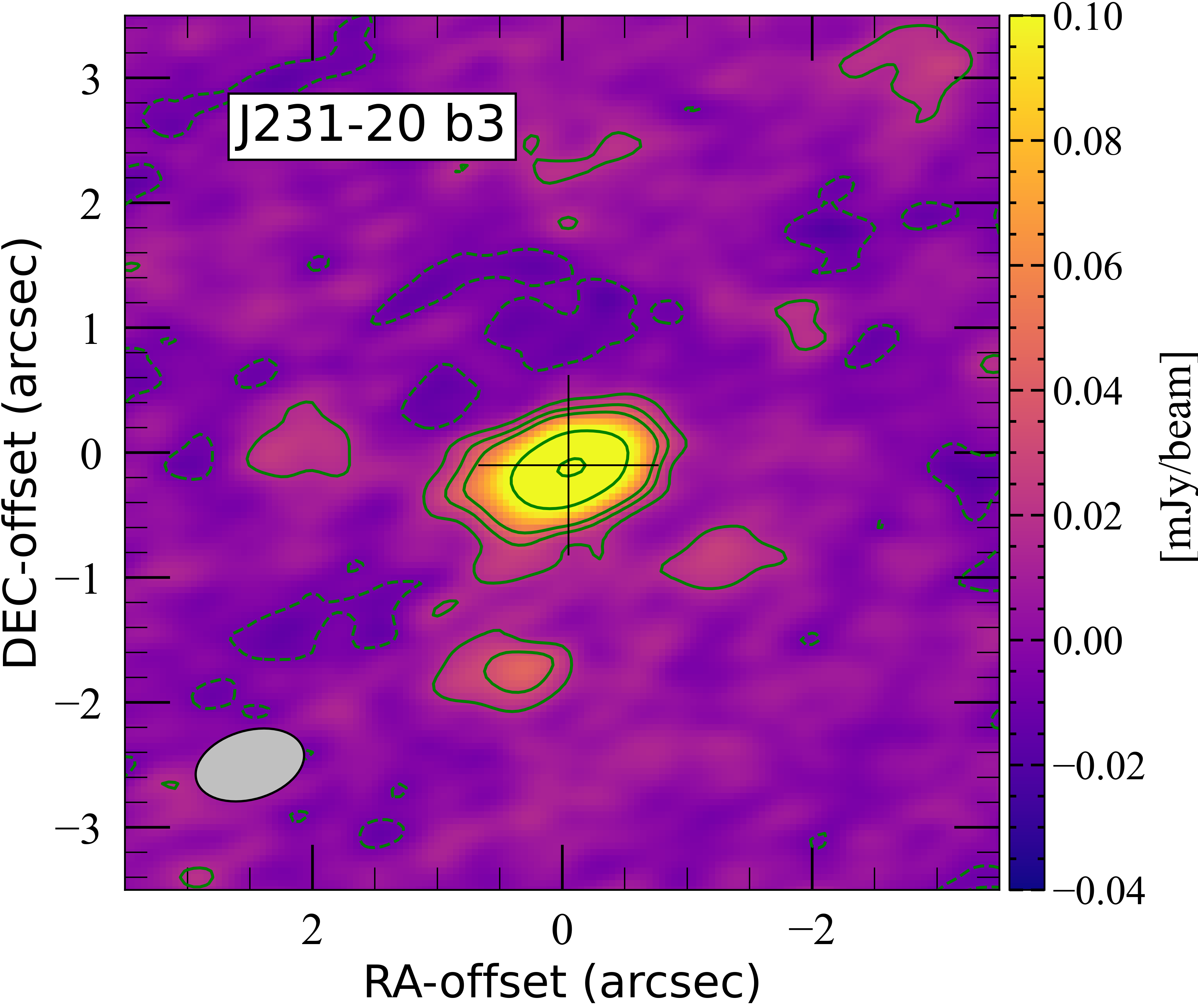}
    \includegraphics[width=0.24\linewidth]{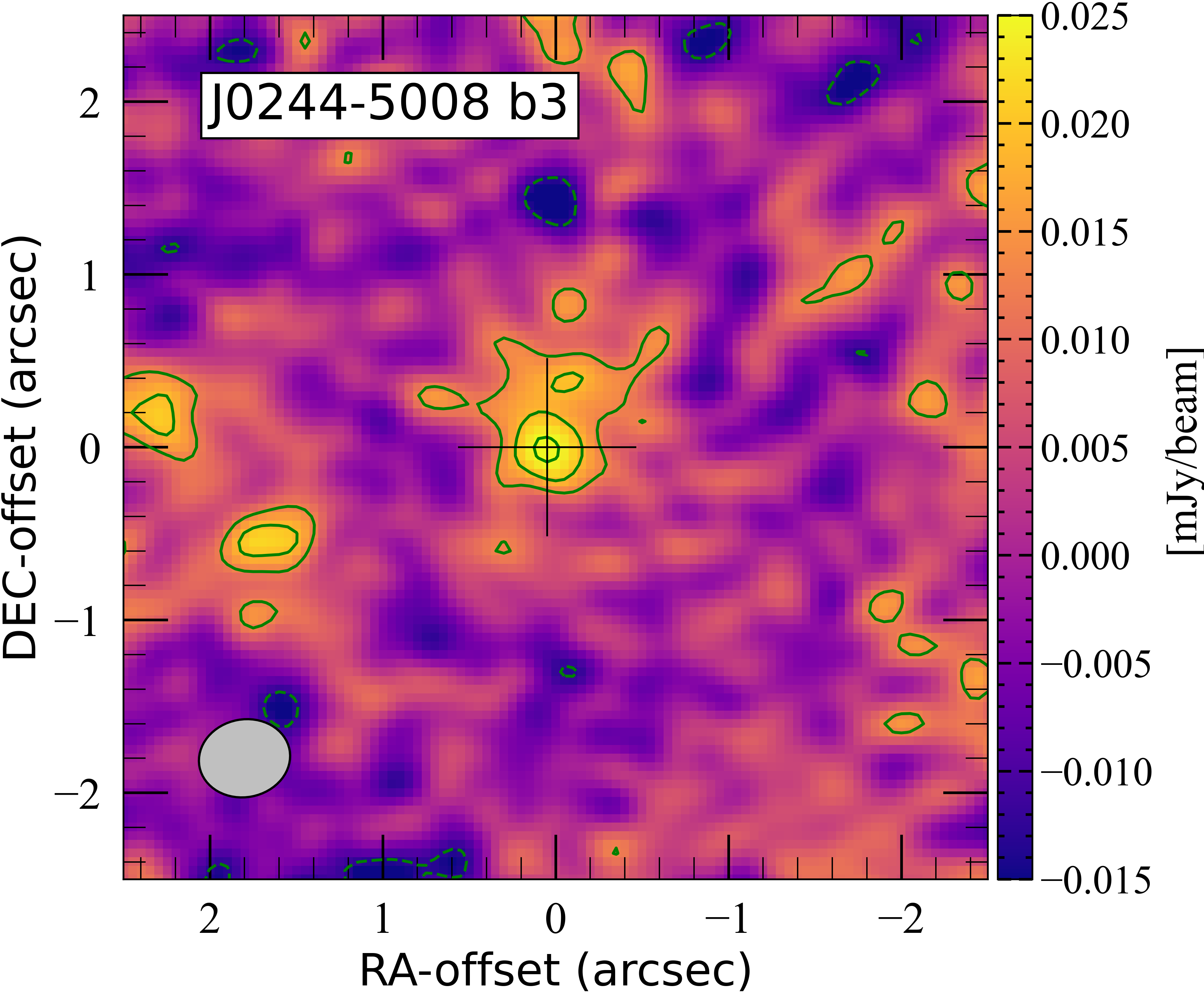}
    \includegraphics[width=0.24\linewidth]{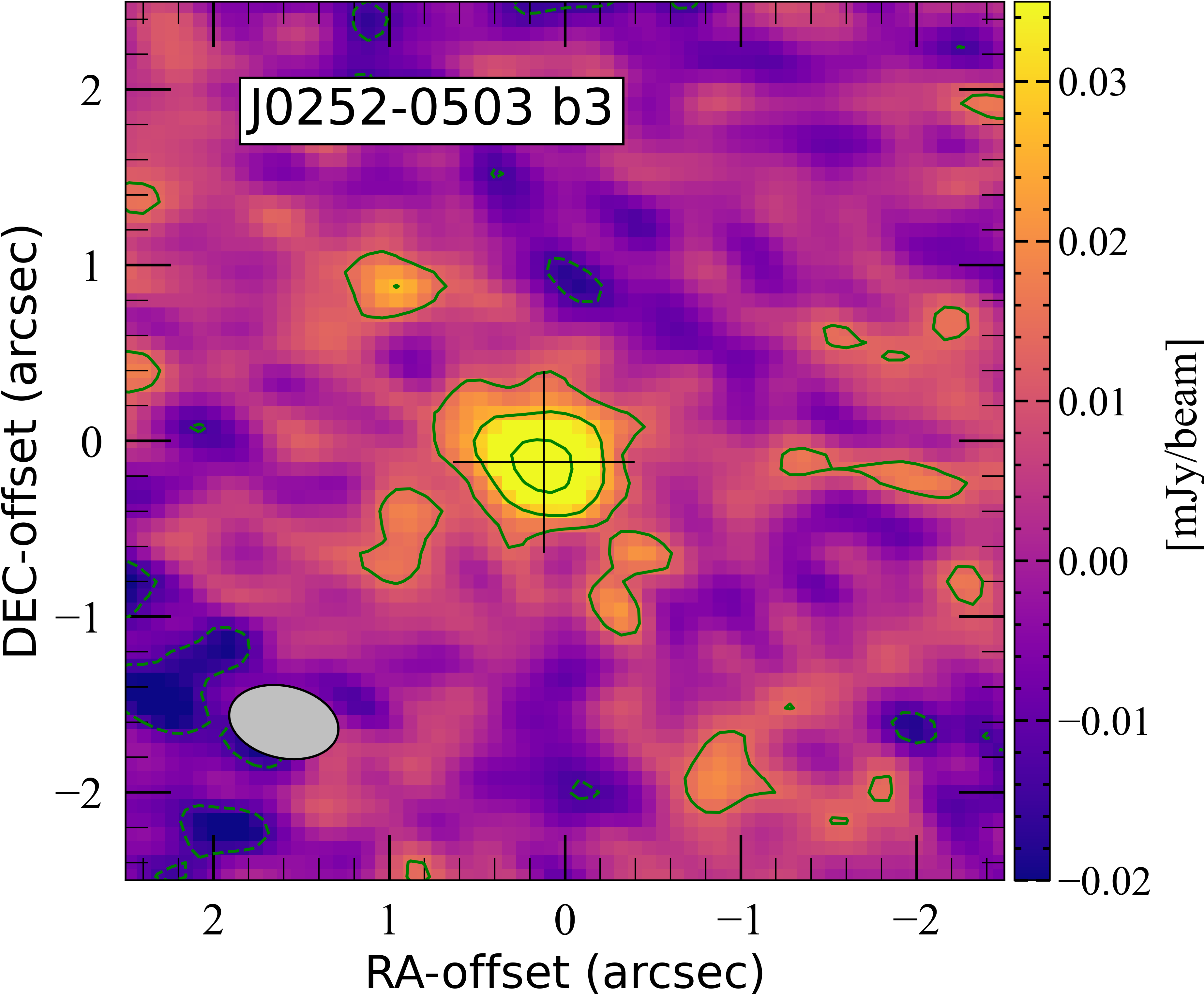}
    \includegraphics[width=0.24\linewidth]{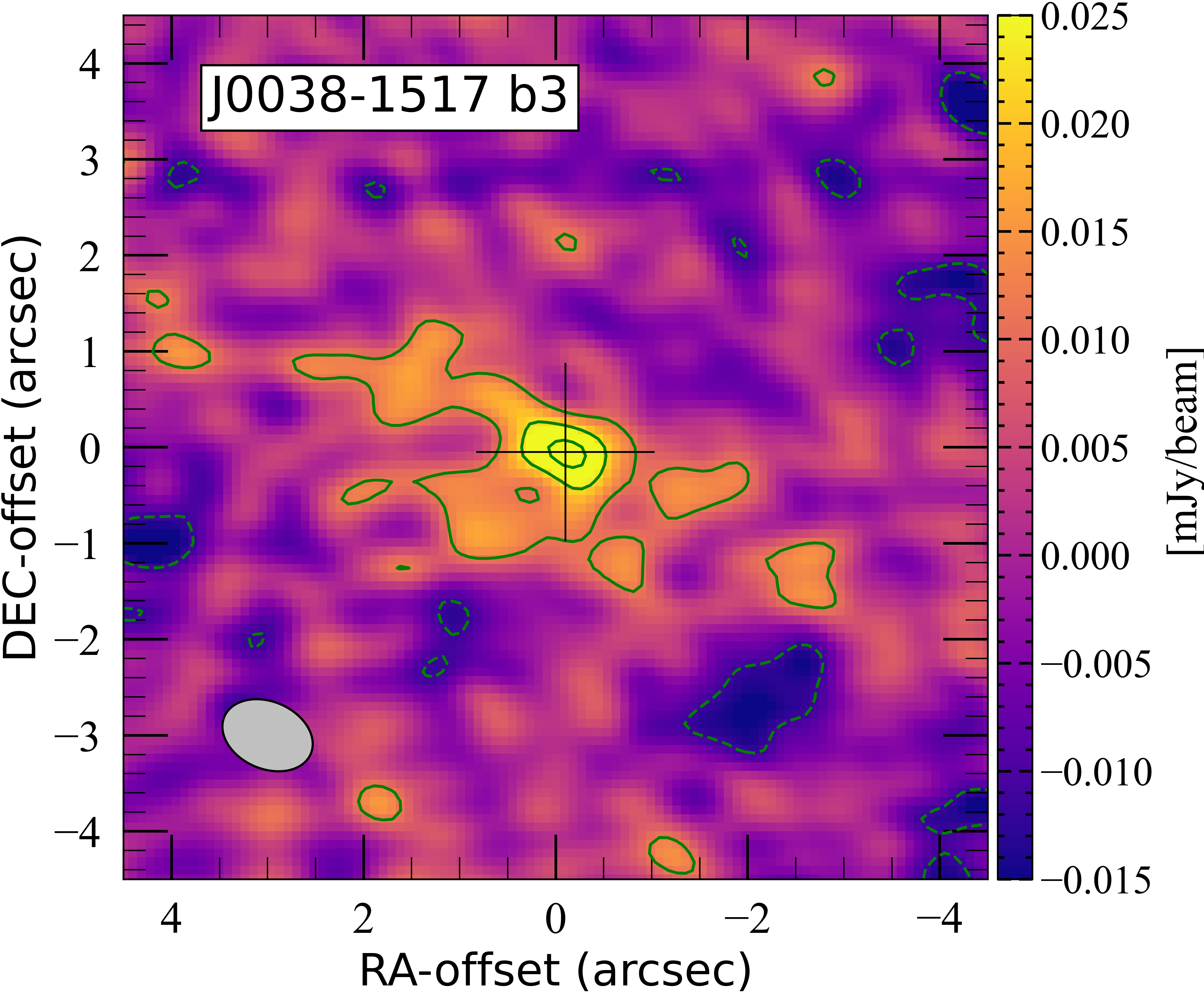}\\
    \caption{Dust continuum maps in B3. From left to right, top to bottom, $\sim 100$ GHz dust continuum map of QSOs J029-36, J025-33, J083+11, J011+09, J0244-5008, J0252-0503 and J0038-1527. The clean beam is shown in gray in the lower left corner of each panel. The size of the beam and the rms of each map is reported in Tab. \ref{tab:res-cont}. Contour levels are at: $2,6,4,10~\sigma$ for J011+09, J025-33, J029-36, J0038-1517, $2,4,10,15,25~\sigma$ for J083+11, $2,4,6,15,30~\sigma$ for J231-20, $2,3,4~\sigma$ for J0244-5008, $2,4,8~\sigma$ for J0252-0503.}
    \label{fig:cont-b3}
\end{figure*}

Targeting quasi-stellar object (QSO) host galaxies at $z>5$--6 provides a unique opportunity to investigate the concurrent assembly of the first super-massive black holes (SMBHs) and their host galaxies, and to characterize the physical properties of the interstellar medium (ISM) within the first Gyr of cosmic history \citep[e.g.,][]{decarli2018,venemans2020,neeleman2021, walter2003, walter2004}. These systems host rapidly accreting SMBHs with masses already reaching $\gtrsim10^9\,M_\odot$, implying extremely efficient early growth. At the same time, their host galaxies are often rich in gas and dust, with intense star formation rates (SFRs) of several hundred to thousands of $M_\odot\,{\rm yr}^{-1}$. Understanding whether such extreme nuclear activity is accompanied by equally rapid host-galaxy growth, and whether the SMBH already affects the gas and dust reservoirs of the host, is therefore central to constraining the earliest phases of black-hole--galaxy co-evolution.

In recent years, major progress has been achieved thanks to sensitive far-infrared and millimetre observations with the Atacama large mm/sub-mm array (ALMA) and the Northern extended millimeter array (NOEMA). Rest-frame far-infrared (FIR) fine-structure lines, in particular [CII]$_{158\,\mu{\rm m}}$ and [OIII]$_{88\,\mu{\rm m}}$, have revealed that high-redshift quasar hosts are typically embedded in compact, gas-rich, and vigorously star-forming galaxies \citep[e.g.,][]{tripodi2023a, tripodi2023c, wang2024, wu2025}. These observations have provided measurements of systemic redshifts, dynamical masses, gas kinematics, and, in some cases, evidence for disturbed morphologies, mergers, companions, or large-scale outflows. In parallel, molecular gas studies based on low- and mid-$J$ CO transitions have started to directly probe the cold gas reservoirs that fuel star formation and SMBH growth \citep[e.g.,][]{vallini2018,madden2020,salvestrini2025,decarli2022, tadaki2025,kaasinen2024}. Although [CII]$\lambda 158\,\mu$m is the brightest and most widely used FIR line in high-redshift QSO hosts, its use as a molecular-gas tracer is not straightforward. The [CII] emission can arise from multiple ISM phases, including photodissociation regions, diffuse neutral gas, ionized gas, and CO-dark molecular gas, and can therefore be affected by the geometry, metallicity, radiation field, and possible outflows of the host galaxy. CO rotational transitions provide a more direct tracer of the cold molecular reservoir. Once an excitation correction (for transitions higher than $J=1$) and a CO-to-H$_2$ conversion factor are assumed, CO therefore offers a more robust reference estimate of $M_{\rm H_2}$ against which indirect [CII]-based calibrations can be tested. However, CO detections remain available only for a limited number of $z>6$ quasars, because the lines are faint and often require deep observations. As a result, the molecular gas content of most high-redshift quasar hosts is still inferred indirectly, with significant systematic uncertainties.

Another key source of uncertainty is the conversion between dust continuum emission and physical quantities such as dust mass, infrared luminosity, and SFR. In many high-$z$ quasar studies, the dust spectral energy distribution (SED) is poorly sampled and the dust temperature is assumed, usually adopting values of $T_{\rm dust}\sim 40$--50 K. This assumption has a direct impact on the inferred infrared luminosities and SFRs, and an even stronger effect on the dust masses. When multi-frequency mm ALMA and/or NOEMA observations are available in the rest-frame FIR, sampling both the peak and the Rayleigh--Jeans tail of the dust SED, the dust temperature and mass can instead be constrained directly, with statistical uncertainties often below $\sim10$--20\% \citep[e.g.,][]{carniani2019,Tripodi2022,tripodi2023a,witstok2023,tripodi2024,costa2026}. Such measurements are essential to obtain accurate SFRs and to avoid biases introduced by assuming a single, fixed dust temperature for all high-redshift quasar hosts.

Accurate dust masses also provide an independent route to estimate the molecular gas mass through the gas-to-dust ratio (GDR). This approach is particularly valuable at high redshift, where direct CO-based gas masses are still scarce. However, the GDR itself remains uncertain. Studies of hyper-luminous quasars at $z\sim2.4$--4.7 indicate a broad range of values, ${\rm GDR}\sim100$--300, with an average ${\rm GDR}\sim180$ \citep{bischetti2021}, broadly consistent with estimates for sub-millimeter galaxies at $z\sim3$--5, where typical values of ${\rm GDR}\sim150$--250 have been reported \citep[e.g.,][]{saintonge2013,miettinen2017}. Other recent studies suggest that the GDR of some dusty star-forming galaxies may be lower than 100 \citep{birkin2021,liao2023, algera2026}, as well as for QSO hosts \citep{salvestrini2025}. In the local Universe, normal star-forming galaxies typically show ${\rm GDR}\gtrsim100$ \citep{draine2007,leroy2011,salvestrini2025b}, although with a strong dependence on metallicity and ISM conditions. Establishing whether the GDR evolves with redshift, galaxy type, or nuclear activity therefore requires systems for which both the gas mass and the dust mass are measured as accurately and consistently as possible.

These advances have opened a new window on the host galaxies of the first quasars, but they have also highlighted a fundamental open question: how does the rapid growth of the SMBH affect the early evolution of its host galaxy? A direct way to address this issue is to compare the host-galaxy properties of the most extreme quasars, in terms of SMBH mass and accretion history, with those of the broader high-redshift quasar and AGN population. If SMBH growth and galaxy growth are tightly coupled from the earliest epochs, the most rapidly growing black holes should be hosted by galaxies undergoing correspondingly extreme growth. Conversely, departures from this expectation may indicate that black-hole and galaxy assembly proceed on different timescales, or that feedback, gas supply, and star formation regulate the two processes in a more complex way.

The ideal laboratory for investigating the different scenarios is the HYPerluminous QSOs at the Epoch of ReionizatION (EoR) sample, HYPERION \citep{zappacosta2023}. HYPERION includes 18 quasars at $z>6$ selected to host SMBHs with masses of $10^9$--$10^{10}\,M_\odot$ that experienced some of the fastest mass growth histories known at these epochs. The sample has a mean redshift of $z\simeq6.7$, average bolometric luminosity $\log(L_{\rm bol}/{\rm erg\,s^{-1}})\simeq47.3$, and SMBH masses \citep{zappacosta2023}. These quasars were selected as the most extreme among the known $z>6$ quasar population by 2020, namely systems in which the central SMBHs underwent exceptionally efficient growth. They also benefit from extensive multi-wavelength coverage, from X-rays to rest-frame UV/optical and FIR wavelengths, making them uniquely suited to investigate the connection between accretion, star formation, feedback, and ISM properties in the early Universe.

HYPERION is based on a 2.4 Ms \textit{XMM-Newton} Multi-Year Heritage Programme designed to obtain, for the first time for a sizeable sample of $z>6$ quasars, a uniform and high-quality X-ray spectral characterization of the nuclear emission. The first results indicate a genuine redshift evolution of the X-ray spectral slopes, which appear to be steeper than those observed in $z<6$ quasars with comparable luminosities and accretion rates \citep{zappacosta2023}. This suggests that the first quasars may be characterized by a different accretion regime, possibly linked to fast disc-driven winds and to the extreme SMBH growth histories inferred for these systems \citep{tortosa2024}.

A first step toward connecting SMBH and host-galaxy growth in this population was presented by \citet{tripodi2024}, who analysed 10 quasars at $6<z<7$, including six HYPERION QSOs. By combining SMBH accretion rates, $\dot{M}_{\rm BH}$, with SFRs, these systems were placed in the $\dot{M}_{\rm BH}/M_{\rm BH}$ versus ${\rm SFR}/M_{\rm gal}$ plane, where $M_{\rm gal}$ is the  dynamical mass minus the black hole mass. This diagnostic allows one to compare the specific growth rates of the SMBH and of the host galaxy. Surprisingly, despite hosting some of the fastest-growing SMBHs known, only one source was found in a clearly BH-dominated growth phase. This result challenges simple co-evolutionary scenarios in which the most extreme SMBH growth is accompanied by proportionally efficient galaxy assembly, and suggests instead that these processes happen at different times with the galaxy intensively growing already at $z\sim6$ after a phase of fast and efficient BH growth.

In this work, we extend this investigation by analysing new ALMA Band 3 observations of 10 quasars from the HYPERION sample. Nine out of ten QSOs are new with respect to the work done in \citet{tripodi2024}. These data target the CO emission from the host galaxies and provide new constraints on their molecular gas reservoirs, while the combination with existing FIR continuum measurements allows us to refine the dust and gas properties of the sample. The properties of the 10 quasars analysed in this work are summarised in Table~\ref{tab:sample}.

The main goals of this paper are the following. First, we aim to derive molecular gas masses, or meaningful upper limits, for 10 HYPERION QSOs. Second, by combining gas constraints with FIR SED modelling, we investigate the gas-to-dust ratios of these quasar hosts and test whether they are consistent with values measured in other quasars,  and local star-forming systems. Third, we revisit the relative growth of SMBHs and their host galaxies in the HYPERION sample, using the improved constraints on bolometric luminosity, gas, dust, and star formation. Last, we combine the FIR and X-ray data to investigate a possible connection among the physical processes involved in these two regimes.

The paper is organized as follows. In Sect.~\ref{sec:obs}, we describe the ALMA observations and data reduction. In Sect.~\ref{sec:analysis}, we present the data analysis and results involving line modelling, and continuum measurements. We also derive the CO luminosities, molecular gas masses, dust properties, and gas-to-dust ratios. In Sect.~\ref{sec:disc}, we discuss the implications for the ISM properties and SMBH--host galaxy co-evolution in the HYPERION sample. Finally, in Sect.~\ref{sec:conc}, we summarize our main conclusions.

\begin{figure}
    \centering
    \includegraphics[width=1\linewidth]{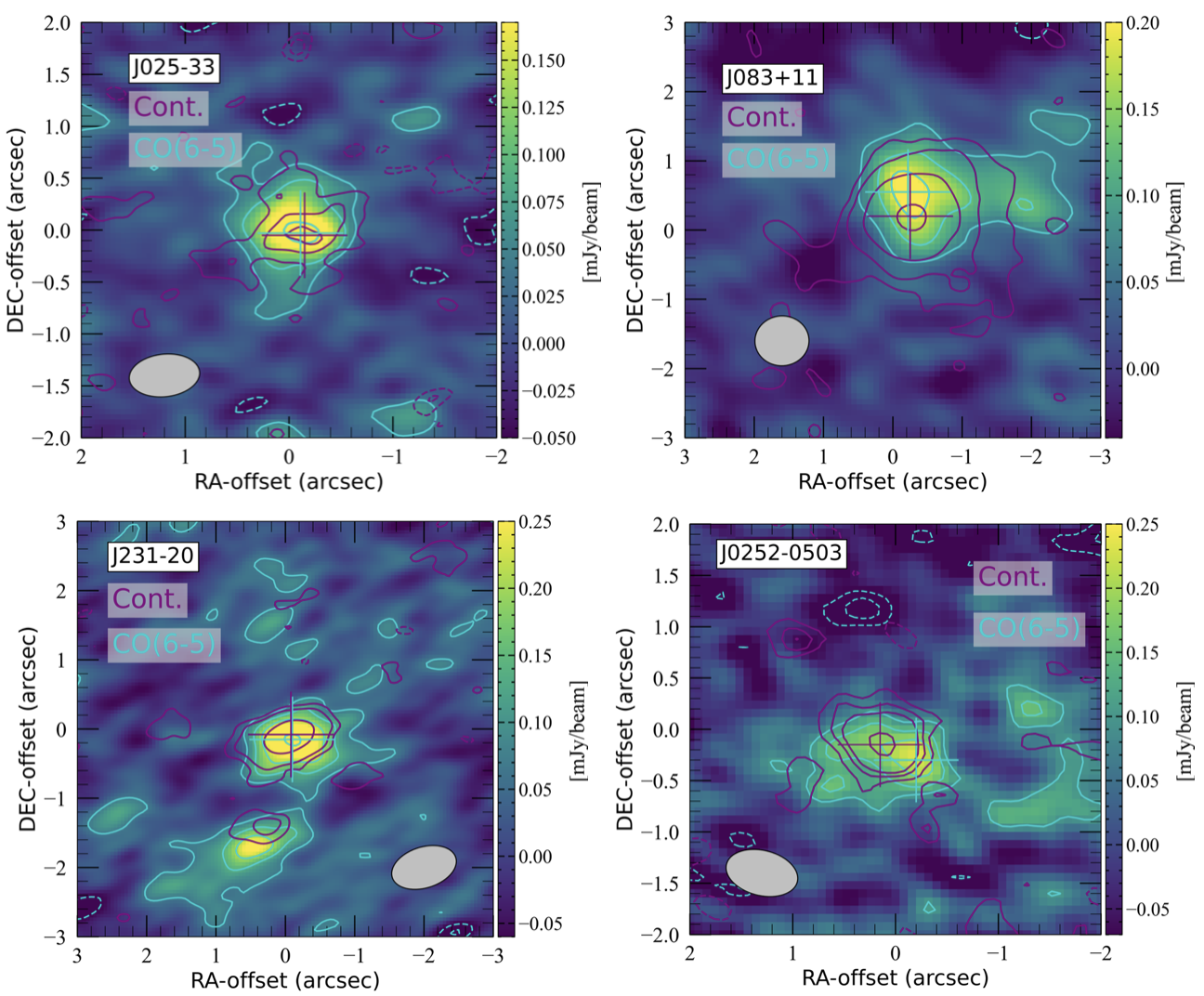}
    \caption{Line emission maps. From left to right, top to bottom, CO(6-5) emission line maps of QSOs J025-33, J083+11, J231-20 and J0252-0503. The contours of their underlying continuum emission shown in Fig.~\ref{fig:cont-b3} are overplotted in purple. The clean beam is shown in gray in the lower corner of each panel. The size of the beam and the r.m.s. of each map is reported in Tab. \ref{tab:res-line}. Contour levels are at $2,4,10~\sigma$ for J025-33, $2,4,6,10~\sigma$ for J083+11 and J231-20, and $2,3,4~\sigma$ for J0252-0503 (dashed contours are negative levels starting at $2,3~\sigma$). The purple (blue) cross marks the emission line (continuum) peak.}
    \label{fig:line-map}
\end{figure}

\begin{table*}[]
    \centering
    \caption{General properties of the sample}
    \begin{tabular}{c|cccccc}
       \hline \hline
       QSO & RA & DEC & $z_{\rm \ion{Mg}{II}}$ & $z_{\rm [\ion{C}{II}]}$ & $\log(M_{\rm BH}/M_\odot)$ & Refs.\\
       \hline
       J029-36      & 01:59:57.97 & -36:33:56.6 & 6.027 & -- & 9.82 & --\\ 
       J025-33      & 01:42:43.70 & -33:27:45.7 & 6.294 & 6.3373$\pm$0.0002 & 9.57 & V20\\ 
       J083+11      & 05:35:20.90 & +09:01:56.9 & 6.346 & 6.3401$\pm$0.0004 & 9.32 & Ak20 \\ 
       J011+09      & 00:45:33.57 & +11:50:53.6 & 6.444 & 6.4694$\pm$0.0002 & 9.32 & E20 \\
       J231-20      & 15:26:37.84 & -20:50:00.7 & 6.587 & 6.58734$\pm$0.0008 & 9.50 & N19 \\ 
       J0244-5008   & 02:44:01.02 & -50:08:53.7 & 6.724 & $6.73050 \pm 0.00003$ & 9.08 &  TW \\
       J0411-0907   & 04:11:28.62 & -09:07:49.7 & 6.824 & -- & 8.80 & --\\
       J0020-3653   & 00:20:31.47 & -36:53:41.8 & 6.834 & -- & 9.24 & --\\
       J0252-0503   & 02:52:16.64 & -05:03:31.8 & 6.99  & 7.0006$\pm$0.0009 & 9.15 & W24 \\ 
       J0038-1527   & 00:38:36.10 & -15:27:23.6 & 7.021 & 7.0340$\pm$0.0003 & 9.14 & W24 \\ 
       \hline \hline
    \end{tabular}
    \flushleft
   \footnotesize{{\bf Notes.} Columns: target name, RA (hms), DEC (dms), redshift from \ion{Mg}{II} emission line, redshift from \ion{C}{II} emission line, BH mass, references for the [CII] redshift. References:  V20 \citep{venemans2020};  Ak20 \citep{andika2020}; E20 \citep{eilers2020}; N19 \citep{neeleman2019}; TW, This work; W24 \citep{wang2024}.}
    
    \label{tab:sample}
\end{table*}

\section{Observations}
\label{sec:obs}

We primarily analysed the dataset 2024.1.01105.S (PI: R. Tripodi) from the ALMA 12m array in B3 targeting the CO(6-5) emission line and underlying continuum in 10 HYPERION QSOs (see Tab.~\ref{tab:sample}). To investigate the dust properties of each source via analysis the cold dust SEDs, we complemented our B3 data set with observations in higher frequency ALMA bands. In this case, we either reduced ALMA archival unpublished data or retrieved the results from published works (see details in Tabs. \ref{tab:res-line-obs} and \ref{tab:res-cont}). When multiple observations for the same source and band were available, we prioritized deeper continuum sensitivity, and a resolution closely matching our data in B3. The latter choice ensures that we probe the same spatial scales of dust emission.

All archival observations have been reduced homogeneously. The visibility calibration of the observations was executed by the ALMA science archive. The imaging was performed through the Common Astronomy Software Applications (CASA; \citealt{mcmullin2007}), version 5.1.1-5. We applied \texttt{tclean} using natural weighting and a $3\times$r.m.s (i.e. root mean square or $\sigma$) cleaning threshold to produce cleaned data cubes or maps. The r.m.s has been estimated from a source-free region of the dirty cube or image. We imaged the continuum using the multi-frequency synthesis (MFS) mode in all line-free channels (the average channel width is $\sim 25~\rm km~s^{-1}$). To image the line emissions, we used the CASA task \texttt{uvcontsub} to fit the continuum visibilities in the line-free channels with a zeroth-order polynomial. We then obtained continuum-subtracted cubes to be used in our line analyses. We produced line maps using the MFS mode in the channels fully enclosing the emission lines. The residuals of cube and map cleaning processes have been checked to avoid over-cleaning. 

Information about the project ID, synthesized beam, fluxes, and r.m.s noise of the line and continuum maps is reported in Tabs.~\ref{tab:res-line-obs}, \ref{tab:res-line} and \ref{tab:res-cont} for each observation analysed in this work. 

\begin{figure*}
    \centering
    \includegraphics[width=1\linewidth]{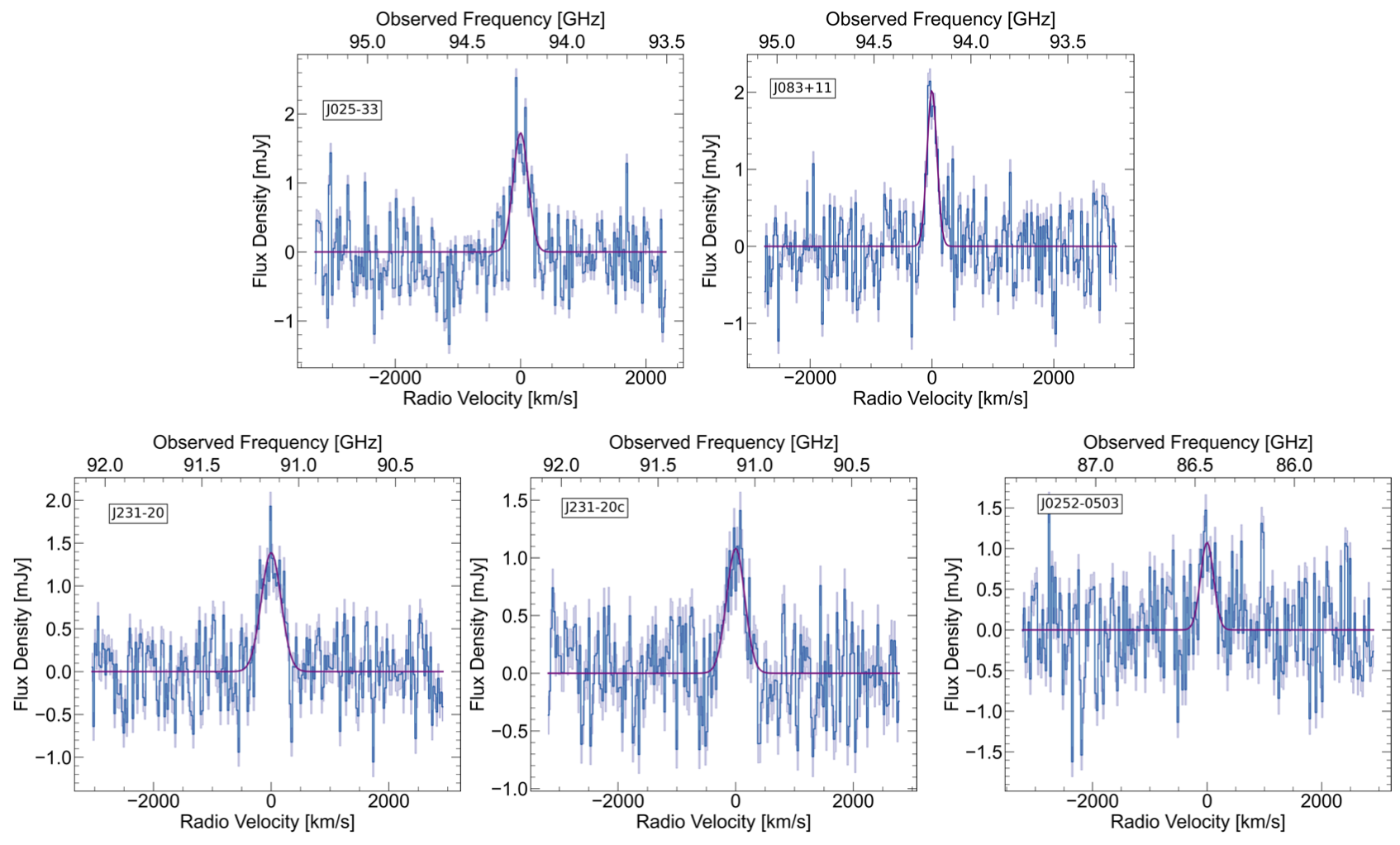}
    \caption{Spectra of CO(6-5) emission line. From left to right, top to bottom, observed spectra of the CO(6-5) emission line with their best-fitting Gaussian models (purple solid line) for QSOs J025-33, J083+11, J231-20 and its companion, and J0252-0503. The shaded region marks the $1\sigma$ noise level of the spectrum. All spectra have been extracted from a circular aperture enclosing the full line emission (see Fig.~\ref{fig:line-map} and Sect.~\ref{sec:analysis}).}
    \label{fig:spectra}
\end{figure*}

\section{Data analysis and results}
\label{sec:analysis}
In the following sections, we report the results of our proprietary data targeting the CO(6-5) emission line and its underlying continuum, as well as the results from archival published and unpublished data at higher frequencies. We also describe the methodology adopted for fitting the dust SED of our targets. Line luminosities are computed based on Eq. 1 in \citet{solomon2005}, and molecular gas masses are obtained from the CO(6-5) line luminosity assuming a luminosity ratio between the CO(6-5) and (1-0) transitions of $r_{\rm 6,1}=\rm CO(6-5)/CO(1-0)=0.92$ \citep[based on the CO Spectral line energy distribution of high-z QSOs, see][]{kaasinen2024}, and a conversion factor $\alpha_{\rm CO}=0.8~\rm M_\odot(K~km~s^{-1}~pc^2)^{-1}$ \citep{Bolatto2013, carilli2013}. This is a common approach for high-z QSOs and is consistent with the methodology adopted for the analysis of other HYPERION QSOs in previous works \citep[see][and refs therein]{tripodi2024}. Upper limits on the CO flux are derived assuming a FWHM of $300\rm ~km~s^{-1}$, which is the typical line width found for this line in $z\gtrsim 6$ QSOs (see also Tab.~\ref{tab:res-line}). All results are summarized in Tabs.~\ref{tab:res-line} and \ref{tab:res-cont}.\vspace{0.2cm}

\subsection{Continuum and CO(6-5) line emission}
\label{sec:cont-line}

Among the 10 HYPERION QSOs in our sample, 8 show detectable continuum emission (see Fig.~\ref{fig:cont-b3}) and 4 display significant CO(6-5) line emission (see Fig.~\ref{fig:line-map}). The continuum fluxes and deconvolved sizes have been computed by performing a 2D Gaussian fit on the continuum map with CARTA 2D fitting tools. 1D spectra have been extracted from a circular aperture centered on the source and enclosing the full line emission to maximise the S/N following the moment-0 morphology, with the following aperture radii: 1.0$''$ for J025-33, 1.2$''$ for J083+11, 0.95$''$ for J231-20, 0.85$''$ for J231-20c, 0.8$''$ for J0252-0503. The integrated line emission is derived integrating the 1D spectrum over the full line width (see Fig.~\ref{fig:spectra}).\\

\textbf{J029-36.} We find that the emission is barely resolved with a deconvolved size of $(0.37''\times0.68'')$ and has a total flux of $60\pm 20 ~\mu$Jy at 104 GHz. The CO(6-5) line emission is undetected in the continuum-subtracted data cube yielding a 3$\sigma$ upper limit on the CO(6-5) luminosity and gas mass of $L_{\rm CO(6-5)}<4.7\times 10^7~\rm L_\odot$ and $M_{\rm H_2}<4.0\times 10^9~\rm M_\odot$, respectively.

\textbf{J025-33.} We find that the emission is resolved with a deconvolved size of $(0.85''\times0.71'')$ and has a total flux of $80\pm 20 ~\mu$Jy at 99 GHz. The CO(6-5) emission line is detected with an integrated flux of $0.52\pm 0.02\rm ~Jy~km~s^{-1}$, and a FWHM of $283\pm 9\rm~km~s^{-1}$. This implies $L_{\rm CO(6-5)}=(2.0\pm 0.1)\times 10^8~\rm L_\odot$ and $M_{\rm H_2}=(1.65\pm0.05)\times 10^{10}~\rm M_\odot$.

While the results for the [CII] emission and its underlying continuum from ALMA B6 data have been already reported in the literature \citep{ansarinejad2022, decarli2018, venemans2020}, we homogeneously and consistently analysed unpublished observations in B5, B7 and B8. Continuum is detected in all bands, as shown in the top panels of Fig.~\ref{fig:cont-otherbands} and results are reported in Tab.~\ref{tab:res-cont}. \citet{venemans2020} report a [CII] size for J025-33 which is more compact by a factor of 2 than the CO(6-5) sizes measured in our work \citep[see Tab. 3 in][]{venemans2020}. This is mainly due to the higher resolution of their observations ($\sim 5$ times higher than ours), which yields to the loss of the fainter and more extended flux, due to the surface brightness dimming effect \citep[e.g.,][]{carniani2020}. 

\textbf{J083+11.} We find that the emission is resolved with a deconvolved size of $(0.75''\times0.81'')$ and has a total flux of $0.28\pm 0.04$ mJy at 101 GHz. The CO(6-5) emission line is detected with an integrated flux of $0.41\pm 0.02\rm ~Jy~km~s^{-1}$, and a FWHM of $190\pm 9\rm~km~s^{-1}$. This implies $L_{\rm CO(6-5)}=(1.6\pm 0.1)\times 10^8~\rm L_\odot$ and $M_{\rm H_2}=(1.31\pm0.05)\times 10^{10}~\rm M_\odot$. Furthermore, we note that there is a significant spatial offset of $\sim 1.3$ kpc between the continuum and line peak emissions. 

We also analysed an unpublished observation in B5, detecting continuum emission with flux of $2.69\pm0.10$ mJy as shown in the bottom right panel of Fig.~\ref{fig:cont-otherbands}. Results for available observations in  B4, B6, B7, and B8 are presented in \citet{andika2020, spilker2025}. 

\textbf{J011+09.} We find that the emission is resolved with a deconvolved size of $(0.7''\times0.8'')$ and has a total flux of $70\pm 20 ~\mu$Jy at 100 GHz. The CO(6-5) line emission is undetected in the continuum-subtracted data cube yielding a 3$\sigma$ upper limit on the CO(6-5) luminosity and gas mass of $L_{\rm CO(6-5)}<7.0\times 10^7~\rm L_\odot$ and $M_{\rm H_2}<5.9\times 10^9~\rm M_\odot$, respectively. 

\textbf{J231-20.} The continuum emission of J231-20 and its nearby companion (J231-20c) in B3 is shown in Fig.~\ref{fig:cont-b3}. We find that the QSO emission is resolved with a deconvolved size of $(0.5''\times0.9'')$ and has a total flux of $0.25\pm 0.04$ mJy at 97 GHz. The CO(6-5) line emission is detected with an integrated flux of $0.58\rm ~Jy~km~s^{-1}$ and FWHM$=395\pm 16\rm ~km~s^{-1}$. This implies $L_{\rm CO(6-5)}=(2.4\pm0.1)\times 10^8~\rm L_\odot$. J231-20 is one of the most studied QSOs at $z>6$, and multiple CO detection of transitions beyond 6-5 have been reported in the literature \citep{Pensabene2021}. Therefore, we also derive the gas mass by modelling the CO spectral line energy distribution (SLED), as reported in Sect.~\ref{sec:disc-cosled}. Instead, the results for its companion can be found in Tabs.~\ref{tab:res-cont} and \ref{tab:res-line}. The emission line map and spectrum of the CO(6-5) line are shown in Figs.~\ref{fig:line-map} and \ref{fig:spectra}, respectively. Interestingly, by comparing the gas and dust morphology, we find that the gas and dust emissions in the companion are offset by $\sim 0.5$ arcsec (i.e. $\sim 2.8$ kpc), and the gas appears to be more extended than the dust by more than a factor of 2 considering a region at S/N$>2$. Similarly to J025-33, \citep{venemans2020} report a [CII] size for J231-20 that is more compact by a factor of 10 than the CO(6-5) sizes measured in our work \citep[see Tab. 3 in][]{venemans2020}, mainly due to the higher resolution of their observations ($20$ times higher than ours). 

This QSO and its companion have been widely studied in the literature \citep{Pensabene2021, tripodi2024, sun2026}, both in continuum and line emissions. Therefore, further analysis of archival data is unnecessary for this source, since the cold dust properties are already well established \citep{tripodi2024}. Specifically, \citet{Pensabene2021} report the detection of the CO(7–6) and [CI] lines, yielding $M_{\rm H_2} = 5.9^{+0.8}_{-0.7} \times 10^{9} ~ \rm M_\odot$ in good agreement with our estimate from the CO(6-5) and with larger uncertainties. Furthermore, our observation confirms the J231-20 companion to be gas-rich, exhibiting half the molecular gas budget of the QSO host, within the range allowed by the previous upper limit \citep[i.e., $M_{\rm H_2}<1.8 \times 10^{10} ~\rm M_\odot$][]{decarli2017, venemans2020, Pensabene2021}. 

\textbf{J0244-5008.} The continuum emission of J0244-5008 in B3 is shown in Fig.~\ref{fig:cont-b3}. By performing a 2D Gaussian fit on the continuum map, we find that the emission is resolved with a deconvolved size of $(0.6''\times0.9'')$ and has a total flux of $40\pm 20 ~\mu$Jy at 93 GHz. The CO(6-5) line emission is undetected in the continuum-subtracted data cube, yielding a 3$\sigma$ upper limit on the CO(6-5) luminosity and gas mass of $L_{\rm CO(6-5)}<6.4\times 10^7~\rm L_\odot$ and $M_{\rm H_2}<5.3\times 10^9~\rm M_\odot$, respectively. 

We also analysed an unpublished observation in B6, targeting and detecting the [CII] line and its underlying continuum emission. The natural resolution of this observation is $\sim 0.1''$, much lower than our data in B3. Therefore, we manually downgraded the resolution performing the imaging with \texttt{uvtaper}=[0.5 arcsec] to match our B3 resolution. Indeed, it has been shown that high resolution observations ($\theta<0.1-0.2''$), if not sensitive enough, may loose $>20\%$ of the continuum flux from the more extended and fainter emitting regions \citep{wang2019b,tripodi2024}. By performing 2D Gaussian fit on the tapered continuum map (bottom right panel of Fig.~\ref{fig:cont-otherbands}), we find a flux of $0.99\pm0.07$ mJy at 230 GHz. 

\textbf{J0411-0907.} This source is undetected both in continuum at 93 GHz and CO(6-5) line emission, yielding a $3\sigma$ upper-limit of $<18~\mu$Jy on the continuum, and of $0.17~\rm Jy~km~s^{-1}$ on the line. The latter implies a limit luminosity and molecular gas mass of $L_{\rm CO(6-5)}<7.8\times 10^7~\rm L_\odot$ and $M_{\rm H_2}<6.4\times 10^9~\rm M_\odot$, respectively. We also analysed the publicly archival B6 observation, finding a continuum emission of $0.19\pm 0.06$ mJy, which is resolved with a size of $0.72''\times 0.91''$. 

\textbf{J0020-3653.} This source is undetected both in continuum at 93 GHz and in CO(6-5) line emission, yielding a $3\sigma$ upper-limit  of $<18~\mu$Jy on the continuum, and of $0.17~\rm Jy ~km~s^{-1}$ on the line. The latter corresponds to a limit luminosity and molecular gas mass of $L_{\rm CO(6-5)}<7.4\times 10^7~\rm L_\odot$ and $M_{\rm H_2}<6.1\times 10^9~\rm M_\odot$, respectively. We also analysed the publicly archival B6 observation, finding a continuum emission of $0.45\pm 0.10$ mJy, which is resolved with a size of $1.5''\times 2.5''$. 

\textbf{J0252-0503.} The continuum emission of J0252-0503 in B3 is shown in Fig.~\ref{fig:cont-b3}. By performing a 2D Gaussian fit on the continuum map, we find that the emission is resolved with a deconvolved size of $(0.9''\times0.7'')$ and has a total flux of $0.11\pm 0.02$ mJy at 93 GHz. The CO(6-5) line emission is detected with an integrated flux of $0.31\pm 0.02\rm ~Jy~km~s^{-1}$ and FWHM$=268^{+27}_{-24}\rm ~km~s^{-1}$. This implies $L_{\rm CO(6-5)}=(1.4\pm0.1)\times 10^8~\rm L_\odot$ and $M_{\rm H_2}=(1.1\pm0.1)\times 10^{10}~\rm M_\odot$. Similarly as J083+11, J0252-0503 also displays a spatial offset of $\sim 0.75$ kpc between the continuum and line peak emissions. 

\citet{salvestrini2025} also present an analysis of CO(6–5) line emission and its underlying continuum from NOEMA observations. While their estimates of the continuum emission at $\sim 100$ GHz are in good agreement with ours in B3, they do not detect any line emission due to lower sensitivity than our observation. Their upper limit on the molecular gas mass is fully consistent with our estimate.

\textbf{J0038-1527.} The continuum emission of J0038-1527 in B3 is shown in Fig.~\ref{fig:cont-b3}. By performing a 2D Gaussian fit on the continuum map, we find that the emission is resolved with a deconvolved size of $(0.63''\times0.99'')$ and has a total flux of $30\pm 10~\mu$Jy at 92 GHz. The CO(6-5) line emission is undetected in the continuum-subtracted data cube yielding a 3$\sigma$ upper limit on the CO(6-5) luminosity and gas mass of $L_{\rm CO(6-5)}<6.3\times 10^7~\rm L_\odot$ and $M_{\rm H_2}<5.3\times 10^9~\rm M_\odot$, respectively. Both our estimates of continuum and line emission are fully consistent with previous upper limits placed by \citet{salvestrini2025} with lower sensitivity NOEMA observations.

\begin{table*}
\centering
\caption{CO(6--5) line and molecular gas properties.}
\begin{tabular}{lccccccc}
\hline
Target 
& Source Size
& $\nu_{\rm obs,line}$ 
& $z_{\rm CO(6-5)}$ 
& \makecell{Int. Flux} 
& \makecell{FWHM} 
& \makecell{$L_{\mathrm{CO(6-5)}}$} 
& \makecell{$M_{\rm H_2}$} \\
& [arcsec$^2$]
& [GHz] 
& 
& [Jy km/s] 
& [km/s] 
& $[10^8\,L_{\odot}]$ 
& $[10^{10}\,M_{\odot}]$ \\
\hline
J029-36 
& -- 
& -- 
& -- 
& -- 
& -- 
& $<0.47$ 
& $<0.40$ \\

J025-33 
& $0.63 \times 0.66$
& $94.234 \pm 0.005$ 
& $6.3379^{+0.0001}_{-0.0001}$ 
& $0.52^{+0.02}_{-0.02}$ 
& $283\pm 9$ 
& $2.01\pm 0.06$ 
& $1.65\pm 0.05$ \\

J083+11 
& $0.55 \times 0.84^\dag$
& $94.204 \pm 0.004$ 
& $6.3404\pm 0.0001$ 
& $0.41\pm 0.02$ 
& $190\pm 9$ 
& $1.59\pm 0.07$ 
& $1.31\pm 0.05$ \\

J011+09 
& -- 
& -- 
& -- 
& -- 
& -- 
& $<0.70$ 
& $<0.59$ \\

J231-20 
& $0.95 \times 0.68$
& $91.144 \pm 0.004$ 
& $6.5867\pm 0.02$ 
& $0.58\pm 0.02$ 
& $395\pm 16$ 
& $2.39\pm 0.10$ 
& $2.65^{+2.06}_{-0.84}$ \\

J231-20c 
& $0.48 \times 1.04$
& $91.096 \pm 0.006$ 
& $6.5903\pm 0.0003$ 
& $0.42\pm 0.02$ 
& $366\pm 21$ 
& $1.72\pm 0.10$ 
& $1.41\pm 0.08$ \\

J0244-5008 
& -- 
& -- 
& -- 
& -- 
& -- 
& $<0.64$ 
& $<0.53$ \\

J0411-0907
& -- 
& -- 
& -- 
& -- 
& -- 
& $<0.78$ 
& $<0.64$ \\

J0020-3653
& -- 
& -- 
& -- 
& -- 
& -- 
& $<0.74$ 
& $<0.61$ \\

J0252-0503 
& $0.61 \times 1.01$
& $86.438 \pm 0.007$ 
& $6.9996\pm 0.03$ 
& $0.31\pm 0.02$ 
& $268^{+27}_{-24}$ 
& $1.37\pm 0.11$ 
& $1.13 \pm 0.10$ \\

J0038-1527 
& -- 
& -- 
& -- 
& -- 
& -- 
& $<0.63$ 
& $<0.53$ \\
\hline
\end{tabular}
\flushleft
\footnotesize{
{\bf Notes.} Columns: target name, deconvolved source size, observed frequency of the CO(6--5) emission line, CO-based redshift, integrated line flux, FWHM, CO(6--5) luminosity, and molecular gas mass. The line frequencies, redshifts, integrated fluxes, and FWHM values are derived from the best-fit Gaussian model of the emission line. 
$^\dag$ J083+11 has been fitted with two Gaussian components; we report the size of the primary one, centered on the QSO position, since the secondary component is not resolved.
}
\label{tab:res-line}
\end{table*}

\subsection{CO(6-5) line moments}

\begin{figure}
    \centering
    \includegraphics[width=1\linewidth]{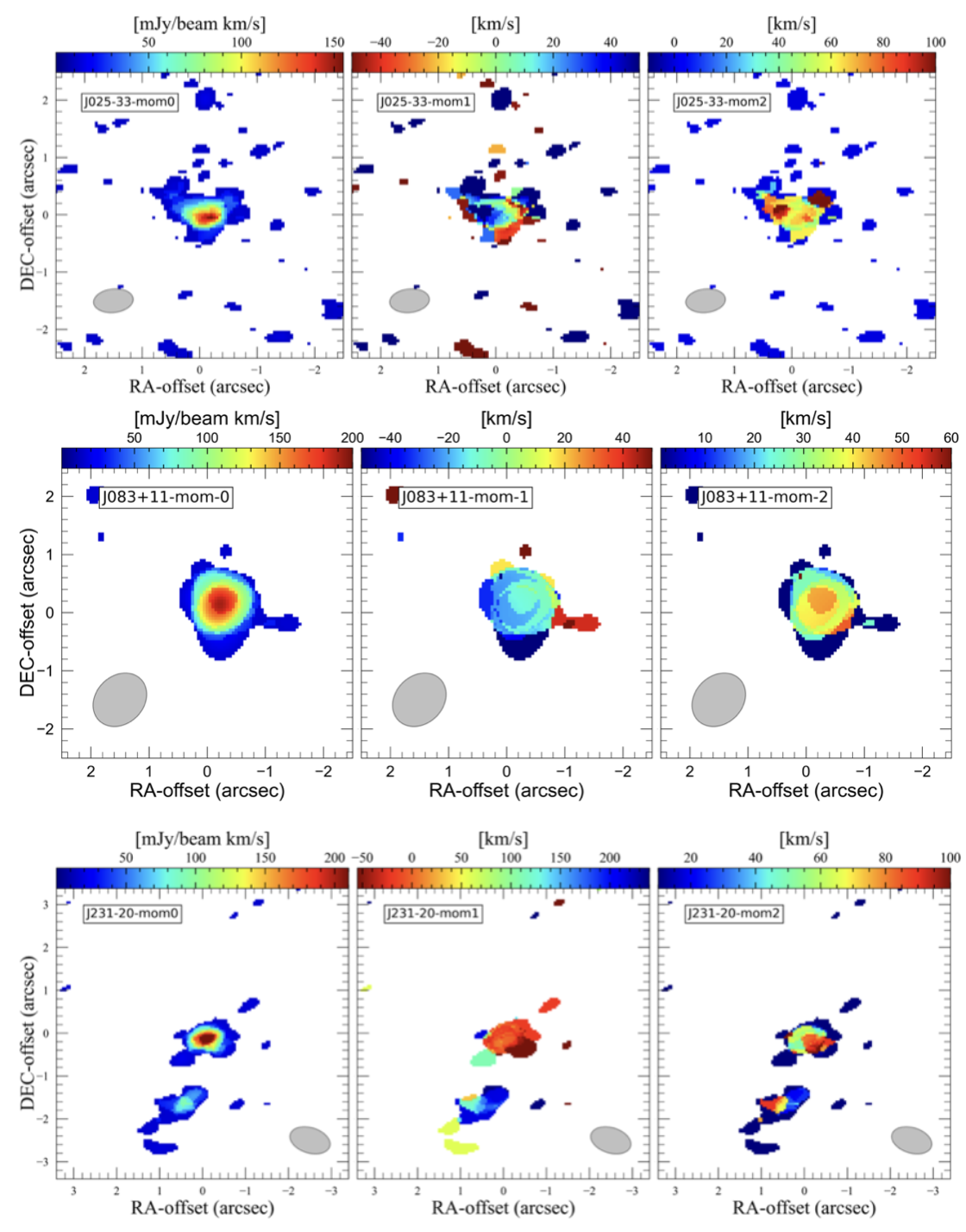}
    \caption{Moment maps of J025-33, J083+11, J231-20 (top to bottom). From left to right: integrated flux (zero-th moment map), mean velocity map (first moment map), and velocity dispersion map (second moment map) of continuum-subtracted data cube centered on the CO(6-5) line emission. The beam is shown in the bottom corner of each panel as gray circle.}
    \label{fig:moments}
\end{figure}

In Fig.~\ref{fig:moments}, we show the moment maps for J025-33, J083+11, and J231-20 produced with \texttt{immoments} CASA task using a $3\sigma$ threshold. Unfortunately, J0252 has a too low S/N ratio to produce reliable moments maps.

As expected, the zero-th moment maps of all sources provides similar information on the source morphology as the line emission maps. J025-33's velocity map shows a clear velocity gradient, aligned East-West, which implies the presence of a rotating gas disk. 

Interestingly, the velocity map of J231-20 also reveals a pronounced velocity gradient ($\Delta v\sim 250~\rm km~s^{-1}$) but between the source and its companion, likely indicating relative motion between the two objects, only partially consistent with the redshift difference between the two sources ($\Delta z=0.0039$, implying $\Delta v=160~\rm km~s^{-1}$). Moreover, the peak of the velocity dispersion of the two sources appears to be displaced from the center, and it is suggestive of an ongoing interaction between the QSO and its companion galaxy. Unfortunately, the low angular resolution of our observation prevents us performing further analysis, since we are not able distinguish any internal motion in the velocity map, and to constrain with high precision the peak position of the velocity dispersion. 

The velocity map of J083+11, in contrast, does not show a clear velocity gradient. This, combined with the high velocity dispersion observed in the second moment map, suggests that the source is dispersion-dominated, with the gas kinematics likely governed by turbulent motions rather than ordered rotation. However, even in this case, beam-smearing effects limit our ability to draw definitive conclusions on the cold-gas kinematics of the QSO host galaxy.

\subsection{[NII]$\lambda ~205\mu$m emission line}
\label{sec:NII-analysis}

\begin{figure}
    \centering
    \includegraphics[width=0.95\linewidth]{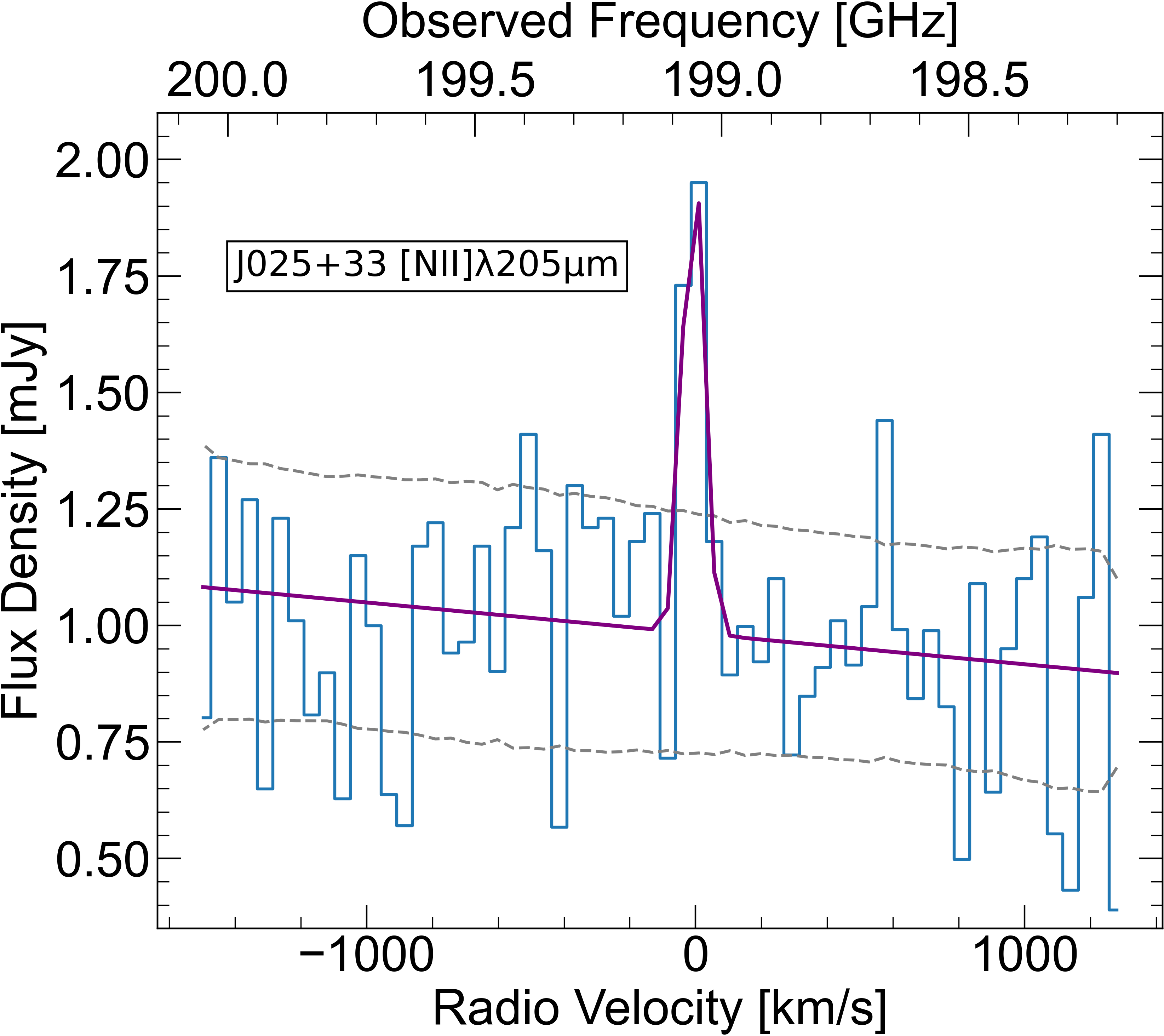}\\
    \includegraphics[width=0.95\linewidth]{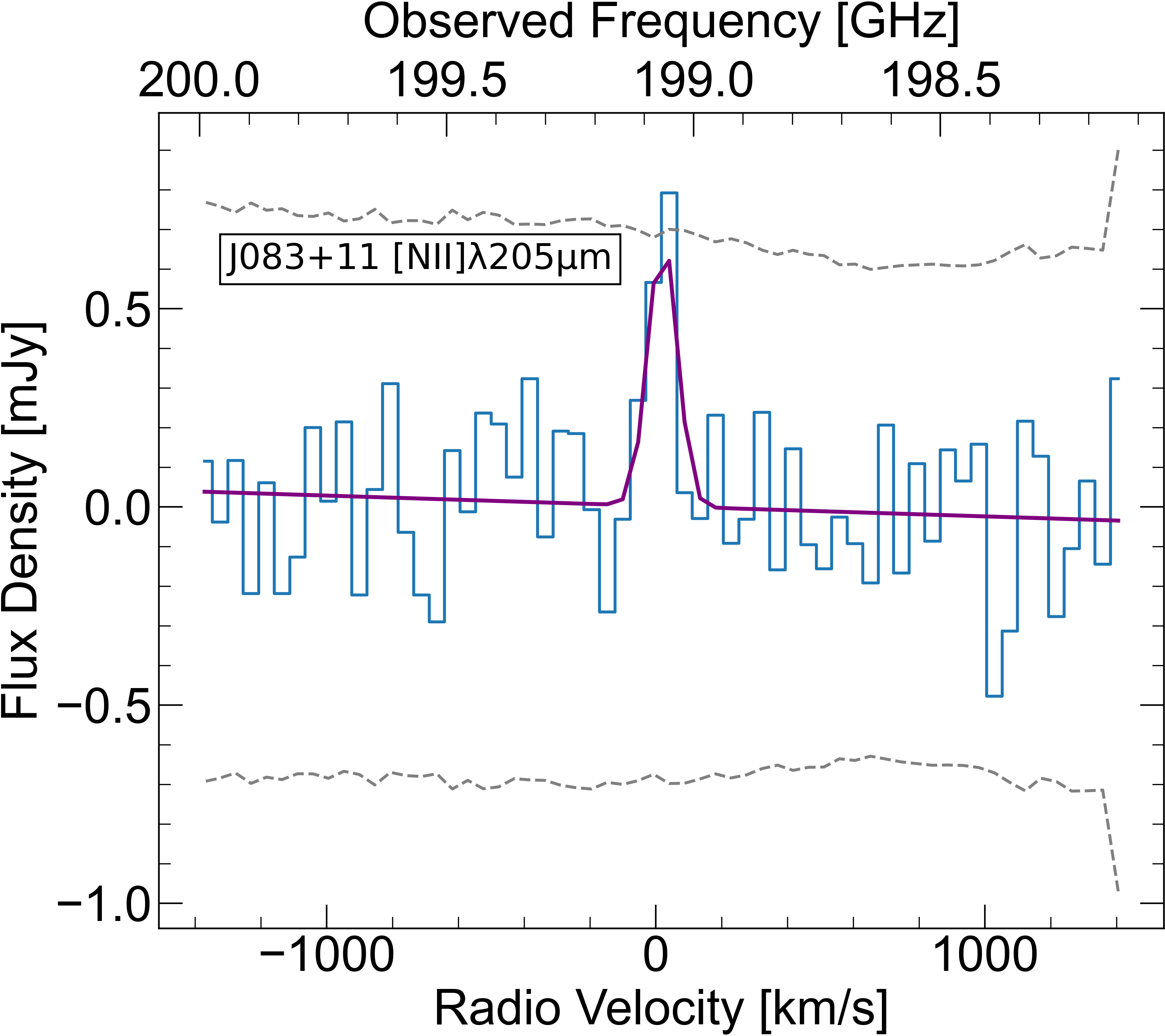}
    \caption{Spectra of [NII]$\lambda205\mu$m emission line in B5 for J025+33 (top) and J083+11 (bottom). The best-fit model (Gaussian + first order polynomial) is shown as a purple line. Gray dashed lines mark the noise level.}
    \label{fig:NII}
\end{figure}

For QSOs J025-33 and J083+11, we detected line emission in B5 arising from [NII]$\lambda205\mu$m. Fig.~\ref{fig:NII} show the observed spectra and relative Gaussian model of [NII] for both sources. We modelled the underlying continuum as a first order polynomial. The [NII]$\lambda205\mu$m of J025-33 is detected at S/N$\sim 5$ over the integrated line, and has an integrated flux of $F_{\rm [NII]\lambda205\mu m}= 0.09\pm 0.02\rm ~Jy~km~s^{-1}$, FWHM$=74_{-25}^{+30}~\rm km~s^{-1}$, corresponding to a luminosity of $L_{\rm [NII]\lambda205\mu m}=(7.0\pm 2.0)\times 10^{7}~\rm L_\odot$. For J083+11 the data are much noisier than for J025-33, and thus we are able to detect [NII]$\lambda~205\mu$m at S/N$\sim 3$ over the integrated line, with an integrated flux of $F_{\rm [NII]\lambda205\mu m}= 0.07^{+0.03}_{-0.02}\rm ~Jy~km~s^{-1}$, FWHM$=103_{-40}^{+70}~\rm km~s^{-1}$, corresponding to a luminosity of $L_{\rm [NII]\lambda205\mu m}=(6.1^{+2.3}_{-1.9})\times 10^{7}~\rm L_\odot$.

We did not detect any [NII]$\lambda122\mu$m emission in B7 for J025-33. Assuming the same line width of the [NII]$\lambda205\mu$m line (FWHM$=74~\rm km~s^{-1}$), we derived a 3$\sigma$ flux upper limit of $<0.14~\rm Jy ~km~s^{-1}$, corresponding to a luminosity of $L_{\rm [NII]\lambda122\mu m}<1.9\times 10^{8}~\rm L_\odot$. Similarly, no [NII]$\lambda122\mu$m emission was reported in the analysis of the B7 observation of J083+11 in \citet{spilker2025}.

Given the $3\sigma$ upper limit on the line luminosity ratio [NII]$\lambda122\mu$m/$\lambda205\mu$m$\lesssim3$ for J025-33, we derive an upper limit on the gas electron density of $<100\rm ~cm^{-3}$ \citep[see e.g. Fig. 2 in][for the density dependence of the line ratios]{herrera-camus2016}.

\subsection{[CII] emission of J0244-5008}
\label{sec:cii-J0244}

\begin{figure*}
    \centering
    \includegraphics[width=0.27\linewidth]{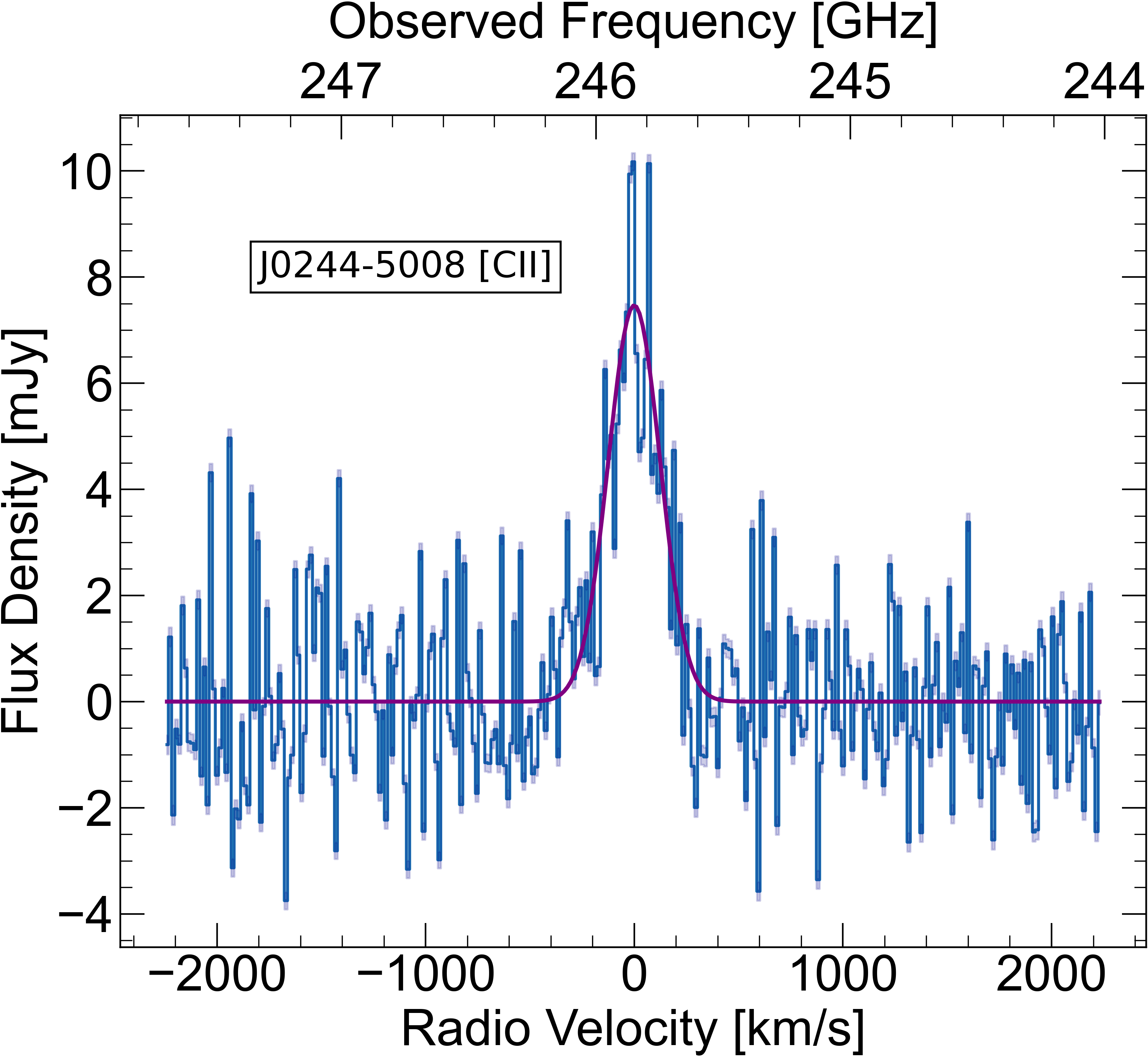}
    \includegraphics[width=0.7\linewidth]{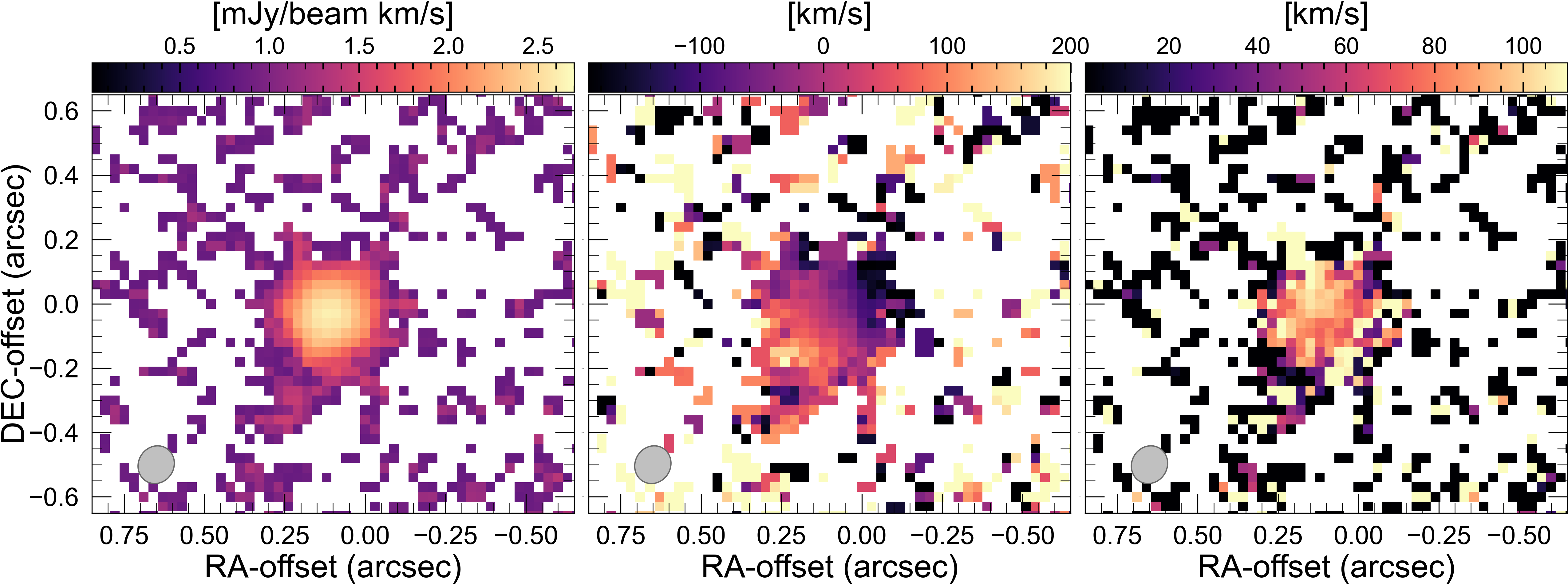}
    \caption{Left: observed spectrum of the [CII] emission line of QSO J0244-5008 (blue) and best-fitting model as a purple solid line. Centrals and right: 0-th, 1-st and 2-nd moment maps of the [CII] emission line of J0244-5008. The 0-th moment is shown in logarithmic scale. The clean beam size is reported in the lower left corner as a gray ellipse.}
    \label{fig:CII-J0244}
\end{figure*}

We analysed the 0.1$''$-resolution observation targeting the [CII] emission line of QSO J0244-5008 to study its properties and derive the dynamical mass of the system. Data have been reduced and cleaned as described in Sect.~\ref{sec:obs} using a natural weighting scheme. The final continuum subtracted data cube has an r.m.s of $0.18$ mJy/beam in a 15$\rm km~s^{-1}$ channel, and a clean beam size of $0.11''\times 0.12''$. Fig.~\ref{fig:CII-J0244} presents the [CII] spectrum extracted in a circular aperture of 0.5$''$ radius centered on the source, and the line moment map created imposing a $2.5\sigma$ threshold. By performing a 2D gaussian fit on the line map, we obtained a [CII] size of $0.18''\times 0.18''$. By modelling the observed spectral emission as a single Gaussian, we obtained an integrated flux of $2.33\pm 0.02\rm ~Jy~km~s^{-1}$, a FWHM$=293\pm 3\rm ~km~s^{-1}$, corresponding to a luminosity of $L_{\rm [CII]}=(2.70\pm 0.02)\times 10^9~\rm L_\odot$ at $z_{\rm [CII]}=6.73050\pm 0.00003$. The velocity map shows a clear velocity gradient of $\Delta v\sim 400\rm ~km~s^{-1}$, suggesting the presence of a rotation disk. 

To derive a robust estimate of the dynamical mass from the analysis of the [CII] kinematics, we derive the rotation curve via kinematic modelling of the observed data with $^{\rm 3D}$Barolo.
$^{\rm 3D}$Barolo fits disc models directly to the 3D cube, considering the effects of spectral and spatial resolution (beam smearing). The galaxy is divided into a series of rings, each one described by nine parameters: coordinates of the kinematic center $(x_0,y_0)$, position angle (PA), inclination ($i$), vertical thickness ($z_0$), systemic velocity ($V_{\rm sys}$), rotation velocity ($V_{\rm rot}$), radial velocity ($V_{\rm rad}$) and velocity dispersion ($\sigma_v$). 

For $(x_0,y_0)$, we choose the position of the peak emission in the 0-th moment while for $V_{\rm sys}$, PA and $i$, we made an initial run letting those parameters free, and we found $V_{\rm sys}\sim 27\rm ~km~s^{-1}$, PA$=130^\circ$, and $i=35^\circ$. We fixed $z_0$ to 300 pc, but this parameter has a negligible effect on our results because  our spatial resolution is $\sim$900 pc and the disc is nearly face-on. Since the kinematic major axis is perpendicular to the minor axis, we fixed the radial velocity to zero. 

Given the 0.1$''$ resolution of the data, we adopted a radial sampling of 0.05$''$ to avoid aggressive over-sampling and we used 5 rings. We then run $^{\rm 3D}$Barolo with $v_{\rm rot}$ and $\sigma_v$ as free parameters. $^{\rm 3D}$Barolo provides asymmetric uncertainties ($\delta_+, \delta_{-}$), corresponding to a variation of 5\% of the residual from the global minimum. We treat them as $1\sigma$ uncertainties and took the mean value of ($\delta_+, \delta_{-}$) to compute symmetric uncertainties.

Fig.~\ref{fig:pv} shows the observed position-velocity (PV) diagram along the major and minor axis and the best-fitting model derived from $^{\rm 3D}$Barolo (purple contours and squares). We find $v_{\rm rot}=185\pm 40\rm ~km~s^{-1}$ at $r=0.2''$, implying a dynamical mass of $M_{\rm dyn}=(9.8\pm4.0)\times 10^9~\rm M_\odot$.

\subsection{Cold dust SED modelling}

\begin{figure}
    \centering
    \includegraphics[width=0.9\linewidth]{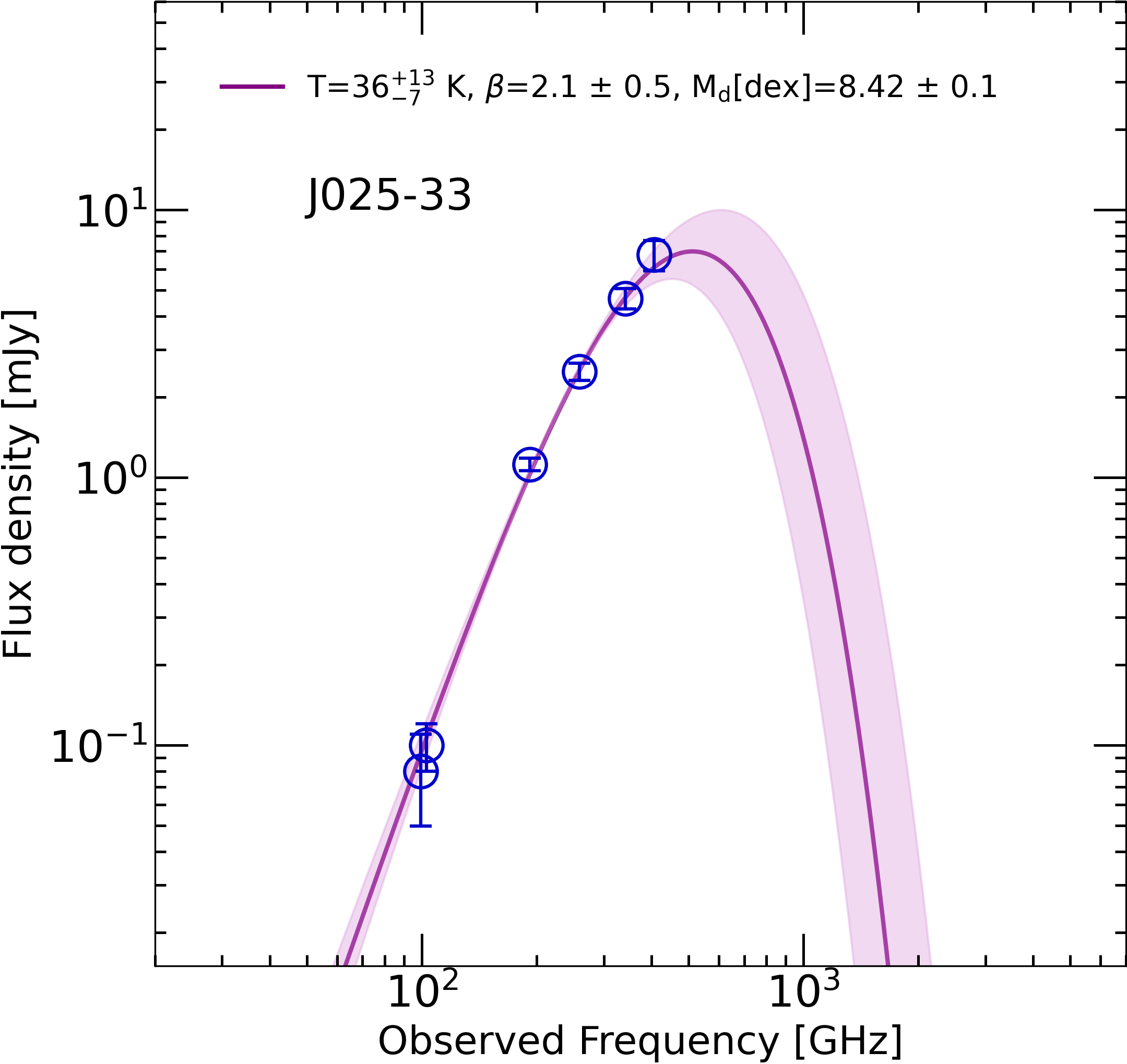}\\[0.1cm]
    \includegraphics[width=0.9\linewidth]{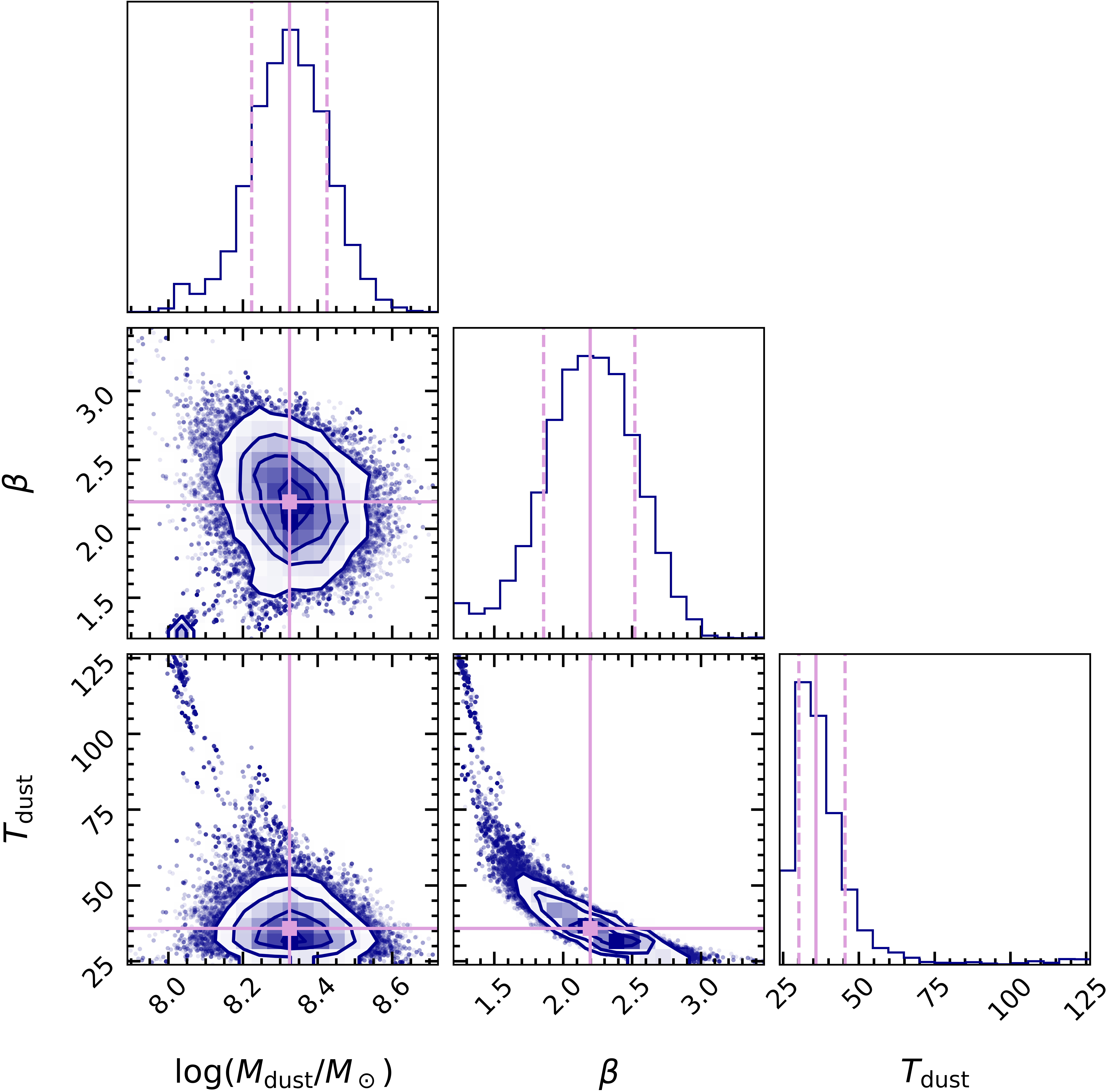}\\
    \caption{Cold dust SED fitting results for J025-33. Top: ALMA observed photometric data are plotted as blue edged circles with errorbars, and NOEMA data as blue edged squares. The best-fitting model and its uncertainty are shown as a solid purple line with shaded region. Bottom: corner plot showing the posterior distributions of the best-fitting parameters.}
    \label{fig:sed-J025}
\end{figure}

\begin{figure}
    \centering
    \includegraphics[width=0.9\linewidth]{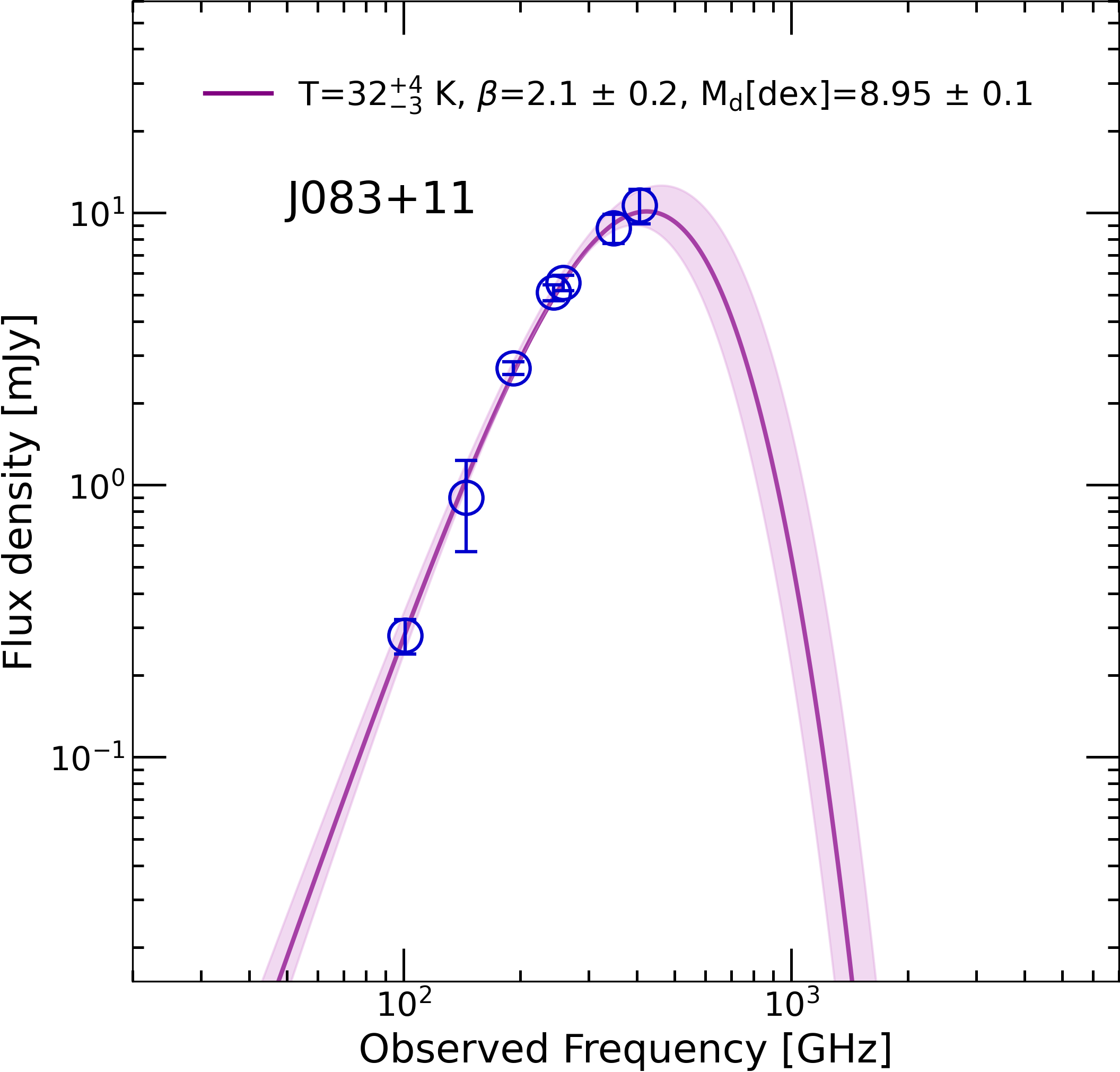}\\[0.1cm]
    \includegraphics[width=0.9\linewidth]{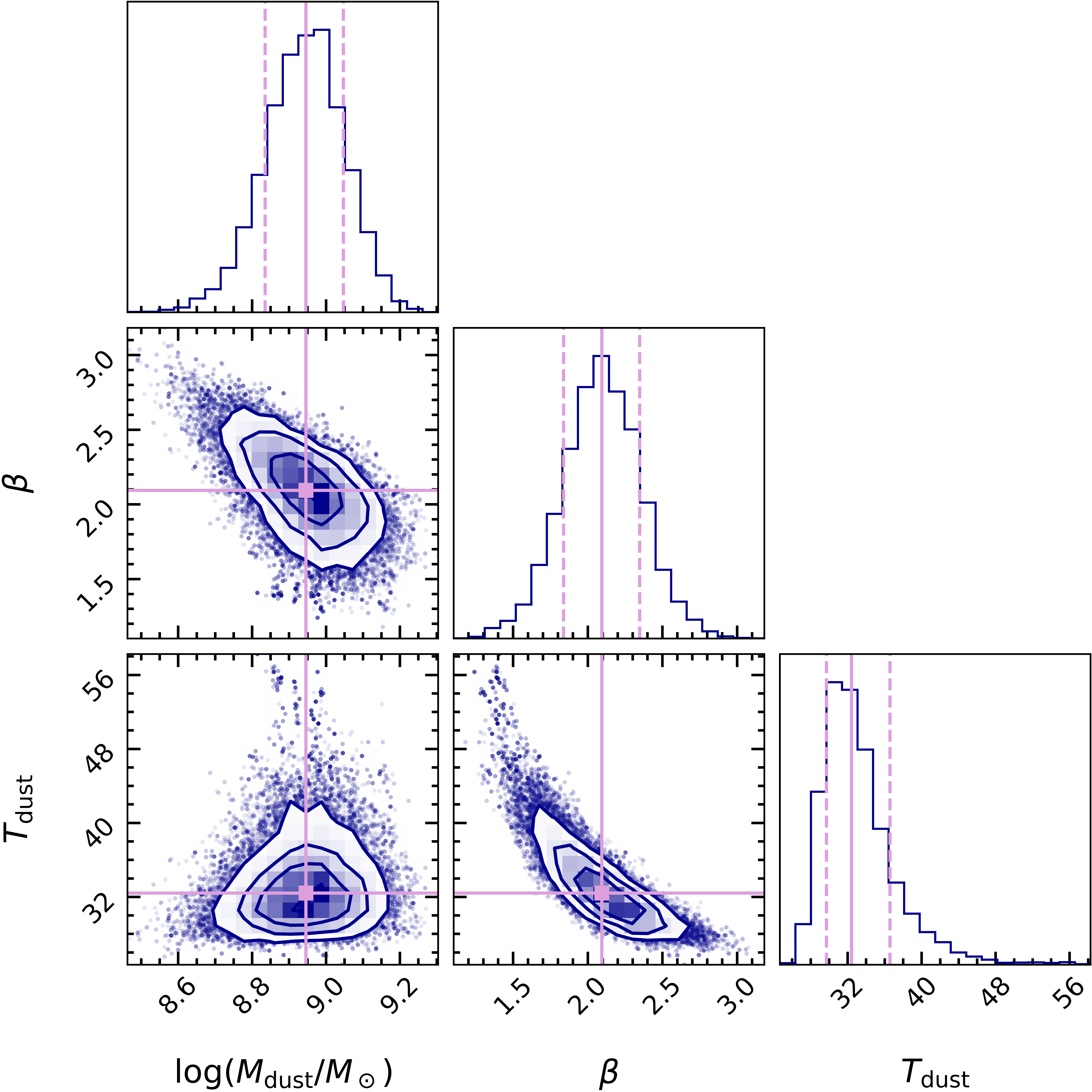}\\
    \caption{Same as Fig.~\ref{fig:sed-J025} for J083+11.}
    \label{fig:sed-J083}
\end{figure}

\begin{table*}[]
    \centering
    \caption{Results of the cold dust SED modelling with EOS-Dustfit.}
    \begin{tabular}{c|ccccc|c}
    \hline
       QSO   & $T_{\rm dust}$ & $\log(M_{\rm dust}/M_\odot)$ & $\beta$ & $L_{\rm TIR}$ & SFR & GDR \\
                & [K]            &  & & [$10^{12}~L_\odot$] & [$M_\odot~\rm yr^{-1}$] \\
                \hline
      J029-36   & $>35$ & $<7.7$ & $<2.9$ & $>2.1$ & $>310$ & --\\ 
      J025-33   & $36^{+13}_{-7}$ & $8.4\pm0.1$ & $2.1\pm0.5$ & $4.9_{-1.6}^{+3.7}$ & $739_{-240}^{+550}$ & $66^{+19}_{-15}$ \\[0.1cm] 
      J083+11   & $32_{-3}^{+4}$ & $8.9\pm0.1$ & $2.1\pm0.2$ & $5.7_{-1.2}^{+2.0}$ & $848_{-180}^{+300}$ & $16_{-4}^{+5}$\\[0.1cm] 
      J011+09   & $>35$ & $<8.3$  & $<2.0$ & $>1.9$ & $>300$ & --\\ 
      J0244-5008 & $>35$ & $<7.6$  & $<2.6$ & $>2.3$ & $>310$ & --\\
      J0411-0907 &  $>35$ & $<8.2$  & $>0.9$ & -- & -- & --\\
      J0020-3653 &  $>35$ & $<7.7$  & $>1.8$ & -- & -- & --\\
      J0252-0503 & $>35$ & $<8.9$  & $<1.3$ & $>1.27$ & $>190$ & $>14$\\ 
      J0038-1527 & $>35$ & $<7.7$  & $<2.6$ & $>2.4$ & $>360$ & --\\ 
      \hline
      \hline
    \end{tabular}
    \flushleft{\footnotesize {\bf Notes.} Columns: target name; dust temperature; dust mass; dust emissivity index; TIR luminosity; SFR computed assuming Kroupa IMF. Lower limits on $L_{\rm TIR}$ and SFR are derived considering the best-fitting models at $T_{\rm dust}=35$ K.}
    \label{tab:res-sed}
\end{table*}

To investigate the cold dust properties of our targets, we performed cold dust SED fitting of the flux densities reported in Tab.~\ref{tab:res-cont} using EOS-Dustfit fitting tool, considering the tapered fluxes for the sources with very high resolution observations (see Sect.~\ref{sec:cont-line}). We do not model the SED of QSO J231-20 since it has already been studied in \citep{tripodi2024}, and our new B3 observation does not provide further constraints. 

\noindent EOS-Dustfit is a publicly available tool for fitting the cold dust SED of galaxies (\href{https://github.com/roberta96/EOS-Dustfit}{https://github.com/roberta96/EOS-Dustfit}\footnote{Details about the model can be found on the GitHub page and in \citet{tripodi2024}.}). The SED is modelled as a modified blackbody (MBB) in the optically thick regime and accounting for the contribution of the CMB heating \citep{tripodi2024}. This model allows for a maximum of three free parameters: dust temperature, $T_{\rm dust}$, dust mass, $M_{\rm dust}$, and emissivity index, $\beta$. EOS-Dustfit explores the parameter space for each SED using a Markov chain Monte Carlo (MCMC) algorithm implemented in the \textit{emcee} package \citep{foreman2013}. We treat continuum emission upper limits as $1\sigma$ detection with $2\sigma$ associated error \citep[see also][]{salvestrini2025, witstok2022, ronconi2024}. 

We let $T_{\rm dust},M_{\rm dust}$, and $\beta$ free for the SED modelling of J025-33 and J083+11, given that the available observations ensure a good coverage of both the Rayleigh-Jeans regime and the peak region of the dust emission. The former regime directly affects the determination of $M_{\rm dust}$, and $\beta$, while the latter of $T_{\rm dust}$. For J0252-0503, which also has ancillary NOEMA continuum detections at 1mm \citep{salvestrini2025}, we let $M_{\rm dust}$, and $\beta$ free to vary and we fix the dust temperature at three different values, $T_{\rm dust}=35,54,70$ K. For the other targets, where only two photometric data points are available in the Rayleigh-Jeans regime (i.e., B3 and B6), we explored the parameter space of viable SED model solutions in a more tailored way. We let $M_{\rm dust}$ be the only free parameter, and we fit the data for every combination of $\beta\in[0.5,3.0]$ with a step of 0.1 and $T_{\rm dust}=35,54,70$ K. The choice of $T_{\rm dust}$ values is motivated by (1) $T_{\rm dust}=35$ K being the lowest dust temperature observed in QSO hosts at $z>6$ \citep{tripodi2024, costa2026}, (2) $T_{\rm dust}=54$ K being the average temperature found in QSO hosts at  $z>6$ \citep{tripodi2024}, and (3) $T_{\rm dust}=70$ K to explore the high temperature regime. For each fixed $T_{\rm dust}$, we choose as the best-fitting model the one with $\beta$ and $M_{\rm dust}$ that minimize the $\chi^2$. The results are shown in Figs.~\ref{fig:sed-J025}, \ref{fig:sed-J083}, \ref{fig:sed} and are reported in Tab.~\ref{tab:res-sed}. 

Dust properties are well constrained for J025-33 and J083+11, thanks to the good ALMA coverage of their SED from low to high frequency. Our results are consistent with those of \citet{costa2026}, who analysed the same sources but could not include the Band 3 observation of J083+11, which in our study provides tighter constraints on its dust properties. For all the other targets, we can only obtain an upper limit on the dust mass, given the available observations, and assuming that $T_{\rm dust}$ cannot be much lower than 35 K. This scenario is physically reasonable considering that the CMB temperature is already $>20$ K at $z>6$. 

Total infrared (TIR) luminosities and SFR are derived by integrating the best-fitting model from 8 to 1000$\mu$m rest-frame, and assuming a Kroupa IMF \citep{kroupa2003}. A Salpeter or Chabrier IMF \citep{salpeter1955, chabrier2003} would imply an SFR that is higher by a factor of 1.16 or lower by a factor 0.67, respectively. In QSO hosts, the SFR is usually corrected for the AGN contribution by a factor of 50\%, on average \citep{duras2017, schneider2015, tripodi2024}. However, J025-33 and J083+11 exhibit dust temperatures that are lower than usually seen in similar sources \citep[see][and Sect.~\ref{sec:disc}]{tripodi2024, costa2026} therefore, we report the uncorrected values in Tab.~\ref{tab:res-sed} and we discuss possible caveats in Sect.~\ref{sec:disc}. Since the dust temperature is uncertain for the other targets, we do not apply any correction on their SFR, which is derived from the best-fitting model at $T_{\rm dust}=35$ K and, therefore, treated as a lower limit.

\section{Discussion}
\label{sec:disc}

\subsection{[NII]$\lambda 205\,\mu$m emission in HYPERION QSOs}
\label{sec:disc-NII}

The far-infrared [NII] fine-structure lines are powerful tracers of the ionized phase of the ISM. Since the ionization potential of nitrogen is higher than that of hydrogen, [NII] emission arises only from ionized gas, unlike [CII], which can originate from both neutral and ionized phases. In particular, the [NII]$\lambda122\,\mu$m/[NII]$\lambda 205\,\mu$m ratio is a well-established electron-density diagnostic, being sensitive to densities in the range $n_{\rm e}\sim10$--$1000~{\rm cm^{-3}}$ \citep{herrera-camus2016}. The two transitions have different critical densities, $n_{\rm crit}\simeq290~{\rm cm^{-3}}$ for [NII]$\,122\,\mu$m and $n_{\rm crit}\simeq44~{\rm cm^{-3}}$ for [NII]$\,205\,\mu$m, making their ratio particularly useful for probing the low-excitation ionized medium. At high redshift, however, detections of [NII] remain rare, especially in QSO host galaxies. Previous measurements in the HYPERION context include [NII]$\,205\,\mu$m detections in J231-20 and its companion \citep{Pensabene2021}, an [NII]$\,122\,\mu$m detection in the QSO J183+05 \citep{decarli2023}, while \citet{meyer2022} reported a tentative [NII]$\,205\,\mu$m detection in J1148+5251. Our detections in J025-33 and J083+11 therefore add two new systems to the still limited sample of $z>6$ QSO hosts with ionized-gas constraints from [NII].

For J025-33, we detect [NII]$\,205\,\mu$m at ${\rm S/N}\sim5$, while for J083+11, the detection is more tentative, at ${\rm S/N}\sim3$ (see Sect.~\ref{sec:NII-analysis}). Unfortunately, [NII]$\,122\,\mu$m is not detected in either source, preventing a direct estimate of $n_{\rm e}$ from the [NII]$\,122/{\rm [NII]}\,205$ ratio. Nevertheless, the position of the two sources in the SFR--$L_{\rm [NII]205}$ plane provides indirect information on the physical conditions of the ionized gas.

\begin{figure*}
    \centering
    \includegraphics[width=0.44\linewidth]{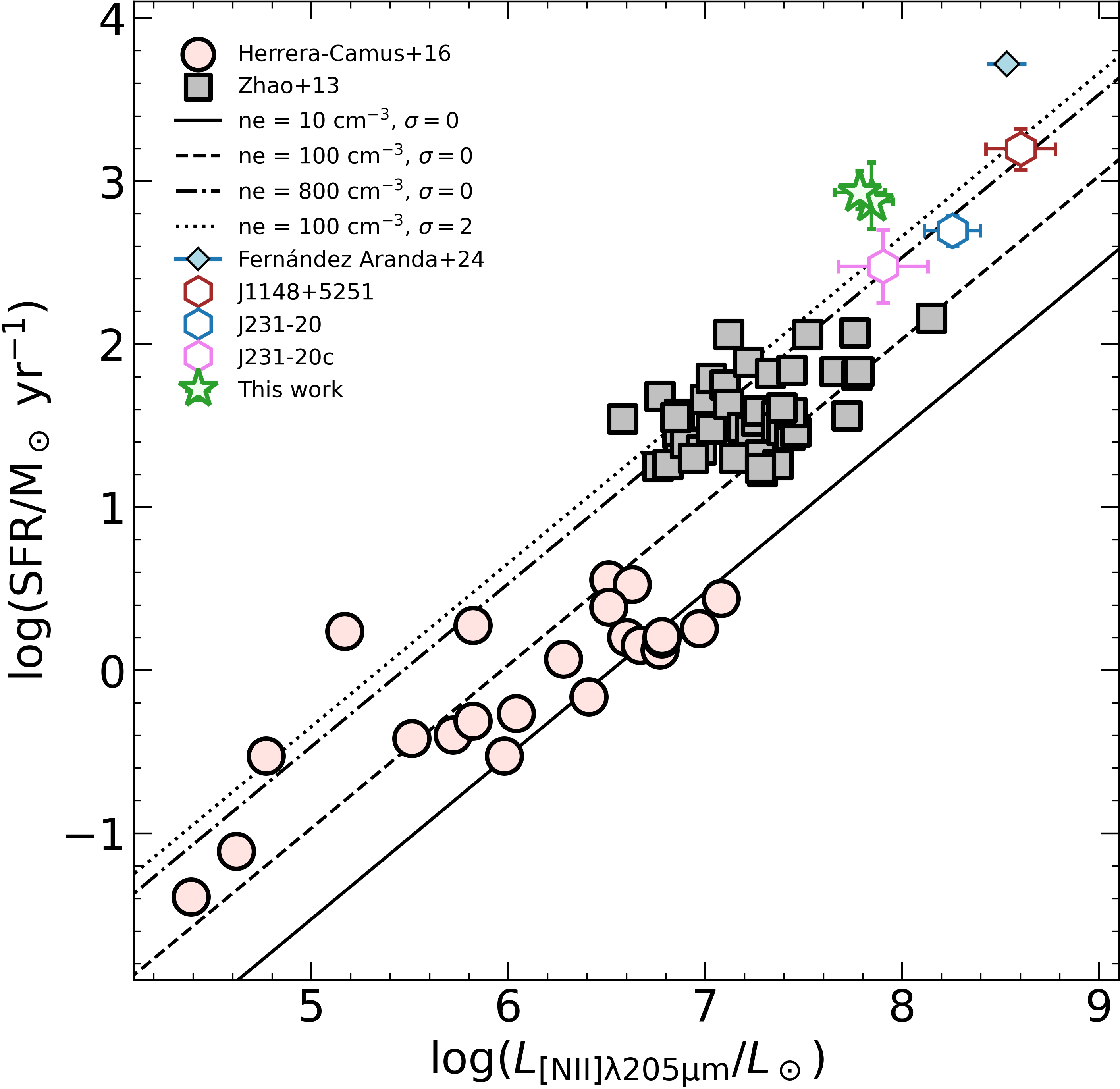}
    \includegraphics[width=0.45\linewidth]{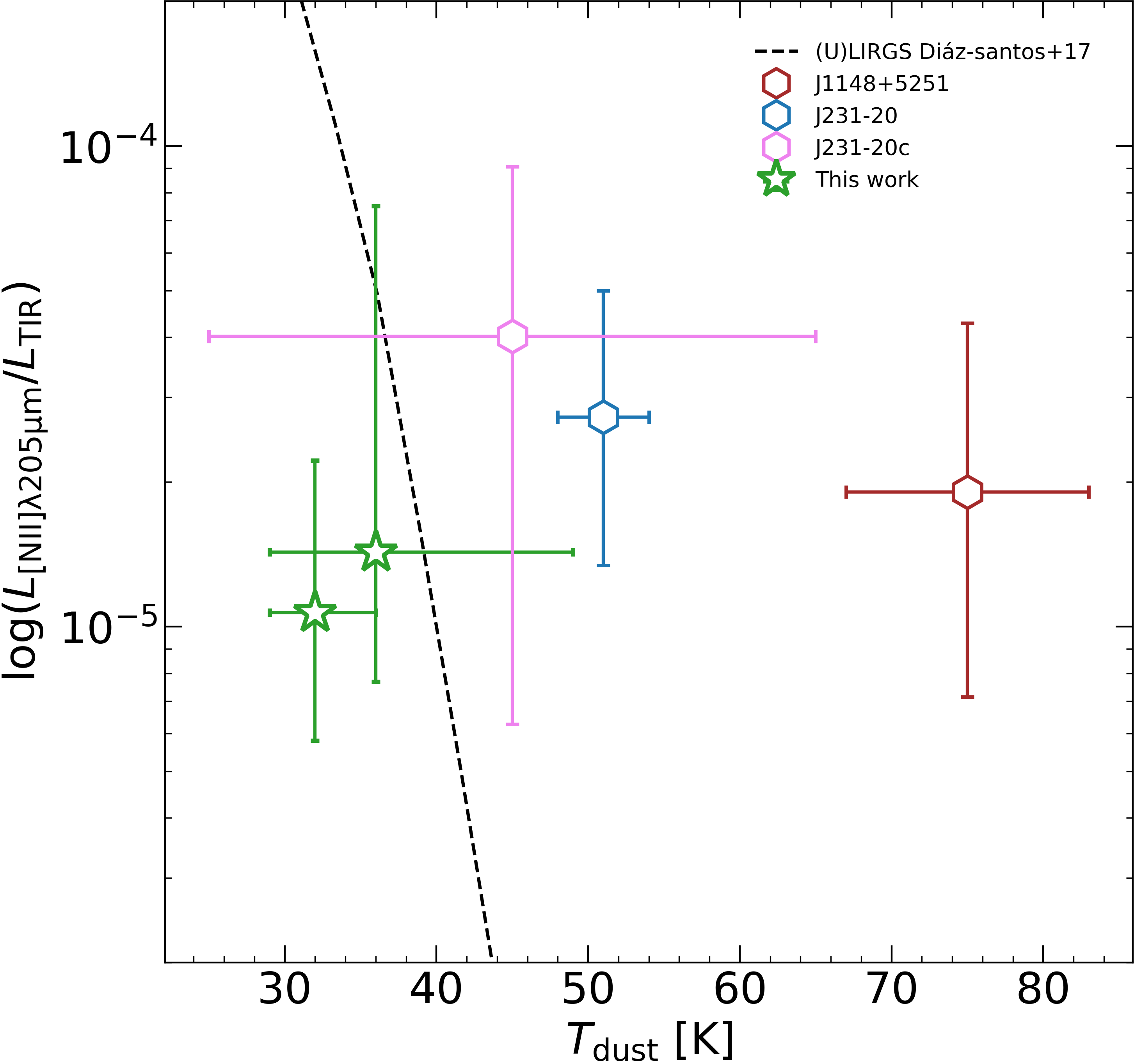}
    \caption{(Left) SFR vs [NII]$\lambda 205\mu$m luminosity for J025-33 and J083+11 (green stars), compared with a sample of local galaxies \citep[pink dots,][]{herrera-camus2016}, a sample of local ULIRGs \citep[silver squares,][]{zhao2013}, the brightest far-IR galaxy \citep[W2246 as a light blue diamond][]{fernandez2024, diazsantos2016}, and other two HYPERION QSOs \citep[J1148+5251 J231-20 and its companion as hexagons; see][respectively]{meyer2022, Pensabene2021}. The different lines correspond to the results from Eq. 18 of \citet{herrera-camus2016} for different assumptions of $n_e$ and $\sigma$. (Right) [NII] deficit by comparing $L_{\rm [NII]205}/L_{\rm TIR}$ with $T_{\rm dust}$. We compare the results for J025-33 and J083+11 (green stars) with the relation for local ULIRGs \citep{diazsantos2017}, and with the results for J1148+5251, J231-20 and its companion \citep{meyer2022, Pensabene2021}.}
    \label{fig:NII-diagnostic}
\end{figure*}

In the left panel of Fig.~\ref{fig:NII-diagnostic}, we compare J025-33 and J083+11 with local galaxies from \citet{herrera-camus2016}, local (U)LIRGs from \citet{zhao2013}, and the few $z>6$ QSOs with available [NII]$\,205\,\mu$m measurements. For the high-redshift comparison sample, we use the [NII] properties of J231-20 and J231-20c from \citet{Pensabene2021}, and those of J1148+5251 from \citet{meyer2022}; the corresponding dust properties and SFRs are taken from \citet{tripodi2024}, \citet{Pensabene2021}, and \citet{carniani2019}, respectively. J025-33 and J083+11 lie above the bulk of the comparison sample, showing higher SFRs at fixed $L_{\rm [NII]205}$. In the framework of the \citet{herrera-camus2016} models, this offset can be interpreted as evidence for either high electron densities, approaching or exceeding the critical density of the [NII]$\,205\,\mu$m transition, or for a broad distribution of electron densities within the beam. In both cases, the [NII]$\,205\,\mu$m luminosity becomes less efficient at tracing the ionizing photon rate, and therefore the SFR, because collisional de-excitation and density-structure effects suppress the emergent [NII] emission.

This interpretation is consistent with the extreme nature of the HYPERION hosts. The two sources are undergoing intense star formation, with ${\rm SFR}\sim750$--$850~M_\odot~{\rm yr^{-1}}$, but their [NII]$\,205\,\mu$m luminosities remain comparable to, or lower than, those measured in other high-redshift QSOs. A possible caveat is that the SFRs are inferred from the TIR luminosity and could be overestimated if a significant fraction of the FIR emission were powered by the AGN. However, J025--33 and J083+11 have cold dust temperatures, $T_{\rm dust}=36^{+13}_{-7}$ K and $32^{+4}_{-3}$ K, respectively, which are lower than typically observed in $z>6$ QSO hosts. This suggests that their FIR emission may be dominated by star formation rather than by AGN-heated dust, making the high SFR/$L_{\rm [NII]205}$ ratios physically meaningful.

In the right panel of Fig.~\ref{fig:NII-diagnostic}, we further investigate the [NII] deficit by comparing $L_{\rm [NII]205}/L_{\rm TIR}$ with $T_{\rm dust}$. Remarkably, the high-redshift QSOs appear broadly consistent with the trend observed in local (U)LIRGs by \citet{diazsantos2017} at $T_{\rm dust}=<50$ K, while they tentatively deviate at higher temperature. However, the statistics is still poor to draw any definitive conclusions. This suggests that the decrease of [NII] line emission relative to the infrared luminosity may be governed by similar ISM conditions in both local compact starbursts and high-redshift QSO hosts. In this picture, the [NII] deficit could be driven by high ionization parameters, dust-bounded HII regions, high gas densities, or a combination of these effects. The fact that J025-33 and J083+11 occupy the cold-dust end of the high-redshift distribution, while remaining compatible with the extrapolation of the local relation, indicates that the [NII] deficit is not necessarily associated only with warm dust or strong AGN heating.

Overall, the detection of [NII]$\,205\,\mu$m in J025--33 and J083+11 provides new evidence that the ionized ISM in HYPERION QSOs is already highly structured and possibly dense less than one billion years after the Big Bang. The current data suggest that [NII]$\,205\,\mu$m may underestimate the SFR in these systems by 0.3 dex if standard low-density calibrations are adopted. At the same time, the consistency between high-redshift QSOs and local (U)LIRGs in the $L_{\rm [NII]205}/L_{\rm TIR}$--$T_{\rm dust}$ plane points toward common physical mechanisms regulating FIR line deficits across cosmic time. A systematic search for both [NII]$\,205\,\mu$m and [NII]$\,122\,\mu$m in larger samples of $z>6$ QSO hosts will be essential to directly constrain electron densities, quantify the scatter in the [NII]--SFR relation, and establish whether the ionized gas properties of HYPERION QSOs are representative of the broader high-redshift QSO population.

\subsection{CO-SLED of J231-20 and its companion}
\label{sec:disc-cosled}

\begin{figure*}
    \centering
    \includegraphics[width=0.44\linewidth]{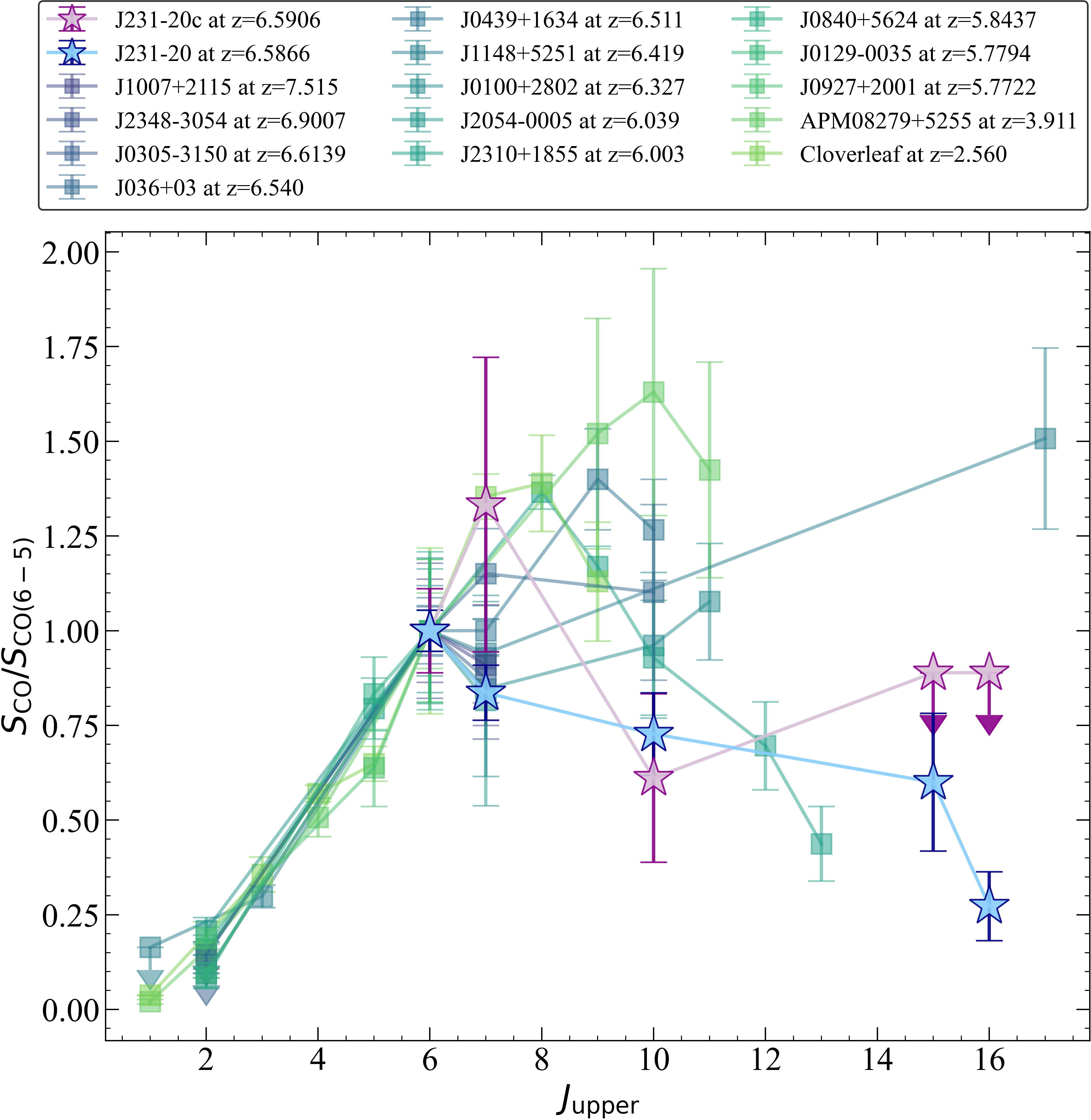}
    \includegraphics[width=0.45\linewidth]{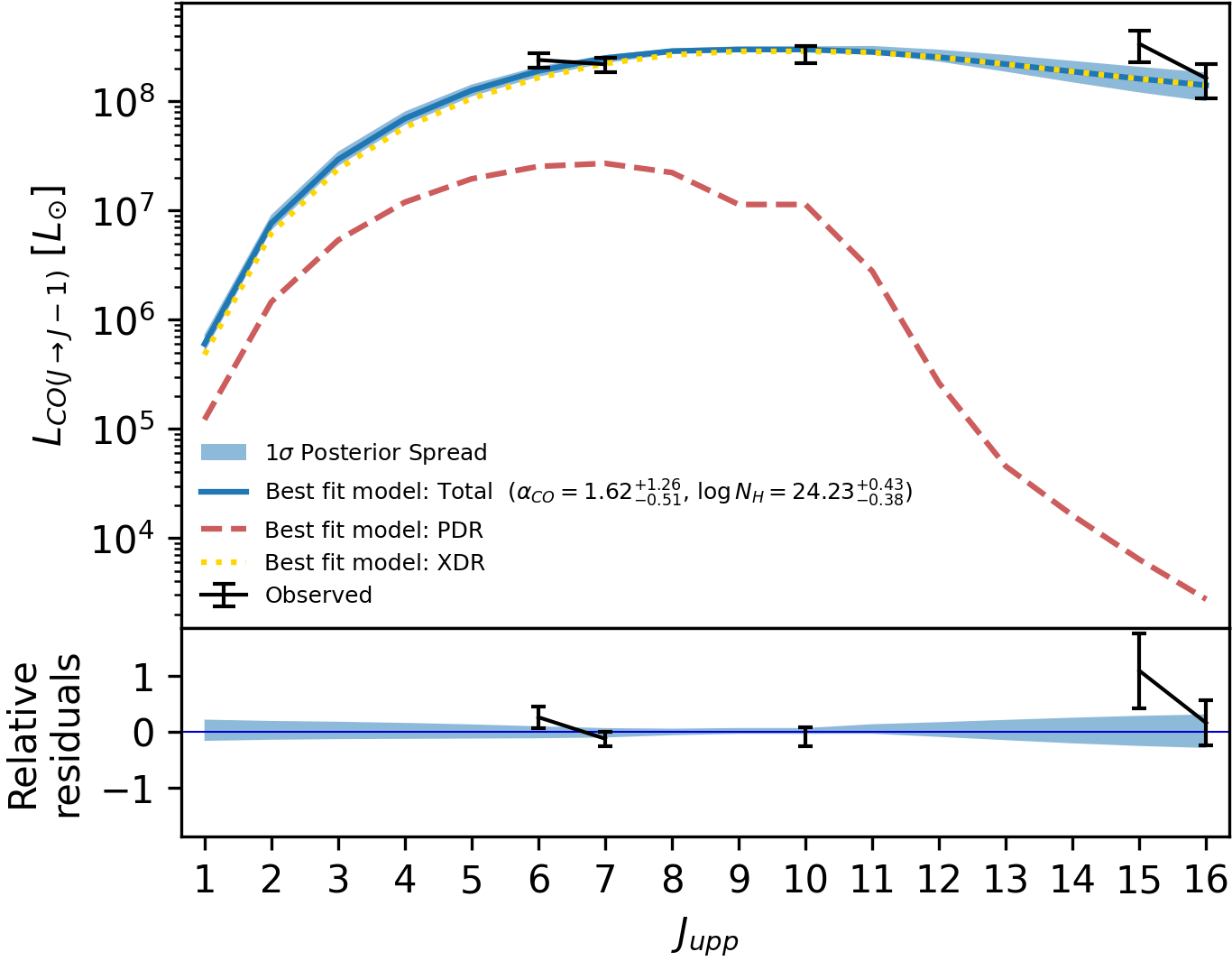}
    \caption{(Left) CO SLED for a sample of high-redshift QSOs and star-forming galaxies, normalized to the CO(6-5) transition. The $J_{\rm upper}$ value on the x-axis corresponds to the upper level of each CO transition. J231-20 and its companion J231-20c are shown with blue and violet star symbols, respectively. For comparison, CO SLEDs of other $z$>2-7 QSOs and dusty star-forming galaxies from the literature are also displayed \citep[][and refs. therein]{tripodi2024, feruglio2023, kaasinen2024}. The different shapes and amplitudes of the SLEDs illustrate the diversity of molecular gas excitation conditions in the early Universe. J231-20 peaks around the CO(6-5) transition and exhibits a steep decline at higher $J$, indicating relatively moderate excitation compared to other luminous QSOs in the sample \citep[see refs. from][]{tripodi2024}.
    (Right) The best-fit model of the CO SLED of J231-20 obtained with \texttt{galaxySLED}. The observed CO line luminosities are shown as black points, while the blue solid line indicates the best-fit total model, with the shaded region marking the $1\sigma$ posterior spread. The dashed red and dotted yellow lines show the PDR and XDR contributions, respectively. The lower panel shows the relative residuals between the observations and the best-fit model.}
 \label{fig:cosled}
\end{figure*}

CO SLEDs provide a powerful diagnostic for disentangling the processes that regulate the interstellar medium in high-redshift galaxies and QSOs, and are widely used in combination with radiative-transfer models to infer the physical state of the molecular gas. 

In the left panel of Fig.~\ref{fig:cosled}, we present the CO SLED of a sample of high-redshift QSOs and star-forming galaxies \citep{li2020,feruglio2023}, normalized to the CO(6-5) transition, including J231-20 in our sample for which this analysis is possible. This representation enables a direct comparison of the excitation conditions across sources with different luminosities and redshifts. J231-20 and its companion J231-20c are highlighted with the blue and violet star symbols, respectively.

\noindent Both components exhibit CO SLEDs that peak around $J_{\rm upper} \sim 6$ - $7$. The CO(6-5) line in J231-20 therefore traces the bulk of the excited molecular gas and provides a reliable anchor point for assessing the excitation ladder. At higher rotational levels ($J_{\rm upper} > 10$), J231-20 shows a clear decline in the SLED, indicating that the gas is not strongly excited to high-$J$ transitions. The companion galaxy J231-20c displays a steeper decline towards the CO(10-9) transition. This suggests that the excitation conditions in the companion may be different from those in the QSO host, possibly reflecting different physical conditions. \\
\noindent When compared to the other high-redshift QSOs in the plot (e.g.\ J1148+5251, J0100+2802, and J0439+1634), J231-20 occupies the low-excitation end of the distribution: its SLED declines more steeply beyond the CO(6-5) transition. This indicates that, despite hosting a luminous QSO, the molecular gas in J231-20 is not as highly excited as in other $z>6$ quasar hosts. 

We modelled the CO SLED of J231-20 with the \textit{galaxySLED} tool\footnote{\url{https://github.com/federicoesposito/galaxySLED}} by \cite{Esposito2024}. 
\texttt{galaxySLED} populates the galaxy volume with a mass distribution of giant molecular clouds (GMCs) whose internal density structure is resolved, and accounts for the dual heating of this gas by stellar radiation and by the AGN, which respectively give rise to photodissociation regions (PDRs) and X-ray-dominated regions (XDRs) within each cloud. Given the radial profiles of molecular mass, far-UV flux (scaled from the observed SFR of the host galaxy), and the measured X-ray flux of the AGN as input, the code returns a synthetic CO SLED that depends on two free parameters, the CO-to-H$_{\rm 2}$ conversion factor $\alpha_{\mathrm{CO}}$ and the X-ray attenuation column density $N_{\mathrm{H}}$. Observed CO line luminosities (including upper limits) are fitted to constrain these parameters through a MCMC sampling of their posterior distributions.

In this work, we used the updated version (v0.1.9) of \texttt{galaxySLED}, which includes a new suite of clouds designed to reproduce the denser and more turbulent ISM observed in high-redshift quasars. In particular, the cloud models adopted here are characterized by higher gas surface density ($\Sigma=500$ M$_{\odot}\ pc^{-2}$) and Mach number 20.
The same version also includes a treatment for the contribution to the excitation of the cold molecular gas due to the CMB at the redshift of P231-20.
Further details on this version of the code will be provided in Salvestrini et al. in prep.

We provided \texttt{galaxySLED} the following inputs: SFR \citep{tripodi2024}, X-ray luminosity ($\log(L_X/{\rm erg\ s^{-1}}) = 45.09$; see \citealt{tortosa2024}), dust continuum source size (see Tab.~\ref{tab:res-cont}), and total cold-gas mass.
Regarding the latter, the total cold-gas mass is used to populate the mock galaxy with a distribution of molecular clouds. These clouds are randomly drawn from an input cloud mass function, adopting the standard approximation for the giant molecular cloud population in the Milky Way, described by a power-law distribution with slope 1.64 (\citealt{Esposito2024}; see also \citealt{Chevance2023} for a review). 

Since there is no observations for low-J CO transition for P231-20, to provide an input total cold-gas mass we explored different prescriptions: first, we inferred it from the dust mass by assuming gas-to-dust ratios of 100, 75, and 50; second, we adopted the mean CO-ladder proposed by \cite{kaasinen2024} to derive the ground transition from the CO(6-5) luminosity, and then assumed a typical $\alpha_{\rm CO} = 0.8\ M_\odot\,({\rm K\,km\,s^{-1}\,pc^2})^{-1}$ ; third, we used the [CII]-to-H$_2$ relation calibrated on CO measurements derived by \citet{salvestrini2025}.
These different prescriptions yield cold-gas masses in the range $M_{\rm H_2}=(1.3-5.3)\times10^{10},M_{\odot}$ (see Tab.~\ref{tab:co_sled}), with the highest value obtained by assuming a gas-to-dust ratio of 100 and the lowest from the calibration of \cite{salvestrini2025}.

The \texttt{galaxySLED} code also requires, as input, the $\alpha_{\rm CO}$ value adopted to derive the molecular gas mass. Therefore, the $\alpha_{\rm CO}$ obtained as a best-fit parameter from the CO SLED should be interpreted as the expected CO-to-H$2$ conversion factor to be applied to the CO(1--0) luminosity extrapolated from the best-fit CO SLED, in order to derive the best-fit gas mass. In other words, \texttt{galaxySLED} provides the posterior distribution of $\alpha_{\rm CO}$, which should be treated as a correction factor with respect to the input value, yielding the posterior distribution of $M_{\rm gas}$ that best reproduces the observed CO SLED.

Since we use the same CO SLED in all runs, varying only the input cold gas mass, the model CO SLED with the highest likelihood is the same for all cases. This solution has a reduced $\chi^2=1.79$ and predicts $M_{\rm H2}=1.92 \times 10^{10},M_{\odot}$. Considering instead the values corresponding to the 50th percentile of the posterior distributions, all estimates are fully consistent within the $1\sigma$ uncertainties. The resulting corrected gas masses span the range $M_{\rm H2}=2.65$--$2.91 \times 10^{10},M_{\odot}$.

The run adopting the input $M_{\rm H2}$ derived from the calibration of \cite{salvestrini2025} yields the lowest median $\chi^2=1.96$, very similar to the values obtained using the calibration of \cite{kaasinen2024} and assuming a gas-to-dust ratio of GDR=50, for which we find $\chi^2=1.97$ and $\chi^2=1.98$, respectively. The runs assuming GDR=75 and GDR=100 yield slightly higher $\chi^2$ values of 2.05 and 2.18, respectively. In Table~\ref{tab:co_sled}, we present the full list of $\alpha_{\rm CO}$ and $N_{\rm H}$ values obtained from the different runs.

In the right panel of Fig.~\ref{fig:cosled}, we show the best-fitting model obtained assuming the calibration by \cite{salvestrini2025} for the input $M_{\rm H2}$ in blue with uncertainties as shaded region (derived as the 50th, and 16-84th percentiles of the posterior distributions, respectively). The XDR and PDR components are also reported separately. The XDR clearly dominates the SLED of J231-20 at all CO transitions: specifically, for the transition 6-5, the ratio XDR/PDR$=8.8$. From the best-fitting COSLED, we obtain an $\alpha_{\rm CO}=1.62^{+1.26}_{-0.51}$ $M_\odot\,({\rm K\,km\,s^{-1}\,pc^2})^{-1}$, and $N_{\rm H}=24.23^{+0.43}_{-0.38}$ $\rm cm^{-2}$ (see the posterior distribution in App.~\ref{app:corner-COSLED}). The derived conversion factor is about a factor 2 higher than what usually adopted for high-z QSOs (see also first paragraph of Sect.~\ref{sec:analysis}), therefore the molecular gas mass of J231-20 becomes
$M_{\rm H_2}=(2.65^{+2.06}_{-0.84})\times 10^{10}\rm M_\odot$ and the gas-to-dust ratio GDR$=50_{-19}^{+48}$ \citep[revised from][]{tripodi2024}.

Since the scope of this work is to derive solid molecular gas masses, we do not deep into the intepretation of the the best fit value for $N_{\rm H}$ (see Tab.~\ref{tab:co_sled}), which suggest that the cold gas need relatively high columns density to shield the X-ray photons emitted by the central AGN. This point will be further discussed in a more general context in Salvestrini et al. in prep. 

Modelling the CO SLED of the companion galaxy of J231-20 is beyond the scope of this work and will be presented in a forthcoming paper (Salvestrini et al. in prep.) dedicated to the study of CO SLEDs in high-redshift quasars, SMGs, and quasar companion galaxies. However, we stress that the detection of a bright CO(10--9) transition in the companion already suggests the presence of an additional excitation mechanism beyond PDRs, possibly requiring a contribution from XDRs.

\subsection{Molecular gas mass of HYPERION QSOs}
\label{sec:disc-molecmass}

\begin{figure*}
    \centering
    \includegraphics[width=0.48\linewidth]{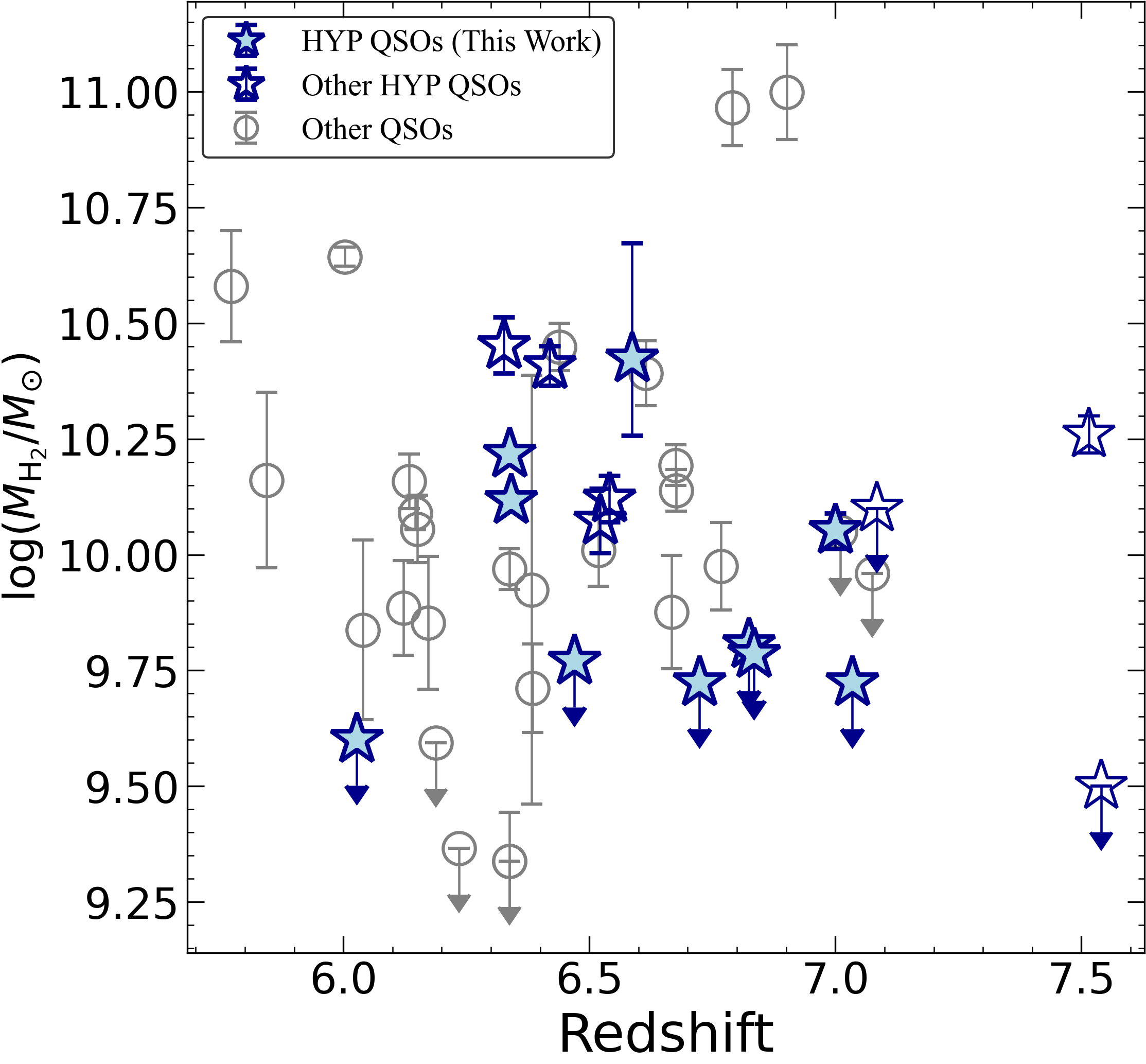}
    \includegraphics[width=0.45\linewidth]{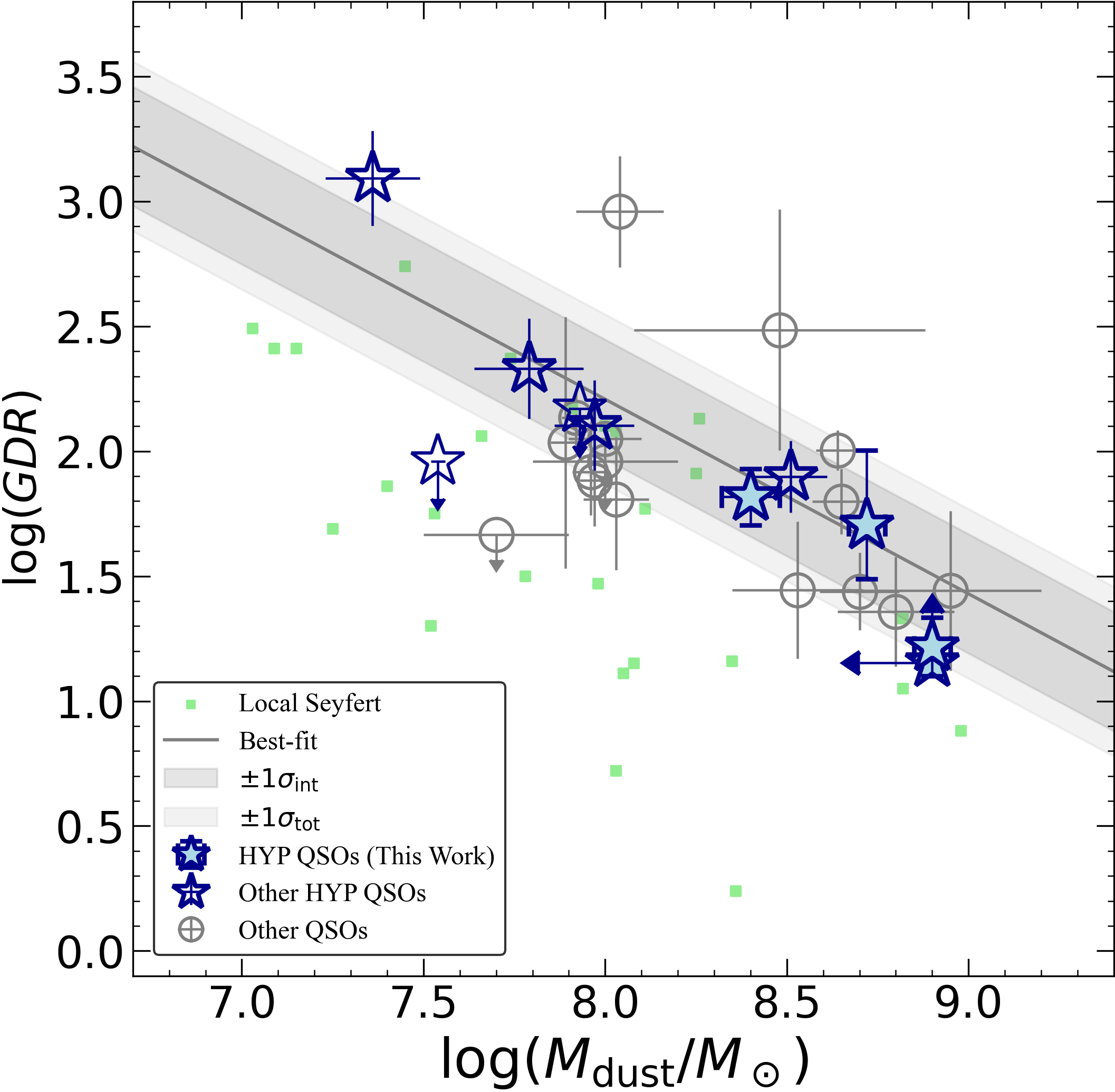}
    \caption{Left: molecular gas mass as a function of redshift for our targets (blue filled stars) along with other HYPERION QSOs \citep[hollow stars][]{tripodi2024} and other QSOs from the literature \citep[hollow gray dots][]{salvestrini2025, xu2026} Right: gas-to-dust ratio vs dust mass for our targets (blue filled stars) along with other HYPERION QSOs (hollow stars) and other QSOs from the literature (hollow gray dots). This high-z population is compared with a sample of local Seyfert, shown as green squares \citep{salvestrini2022}. The best-fit linear relation is shown as solid gray line. The gray shaded regions represent the intrinsic and total scatter as color coded in the legend.}
    \label{fig:Mgas_z}
\end{figure*}

The molecular gas content of high-redshift QSOs is expected to be governed by the interplay between gas accretion, star formation, black-hole fueling, and feedback. The large accretion rates required to power luminous QSOs imply the presence of substantial cold-gas reservoirs, which can simultaneously sustain SMBH growth and vigorous star formation in the host galaxy. At the same time, these reservoirs can be rapidly depleted by star formation, redistributed by gravitational interactions, or affected by AGN- and stellar-driven feedback, which may heat, disrupt, or expel part of the cold molecular phase. The observed molecular gas mass therefore reflects a balance between gas supply, consumption, and removal.

The inferred value of $M_{\rm H_2}$ also depends on a number of observational and modelling assumptions. These include the adopted CO-to-H$_2$ conversion factor, the CO excitation correction used to infer the ground-state CO luminosity from mid- or high-$J$ transitions, and the specific gas tracer employed \citep[e.g.,][]{decarli2022, salvestrini2025}. In this work, and for the other QSOs shown in Fig. \ref{fig:Mgas_z}, all molecular gas masses have been recomputed in a homogeneous way (see Sect.~\ref{sec:analysis}). This allows a more meaningful relative comparison between the HYPERION QSOs and other $z>6$ quasars \citep{decarli2022, kaasinen2024, salvestrini2025, xu2026}, although the absolute normalization remains subject to the usual systematic uncertainties associated with $\alpha_{\rm CO}$ and the CO SLED.

The molecular gas reservoirs of HYPERION QSOs have been investigated in several previous works. In particular, \citet{salvestrini2025} derived molecular gas constraints from CO(6--5), CO(7--6), and [CII] observations for a sample of $z>7$ quasars, including two sources also belonging to HYPERION, namely J0038-1527 and J0252-0503. They reported upper limits of $M_{\rm H_2}\lesssim10^{10}\,{\rm M_\odot}$. Thanks to our deeper observations, we improve the constraints on both systems. For J0038-1527 we obtain a tighter upper limit, while for J0252-0503 we measure $M_{\rm H_2}=(1.1\pm0.1)\times10^{10}\,{\rm M_\odot}$, consistent with the previous constraint by \citet{salvestrini2025}. For J231-20 and its companion, the detection of CO(6--5) emission and the analysis of J231-20 CO SLED (see Sect.~\ref{sec:disc-cosled}) provides the most robust estimate of their molecular gas masses to date, while remaining consistent with values previously inferred from CO(7--6) observations \citep{Pensabene2021, tripodi2024}. Molecular gas estimates for six additional HYPERION QSOs have also been presented in the literature \citep{stefan2015, novak2019, decarli2023, tripodi2024}. Combining these results with the measurements presented here, molecular gas constraints are now available for about $60\%$ of the HYPERION sample.

As shown in the left panel of Fig. \ref{fig:Mgas_z}, the HYPERION QSOs span a broad range of molecular gas masses, from values comparable to the typical reservoirs observed in other luminous $z>6$ QSOs\footnote{To enable a proper comparison, we recomputed all the molecular gas masses from literature with the same assumptions used in this work, i.e., $r_{61}=0.92$ and $\alpha_{\rm CO}=0.8 ~\rm M_\odot(K~km~s^{-1}~pc^{1})^{-1}$.} to significantly lower measurements or upper limits. Literature studies of high-redshift quasar hosts generally find molecular gas masses of order $M_{\rm H_2}\sim10^{10}\,{\rm M_\odot}$, with substantial object-to-object scatter \citep[e.g.,][]{decarli2022}. The HYPERION sources are broadly consistent with this distribution. However, the homogeneous comparison shown in Fig. \ref{fig:Mgas_z} suggests that part of the HYPERION population may occupy the lower-$M_{\rm H_2}$ tail, with several objects lying below, or constrained to lie below $\sim10^{10}\,{\rm M_\odot}$.

This possible diversity is particularly interesting because HYPERION targets were selected as among the most rapidly accreting and massive SMBHs known at these epochs. The presence of relatively modest molecular reservoirs in some of these systems could indicate that they are caught in a phase where a large fraction of the available cold gas has already been consumed by star formation and SMBH accretion, or has been affected by feedback and environmental processes. Alternatively, the low inferred $M_{\rm H_2}$ values may reflect compact, highly excited, or otherwise complex ISM conditions that are not fully captured by standard excitation and conversion assumptions. Therefore, while the current data indicate that HYPERION QSOs are not uniformly associated with the most massive molecular reservoirs known at $z>6$, this result should still be interpreted with caution.

Overall, the updated census supports a picture in which luminous QSOs at the epoch of Reionization inhabit gas-rich but highly diverse host galaxies. Some HYPERION QSOs contain molecular gas reservoirs comparable to those of other high-$z$ quasars, while others appear to host comparatively smaller cold-gas masses. Confirming whether this reflects a real physical diversity in the gas content of rapidly growing SMBH hosts will require deeper and more sensitive CO observations, ideally targeting multiple transitions to better constrain the CO excitation and reduce the systematic uncertainty on $M_{\rm H_2}$.

\subsection{Dust properties and gas-to-dust ratio}
\label{sec:disc-gdr}

The cold-dust properties of the HYPERION QSOs provide an additional view of the ISM conditions in the host galaxies of the most rapidly growing SMBHs at the epoch of Reionization. Modelling the cold-dust SED allows us to constrain the dust temperature, dust mass, infrared luminosity, and star formation rate, while the combination of $M_{\rm dust}$ with the molecular gas mass provides the gas-to-dust ratio, ${\rm GDR}=M_{\rm H_2}/M_{\rm dust}$. This quantity is a useful tracer of the relative enrichment level of the ISM, although it remains affected by the systematic uncertainties associated with both the dust opacity model and the molecular-gas calibration.

The results of the SED modelling are reported in Table \ref{tab:res-sed}. For two sources, J025-33 and J083+11, the available continuum data allow us to constrain the dust temperature and dust mass. We find $T_{\rm dust}=36^{+13}_{-7}$ K and $\log(M_{\rm dust}/M_\odot)=8.4\pm0.1$ for J025-33 (see Fig.~\ref{fig:sed-J025}), and $T_{\rm dust}=32^{+4}_{-3}$ K and $\log(M_{\rm dust}/M_\odot)=8.9\pm0.1$ for J083+11 (see Fig.~\ref{fig:sed-J083}). These dust masses further confirm that at least part of the HYPERION population has already assembled substantial dust reservoirs \citep[see also][]{tripodi2024}. The remaining targets are less tightly constrained, mostly providing lower limits on $T_{\rm dust}$ and upper limits on $M_{\rm dust}$, but they are still consistent with dust reservoirs of order $\sim10^{8}\,M_\odot$ or below.

A particularly interesting result is the low dust temperature measured in J025-33 and J083+11. Both values are well below the mean dust temperature of high-redshift QSOs, $T_{\rm dust}\simeq54$ K, derived from the mean cold-dust SED of $z>6$ quasars by \citet{tripodi2024}. They are also lower than the typical values often adopted when the FIR SED peak is not directly sampled, $T_{\rm dust}\simeq55$ K. Such low temperatures may indicate that, in these two systems, the rest-frame FIR emission is dominated by dust heated by star formation rather than by a strong AGN-heated component. This interpretation is consistent with the idea that the AGN contribution to the FIR luminosity can vary significantly among luminous QSOs, and that assuming a fixed contribution may not be appropriate for all sources. In this scenario, the SFRs inferred from the IR luminosity of J025-33 and J083+11 would be less affected by AGN contamination than in hotter systems. Alternatively, the low $T_{\rm dust}$ may reflect a more extended dust distribution or a different geometry between the star-forming regions, dust, and central AGN, or a low SFR density \citep{parente2026}.  Higher-angular-resolution FIR observations will be required to distinguish between these possibilities.

The inferred emissivity indices for J025-33 and J083+11, $\beta=2.1\pm0.5$ and $\beta=2.1\pm0.2$, respectively, are broadly consistent with values measured in dusty star-forming galaxies and high-redshift QSOs, although they are somewhat higher than the average $\beta\simeq1.6$ commonly adopted for high-$z$ quasar hosts \citep{witstok2023, tripodi2024}. This may suggest differences in dust-grain properties, optical depth, or a broad temperature distribution within the host galaxies. 

The GDRs derived for high-z QSOs with both dust and molecular-gas constraints show a wide range of values \citep{tripodi2024, salvestrini2025}. In this context, J083+11 exhibits an exceptionally low value (${\rm GDR}=16^{+5}_{-4}$). To our knowledge, the latter is possibly the lowest GDR measured so far in a high-redshift QSO host. This is well below the canonical local value ${\rm GDR}\sim100$ and below the average values typically reported for luminous QSOs at both high redshift and Cosmic Noon. Such a low GDR implies either a very dust-rich ISM, a relatively small molecular reservoir, or a combination of both. In the case of J083+11, the high dust mass, $\log(M_{\rm dust}/M_\odot)=8.9$, together with the modest molecular gas mass, naturally drives the source towards the low-GDR regime. As a word of caution, we recall that systematics related to the assumptions of the same $r_{61}$ and $\alpha_{\rm CO}$ for all objects may affect the estimate of $M_{\rm H_2}$, and therefore the GDR. Future follow-up observations targeting lower CO transitions will allow precise CO SLED modelling, definitely alleviating the uncertainties on both $M_{\rm H_2}$ and GDR.

As shown in the right panel of Fig. \ref{fig:Mgas_z}, our measurements follow the anti-correlation between GDR and $M_{\rm dust}$ observed in other QSO and Seyfert samples \citep{salvestrini2025}. We performed a linear regression on the data via MCMC, excluding the upper and  lower-limits. The fit supports the strong anti-correlation with $r_{\rm Pearson} =-0.66_{-0.09}^{+0.12}$ and $\rho_{\rm Spearman}=-0.64\pm0.12$. This trend has been interpreted as a consequence of increasing metal enrichment in dust-rich systems: galaxies with larger dust reservoirs are expected to have already converted a significant fraction of their metals into dust grains, thereby lowering the gas-to-dust ratio \citep{salvestrini2025, remy2014}. J083+11 lies at the extreme low-GDR, high-$M_{\rm dust}$ end of this relation, extending the trend observed in previous samples. J025-33 also falls below the canonical ${\rm GDR}\sim100$ value, supporting the idea that some HYPERION hosts may already be highly enriched systems despite being observed less than one billion years after the Big Bang.

These results suggest that the HYPERION population is not homogeneous in terms of dust enrichment and gas-to-dust content. While \citet{tripodi2024} found that HYPERION QSOs tend, on average, to have lower dust masses and higher GDRs than other $z>6$ QSOs, our new measurements reveal that at least some members of the sample can instead host massive dust reservoirs and very low GDRs. This diversity may reflect different evolutionary stages within the host: some systems may still be gas-rich and relatively dust-poor, while others, such as J083+11, may have already undergone rapid metal and dust enrichment, possibly accompanied by efficient gas consumption through star formation. Overall, the dust and GDR measurements reinforce the picture of HYPERION QSOs as a heterogeneous population of rapidly evolving galaxies, in which the relative timing of SMBH growth, star formation, gas consumption, and dust enrichment can vary substantially from source to source.

\begin{figure}
    \centering
    \includegraphics[width=1\linewidth]{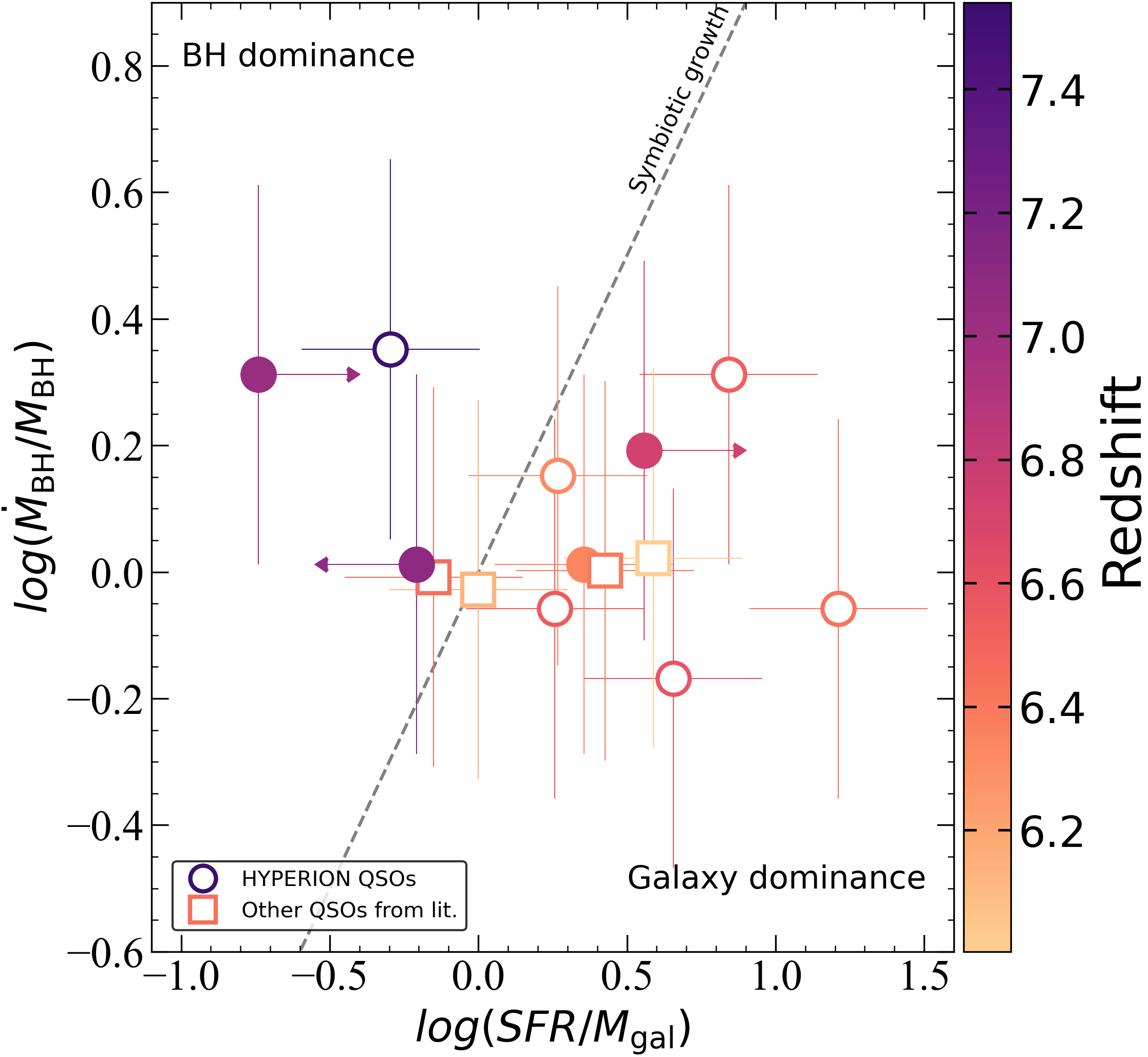}
    \caption{Black hole growth rate vs galaxy growth rate. Errors on both x and y-axis are of 0.3 dex, to account for systematics on the BH and galaxy properties estimates. Circles are HYPERION QSOs and squares are the only other QSOs from literature for which this study is feasible \citep{tripodi2024}. Filled circles are the four new QSOs added with this work. The dashed line marks the symbiotic growth phase space where $\dot{M}_{\rm BH}/M_{\rm BH}=\rm SFR/M_{\rm gal}$. Given this definition, the region above the dashed line is the BH dominance regime, and below the dashed line is the galaxy dominance regime.}
    \label{fig:coevol}
\end{figure}

\begin{figure*}
    \centering
    \includegraphics[width=0.50\linewidth]{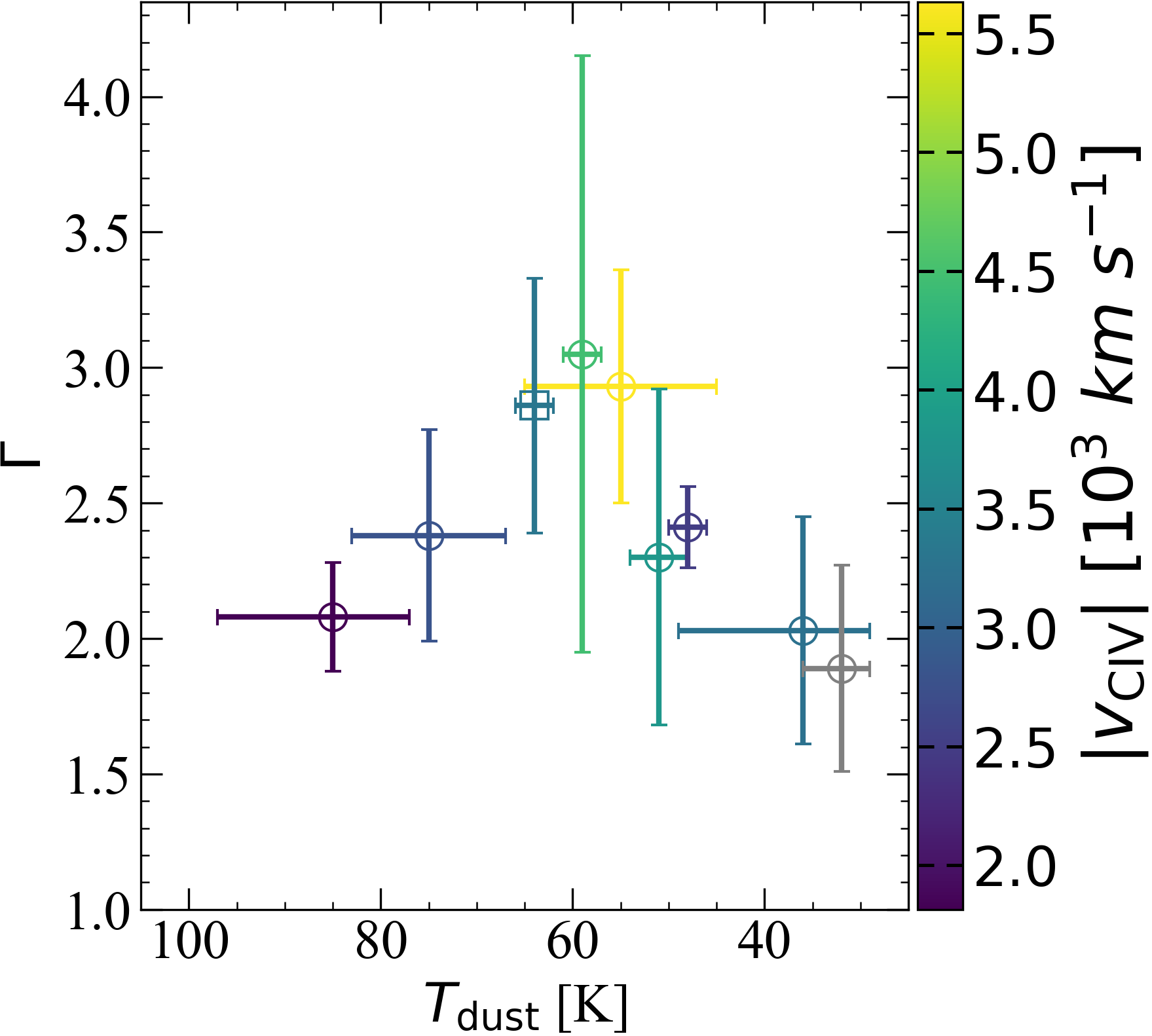}\hspace{0.1cm}
    \includegraphics[width=0.45\linewidth]{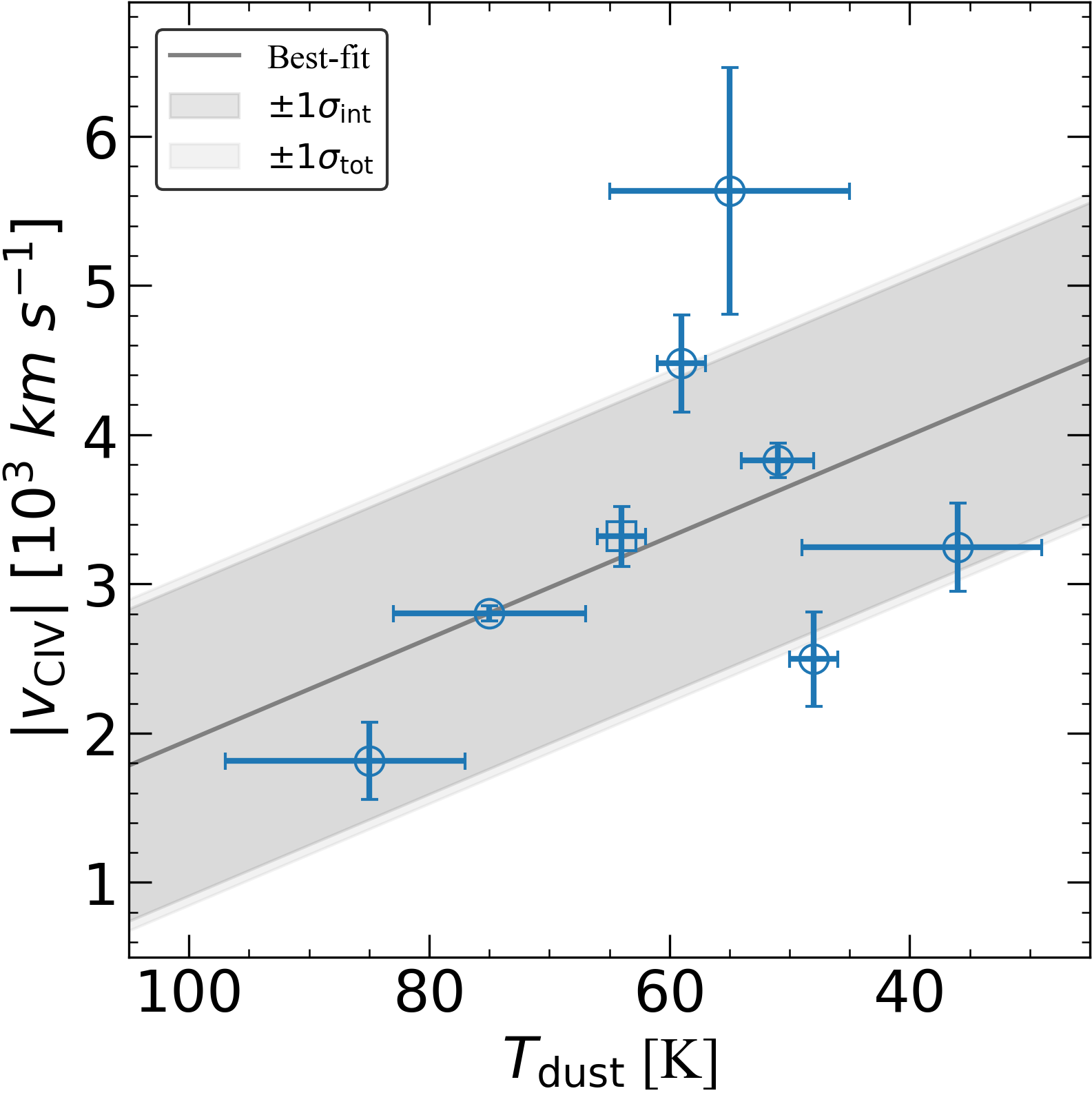}
    \caption{(Left) $\Gamma$ vs $T_{\rm dust}$ in the HYPERION sample (dots) and for J2054-0005 (square). Data are color coded by the velocity shift of CIV]. (Right) $v_{\rm CIV}$ vs $T_{\rm dust}$ in the HYPERION sample (dots) and for J2054-0005 (square). The best-fit linear relation is shown as solid gray line. The gray shaded regions represent the intrinsic and total scatter as color coded in the legend. The x-axis is inverted for visualization purposes only.}
    \label{fig:gammaT}
\end{figure*}

\subsection{Galaxy dominance at $z\sim6$}
\label{sec:disc-coevol}

\noindent In Fig. \ref{fig:coevol}, we compare the instantaneous growth rate of the SMBH with that of the host galaxy for HYPERION QSOs and for other luminous $z\gtrsim6$ QSOs for which this analysis is currently possible. The horizontal axis shows the specific growth rate of the galaxy, $\log({\rm SFR}/M_{\rm gal})$, where $M_{\rm gal}=M_{\rm dyn}-M_{\rm BH}$, while the vertical axis shows the specific SMBH accretion rate, $\log(\dot{M}_{\rm BH}/M_{\rm BH})$. The dashed line marks the locus where the two components grow at comparable relative rates, corresponding to the symbiotic-growth regime. Sources above this line are preferentially in a BH-dominated phase, while sources below it are in a galaxy-dominated phase. The points are color-coded by redshift, and typical uncertainties of 0.3 dex are shown on both axes.

Following the framework introduced by \citet{tripodi2024}, this diagram provides a quantitative way to identify different stages of SMBH--host-galaxy coevolution. In a BH-dominated regime, the SMBH is still increasing its mass more efficiently than the host. In the galaxy-dominated regime, the host galaxy grows faster than the SMBH, potentially catching up with the already assembled black hole. The symbiotic regime represents an intermediate phase in which the two components grow at comparable relative rates. We stress that these classifications refer to relative growth efficiencies: even in the BH-dominated regime the host galaxy is still forming stars, and even in the galaxy-dominated regime the SMBH is still accreting.

Compared to the sample discussed in \citet{tripodi2024}, our analysis includes four additional HYPERION QSOs (filled dots; J025-33, J0038-1527, J0244-5008 and J1120+0641\footnote{J025-33, J0244-5008 and J0038-1527 are the only ones among our targets to have constraints on both the SFR (this work) and dynamical mass \citep[][and this work for J0244-5008]{neeleman2021, wang2024}. Moreover, we retrieved the same information for J1120+0641 from the literature \citep{neeleman2021,salvestrini2025}.}), extending this test to a larger fraction of the population of the most rapidly growing SMBHs known at these redshifts. Moreover, we recomputed the BH accretion rate of all HYPERION QSOs based on the new estimates of their bolometric luminosities given by \citet{saccheo2025}. Results are still consistent within uncertainties with those reported in \citet{tripodi2024}. Overall, most of the QSOs lie either in the galaxy-dominated region or close to the symbiotic-growth locus, once the typical 0.3 dex uncertainties are considered. This supports a scenario in which many luminous QSOs at $z\sim6$ are no longer observed during the phase of maximum relative SMBH growth, but rather during a subsequent stage in which the host galaxy is undergoing rapid assembly. In this phase, intense star formation increases $M_{\rm gal}$ and may drive the system toward the local $M_{\rm BH}$--host-galaxy relation.

The HYPERION QSOs span a broad range of positions in this plane. Some sources occupy the BH-dominated side of the diagram, suggesting that their SMBHs may still be growing more efficiently than their hosts. Others are consistent, within the uncertainties, with symbiotic or galaxy-dominated growth. This diversity indicates that the HYPERION selection does not isolate a single instantaneous evolutionary phase. Instead, these objects may trace different moments along a rapid sequence in which an early episode of very efficient SMBH growth is followed by a possibly longer phase of intense host-galaxy build-up.

A possible trend with redshift is suggested by the color coding. The three highest-redshift HYPERION QSOs appear preferentially located toward the BH-dominated or near-symbiotic region, while several lower-redshift systems lie closer to the galaxy-dominated regime. If confirmed with a larger sample, this could indicate an evolutionary segregation: the highest-redshift objects may be observed closer to the phase of rapid SMBH assembly, whereas systems at slightly later cosmic times may already be transitioning toward host-dominated growth. However, given the current uncertainties and the limited number of sources, this interpretation remains tentative.

The prevalence of symbiotic and galaxy-dominated systems is consistent with the idea that luminous QSOs at $z\gtrsim6$ are moving closer towards the local $M_{\rm BH}-M_{\rm dyn}$ relation \citep{tripodi2024}. Their SMBHs have already reached very large masses, while their host galaxies are still forming stars at high rates. This late host-galaxy growth may be regulated by feedback processes. Indeed, several objects in this regime show evidence for powerful outflows or broad absorption line systems \citep[see discussion in][]{tripodi2024, spilker2025}, suggesting that AGN feedback may contribute to slowing down SMBH accretion without immediately suppressing star formation across the host. In this picture, feedback may help move the system away from a purely BH-dominated phase and toward a more balanced or host-dominated evolutionary stage.

Overall, Fig. \ref{fig:coevol} supports a scenario in which the coevolution of luminous QSOs at the epoch of Reionization is not strictly simultaneous. Instead, the data favor an evolutionary sequence in which a rapid early SMBH growth phase is followed by substantial host-galaxy assembly. The current sample suggests that many $z\sim6$ QSOs are caught during or after this transition, but a robust test of the possible redshift dependence of the growth regime will require homogeneous SFR measurements from well-sampled dust SEDs for the remaining HYPERION QSOs, together with consistent dynamical-mass estimates.

\subsection{A multi-scale connection among dust, X-ray corona and feedback}
\label{sec:disc-xray}

\noindent Motivated by the recent results of \citet{tortosa2024}, we explored possible connections between the nuclear X-ray properties and the cold-dust properties of the HYPERION QSOs. \citet{tortosa2024} found a statistically significant correlation between the X-ray photon index, $\Gamma$, and the C\,{\sc iv} velocity shift, $v_{\rm C\,IV}$, in luminous $z>6$ QSOs, such that sources with faster C\,{\sc iv} disc winds show steeper X-ray continua. Since $\Gamma$ is linked to the physical state of the X-ray corona, being sensitive to its temperature and optical depth, this result suggests a connection between high-velocity accretion-disc winds and the disc--corona configuration. In particular, a steeper $\Gamma$ can be interpreted, at zeroth order, as a signature of a softer and more efficiently cooled corona, corresponding to lower effective coronal temperatures and/or different optical-depth conditions.

The HYPERION QSOs are ideally suited for this test, since they currently have the best X-ray coverage among luminous $z>6$ QSOs, thanks to the dedicated \textit{XMM-Newton} HYPERION heritage programme. We therefore compared the dust properties derived from our FIR SED modelling with several nuclear quantities, including $L_{\rm X}$, $\alpha_{\rm ox}$, $L_{\rm bol}$, $\Gamma$, and $v_{\rm C\,IV}$. Among the explored relations, the most interesting behaviour is found in the $\Gamma$--$T_{\rm dust}$ plane. As shown in the left panel of Fig. \ref{fig:gammaT}, the data suggest a non-monotonic, inverted-V-shaped trend: moving from high to lower dust temperatures, $\Gamma$ first increases, reaching its highest values around $T_{\rm dust}\sim55$--$65$ K, and then decreases again toward the lowest $T_{\rm dust}$ values. Given the small number of sources and the current uncertainties, this trend should be considered tentative. It may either represent a genuine connection between the coronal state and the large-scale dust-heating conditions, or be the projection of a more fundamental relation involving the C\,{\sc iv} wind velocity.

This second possibility is supported by the colour coding in the left panel of Fig. \ref{fig:gammaT}. In the high-$T_{\rm dust}$ regime, sources with larger C\,{\sc iv} velocities tend to occupy a coherent sequence, suggesting that the apparent $\Gamma$--$T_{\rm dust}$ behaviour may partly reflect the underlying link between $\Gamma$ and $v_{\rm C\,IV}$ reported by \citet{tortosa2024}. Indeed, the relation between $T_{\rm dust}$ and $v_{\rm C\,IV}$ appears even clearer in the right panel of Fig. \ref{fig:gammaT}, where we also show the MCMC best-fit of the observed trend as solid gray line (the shaded regions represent the total and intrinsic scatter). We find a weak anti-correlation between $T_{\rm dust}$ and $v_{\rm CIV}$, with $r_{\rm Pearson} =-0.36_{-0.17}^{+0.21}$ and $\rho_{\rm Spearman}=-0.31\pm0.17$. This suggests a possible trend for sources with colder dust temperatures to show larger C\,{\sc iv} blueshifts, although the correlation is tentative. The C\,{\sc iv} velocity shift is commonly interpreted as a tracer of the energetics of the accretion-disc wind. In a simple outflow framework, the kinetic power scales as $\dot{E}_{\rm kin}\sim\frac{1}{2}\dot{M}_{\rm out}v_{\rm out}^2$; if $\dot{M}_{\rm out}\sim M_{\rm out}v_{\rm out}/R$, then $\dot{E}_{\rm kin}\propto M_{\rm out}v_{\rm out}^3/R$. Therefore, even moderate changes in $v_{\rm C\,IV}$ can correspond to large differences in wind energetics. Since the HYPERION QSOs span a relatively narrow range in optical luminosity, with $L_{5100}\sim10^{46}\,{\rm erg\,s^{-1}}$, and since $L_{5100}$ is linked to the broad-line-region radius, we may expect the characteristic wind-launching radii to be broadly similar across the sample. Under this assumption, variations in $v_{\rm C\,IV}$ could be the dominant driver of differences in wind kinetic power.

One possible physical interpretation of the $v_{\rm C\,IV}$--$T_{\rm dust}$ anti-correlation is that it traces a connection between AGN feedback, SMBH growth, and the dominant dust-heating mechanism. Moving from high to low $T_{\rm dust}$, the C\,{\sc iv} blueshift generally increases, suggesting that more powerful disc winds may displace part of the dust away from the central AGN radiation field. At larger distances from the active nucleus, the dust would experience a weaker and softer radiation field, resulting in lower temperatures. In the coldest systems with evidence of ongoing feedback, the FIR emission could therefore be predominantly, or even almost entirely, powered by star formation rather than by direct AGN heating. Within this scenario, the observed $T_{\rm dust}$ would not only depend on the instantaneous luminosity of the QSO, but also on the geometry and radial redistribution of the dusty ISM caused by feedback.

This interpretation may also explain why the apparent $\Gamma$--$T_{\rm dust}$ sequence becomes less coherent at the lowest dust temperatures. If powerful winds have transported the dust sufficiently far from the nucleus, the thermal state of the large-scale dust reservoir may become progressively decoupled from the properties of the X-ray corona. The breakdown of the $\Gamma$--$T_{\rm dust}$ trend would then reflect a transition from a regime in which AGN heating contributes significantly to the FIR SED to one in which star formation dominates the dust heating. Interestingly, the largest scatter in the $v_{\rm C\,IV}$--$T_{\rm dust}$ relation is also observed at the lowest temperatures. This may arise from the different timescales probed by the two quantities. While $v_{\rm C\,IV}$ traces the current state of the nuclear wind, $T_{\rm dust}$ may retain the imprint of previous feedback episodes. Some of the coldest sources may therefore have experienced a more powerful wind phase in the recent past, which displaced or redistributed the dust, while the currently observed C\,{\sc iv} wind corresponds to a subsequent, less energetic episode. Variations in wind duty cycle, source geometry, and viewing angle could further contribute to the observed scatter. Although this scenario provides a coherent qualitative explanation of the data, spatially resolved observations of the dust and independent constraints on the wind energetics on a larger sample will be required to test whether the cold dust is indeed more extended in QSOs with the largest C\,{\sc iv} blueshifts.

J2054--0005, shown as a square in Fig. \ref{fig:gammaT}, is not part of HYPERION, but it is currently the only additional $z\sim6$ QSOs with sufficiently good X-ray coverage for this analysis. Interestingly, J2054--0005 follows the same trends defined by the HYPERION QSOs in both the $\Gamma$--$T_{\rm dust}$ and $v_{\rm C\,IV}$--$T_{\rm dust}$ planes. Although based on a single object, this agreement suggests that the connection between dust heating, disc winds, and the X-ray corona may not be unique to the HYPERION selection, but could instead be a more general feature of luminous high-redshift QSOs. Confirming this scenario will require homogeneous FIR SED modelling and deep X-ray observations for larger samples of $z>6$ quasars.

\section{Conclusions}
\label{sec:conc}

We have presented new ALMA Band 3 observations of ten HYPERION QSOs at $z>6$, targeting the CO(6--5) emission line and the underlying $\sim100$ GHz dust continuum. These observations were combined with archival and literature ALMA/NOEMA data to investigate the molecular gas reservoirs, cold-dust properties, gas-to-dust ratios, and relative SMBH--host-galaxy growth in one of the most extreme samples of quasars known at the epoch of Reionization. Our main results are summarized below.

\begin{itemize}

\item We detect $\sim100$ GHz continuum emission in eight of the ten targeted HYPERION QSOs. The continuum emission is generally resolved on scales of several kpc, indicating that the cold dust is distributed over the host galaxy. CO(6--5) emission is detected in four QSO hosts, namely J025--33, J083+11, J231--20, and J0252--0503, and also in the companion galaxy of J231--20.

\vspace{0.1cm}

\item The CO(6--5)-based molecular gas masses of the detected QSO hosts are $M_{\rm H_2}\simeq(0.6-2.6)\times10^{10}~{\rm M_\odot}$ under our fiducial assumptions. For the non-detected sources, we derive upper limits of a few $10^9~{\rm M_\odot}$. For J231-20 specifically, we derived a solid estimate of $M_{\rm H_2}$ from the modelling of its CO SLED. When combined with previous measurements, molecular gas constraints are now available for a substantial fraction of the HYPERION sample. The resulting census shows that HYPERION QSOs span a broad range of molecular gas masses: some are consistent with the typical $\sim10^{10}~{\rm M_\odot}$ reservoirs observed in luminous $z>6$ QSOs, while others appear to occupy the lower-$M_{\rm H_2}$ tail.

\vspace{0.1cm}

\item The CO(6--5) kinematics reveal diversity within the sample. J025--33 shows a clear velocity gradient, consistent with rotating gas. J083+11 does not show an ordered velocity field and appears instead to be dispersion dominated, suggesting turbulent or disturbed gas motions. In J231--20, the velocity structure and the offset between the QSO and its companion suggest an ongoing interaction, with the companion contributing a significant fraction of the molecular gas budget of the system.

\vspace{0.1cm}

\item We detect [NII]$\lambda 205\,\mu$m emission in J025--33 and tentatively in J083+11, adding two new HYPERION QSOs to the small sample of $z>6$ QSO hosts with [NII]-based ionized-gas constraints. Their high SFRs relative to $L_{\rm [NII]205}$ suggest that the [NII] emission may arise from dense or highly structured ionized gas, where [NII]$205\,\mu$m becomes inefficient at tracing the ionizing photon rate, and thus the SFR. The comparison with local (U)LIRGs further suggests that similar mechanisms may regulate the [NII] deficit across cosmic time.

\vspace{0.1cm}

\item For J025--33 and J083+11, the available FIR photometry samples both the Rayleigh--Jeans side and the peak of the cold-dust SED, allowing us to constrain the dust properties. We find $T_{\rm dust}=36^{+13}_{-7}$ K and $\log(M_{\rm dust}/M_\odot)=8.4\pm0.1$ for J025--33, and $T_{\rm dust}=32^{+4}_{-3}$ K and $\log(M_{\rm dust}/M_\odot)=8.9\pm0.1$ for J083+11. These dust temperatures are significantly lower than the average value of $T_{\rm dust}\simeq54$ K found for luminous $z>6$ QSOs. This may indicate that the rest-frame FIR emission in these two systems is dominated by star-formation-heated dust, with a smaller contribution from AGN heating than usually assumed.

\vspace{0.1cm}

\item Combining molecular gas and dust masses, we find a low gas-to-dust ratios for J083+11: ${\rm GDR}=16^{+5}_{-4}$. This is among the lowest GDR values measured so far in a high-redshift QSO host, and  indicate that at least some HYPERION hosts have already undergone rapid dust and metal enrichment less than one billion years after the Big Bang.

\vspace{0.1cm}

\item The new measurements follow the anti-correlation between GDR and $M_{\rm dust}$ observed in other QSO and Seyfert samples. We find $r_{\rm Pearson}=-0.66^{+0.12}_{-0.09}$ and $\rho_{\rm Spearman}=-0.64\pm0.12$, supporting the idea that dust-rich systems tend to have lower gas-to-dust ratios, likely reflecting more advanced chemical enrichment and efficient conversion of metals into dust grains.

\vspace{0.1cm}

\item By revisiting the comparison between SMBH and host-galaxy growth, we find that most QSOs for which this analysis is possible lie either in the galaxy-dominated regime or close to the symbiotic-growth locus. This suggests that many luminous $z\sim6$ QSOs are no longer observed during a phase of maximum relative SMBH growth, but rather during a stage in which the host galaxy is rapidly assembling and may be catching up with the already massive SMBH. At the same time, HYPERION QSOs occupy different regions of the SMBH--galaxy growth plane, indicating that the selection does not isolate a single instantaneous evolutionary phase. Instead, these objects may trace different stages of a rapid evolutionary sequence, from BH-dominated growth to symbiotic or galaxy-dominated assembly, with a tentative redshift dependence that will require a larger and more homogeneous sample to test.

\vspace{0.1cm}

\item Finally, we explored possible links between the host-galaxy dust properties, the X-ray corona, and accretion-disc winds. The $\Gamma$--$T_{\rm dust}$ plane shows a tentative inverted-V-shaped behaviour, while the $T_{\rm dust}$--$v_{\rm C\,IV}$ relation suggests a weak anti-correlation. If confirmed, these trends may indicate that more powerful winds redistribute dust away from the central AGN heating source, lowering its temperature and increasing the relative contribution of star formation to the FIR emission. This would point to a multi-scale connection among the X-ray corona, disc winds, feedback, and dust heating in the host galaxy.

\end{itemize}

Overall, our results show that the host galaxies of HYPERION QSOs are highly diverse in their molecular gas content, dust enrichment, and evolutionary state. The most rapidly growing SMBHs at the epoch of Reionization are not always embedded in the most massive molecular reservoirs, and some of their hosts already show large dust masses and very low gas-to-dust ratios. This diversity suggests that SMBH growth, star formation, gas consumption, dust enrichment, and feedback proceed on different timescales in the first luminous quasars. Completing the molecular gas and FIR SED census of the full HYPERION sample, ideally with deeper multi-transition CO observations and homogeneous X-ray constraints, will be essential to determine whether these trends are representative of the broader population of luminous high-redshift QSOs.

\begin{acknowledgements}
    This paper makes use of the following ALMA data: ADS/JAO.ALMA 2024.1.01105.S, 2024.1.00071.S, 2021.2.00064.S, 2021.1.00934.S, 2018.1.01188.S, 2023.1.00443.S, 2019.1.01025.S. ALMA is a partnership of ESO (representing its member states), NFS (USA) and NINS (Japan), together with NRC (Canada), MOST and ASIAA (Taiwan) and KASI (Republic of Korea), in cooperation with the Republic of Chile. The Joint ALMA Observatory is operated by ESO, AUI/NRAO and NAOJ. The project leading to this publication has received support from ORP, that is funded by the European Union’s Horizon 2020 research and innovation programme under grant agreement No 101004719 [ORP]. CMP, RT, and LP acknowledge support from PRIN 2022 MUR project 2022CB3PJ3 - First Light And Galaxy aSsembly (FLAGS) funded by the European Union – Next Generation EU and from the ERC synergy grant 101166930 - RECAP. CF acknowledges financial support from the Ricerca Fondamentale INAF 2023 Data Analysis grant 1.05.23.03.04 ``ARCHIE ARchive Cosmic HI \& ISM  Evolution''. FE acknowledges support from the Spanish grant PID2022-138560NB-I00, funded by MCIN/AEI/10.13039/501100011033/FEDER, EU. MVZ acknowledges financial contribution from INAF project "VLT-MOONS" CRAM 1.05.03.07.

\end{acknowledgements}

\bibliography{biblio}{}
\bibliographystyle{aa}

\appendix
\section{Tables}
\nolinenumbers

Tabs.~\ref{tab:res-line-obs} and \ref{tab:res-cont} report the properties of the line map and the results of the continuum analysis for all the targets, respectively.

\begin{table*}
\centering
\caption{Properties of the line cube.}
\begin{tabular}{lccc}
\hline
Target & Coordinates & Beam & r.m.s \\
       & $[\mathrm{RA, Dec}]$ & [arcsec$^2$] & [mJy/beam] \\
\hline
J029-36 & -- & -- & -- \\

J025-33 & \makecell{1:42:43.71, -33:27:45.43} 
& $0.68 \times 0.41$ 
& 0.02 \\

J083+11 & \makecell{5:35:20.89, 11:50:53.90} 
& $0.78 \times 0.72$ 
& 0.03 \\

J011+09 & -- & -- & -- \\

J231-20 & \makecell{15:26:37.83, -20:50:00.90} 
& $0.96 \times 0.59$ 
& 0.04 \\

J231-20c & \makecell{15:26:37.87, -20:50:02.35} 
& $0.96 \times 0.59$ 
& 0.04 \\

J0244-5008 & -- & -- & -- \\

J0252-0503 & \makecell{2:52:16.65, -5:03:32.07} 
& $0.71 \times 0.44$ 
& 0.05 \\

J0038-1527 & -- & -- & -- \\
\hline
\end{tabular}
\flushleft
\footnotesize{
{\bf Notes.} Columns: target name, source coordinates, beam size, and r.m.s in a channel in the continuum subtracted data cube. The r.m.s is computed from an emission-free region.
}
\label{tab:res-line-obs}
\end{table*}

\begin{table*}[]
    \centering
    \caption{Results for the continuum emission}
    \begin{tabular}{c|cccccccc}
    \hline \hline
    QSO     & band & Freq & Flux density & Size & Beam & r.m.s & Project ID & Ref. \\
            &  & [GHz] & [mJy] & [$''\times''$] & [$''\times''$] & [mJy/beam]  & &\\

    \hline
    J029-36 & 3 &  104 & $0.06\pm 0.02$ & 0.37 $\times $ 0.68  & 0.49 $\times $ 0.35  & 0.006 & 2024.1.01105.S & TW \\
            & 6$^\dagger$ & 262 & $1.99\pm0.09$ & 0.33 $\times$ 0.22 & 0.34 $\times $ 0.31 & & 2016.1.01510.S & A22\\
    \hline
    J025-33 & 3 & 99 & $0.08\pm0.03$ & 0.85 $\times$ 0.71 & 0.62 $\times$ 0.39 & 0.006 & 2024.1.01105.S & TW\\
            & 5 & 192 & $1.12\pm0.04$ & $1.56 \times 1.87$ & $1.66\times 1.40$ & 0.019 & 2024.1.00071.S & TW\\
            & 6$^\dagger$ & 259 & $2.49\pm0.11$ & 0.27 $\times$ 0.19 & 0.24 $\times$ 0.21  & - & 2017.1.01301.S & V20\\
            & 7 & 342 & $4.67\pm0.23$ & $1.10 \times 0.88$ & $0.97\times 0.79$ & 0.043 & 2024.1.00071.S & TW \\
            & 8 & 407 &  $6.00 \pm 0.28 $ & $2.02 \times 1.65 $& $1.95 \times 1.63 $& 0.082 & 2021.2.00064.S & TW\\ 
    \hline 
    J083+11 & 3 & 101 & $0.28\pm0.04$  & 0.75 $\times$ 0.81 & 0.71 $\times$ 0.74 & 0.006 & 2024.1.01105.S & TW\\ 
            & 4$^\dagger$ & 145 & $0.90\pm0.33$  & - & -& - & 2021.2.00151.S & S25\\ 
            & 5 &  192 & $2.69\pm 0.10$ & $1.85\times 1.67 $ & $1.69\times 1.59$& 0.047 & 2024.1.00071.S & TW \\
            & 6$^\dagger$ & 244 & $5.10\pm0.15$ & $0.40\times0.30$ & $0.42\times 0.37$ & - & 2019.1.01436.S & Ak20\\
            & 6$^\dagger$  &  258  &  $5.54\pm0.16$   & $0.40\times0.30$ & $0.42\times 0.37$ & -& 2019.1.01436.S & Ak20\\
            & 7$^\dagger$ & 348 & $8.81 \pm 0.88$ & - & $0.59\times 0.64$& - & 2021.1.00443.S & S25\\
            & 8$^\dagger$ & 407 & $10.7\pm1.3$ & - & -& - & 2021.2.00064.S & S25\\
    \hline
    J011+09 & 3 & 100 & $0.07\pm0.02$ &0.70 $\times $ 0.80 & 0.67 $\times $ 1.03 & 0.006 & 2024.1.01105.S & TW \\
            & 6$^\dagger$ & 248 & $1.20\pm0.01$ & -& -&-& 2017.1.00332.S & E20\\
    \hline
    J231-20 & 3 &  97 & $0.25 \pm 0.04$ & $0.53 \times 0.95$ & $0.89 \times 0.55$ &  0.008 & 2024.1.01105.S & TW\\
    \hline
    J231-20c & 3 & 97 & $0.02 \pm 0.01$ & $0.38 \times 0.70$ & $0.89 \times 0.55$ &   0.008 & 2024.1.01105.S & TW\\
    \hline
    J0244-5008   & 3 & 96 & $0.04\pm0.02$ &0.58 $\times $ 0.90 & 0.63 $\times $ 0.41 & 0.006 & 2024.1.01105.S & TW\\
                 & 6tap & 254 & 1.18 $\pm$ 0.04 & 0.62 $\times$ 0.56 & 0.56 $\times$ 0.54 & 0.020 &2021.1.00934.S& TW\\
    \hline
    J0411-0907  & 3 & 93 & $<0.018$ & - & $0.51\times 0.37$ & $0.006$ & 2024.1.01105.S & TW\\
                & 6 & 236 & $0.19\pm 0.06$ & $0.72\times 0.91$ & $0.71\times 0.56$ & 0.019 & 2018.1.01188.S & TW \\
    \hline
    J0020-3653  & 3 & 93 & $<0.018$ & - & $1.09 \times 1.01 $ & $0.006$ & 2024.1.01105.S & TW \\
                & 6 & 236 & $0.45\pm 0.10$ & $1.46\times 2.52$ & $1.71\times 1.25$ & 0.035& 2023.1.00443.S & TW\\
    \hline
    J0252-0503   & NOEMA$^\dagger$ & 88 & $0.077 \pm 0.015$ & -- & $6.6\times 4.9$ & 0.016 & S23CX & S24 \\
                 & NOEMA$^\dagger$ & 101 & $0.123\pm0.018$ & -- & $5.6\times 4.2$ & 0.019 & S23CX & S24 \\
                 & 3 & 93 & $0.11\pm0.02$ & 0.89 $\times $ 0.68 &0.63 $\times $ 0.41& 0.007 & 2024.1.01105.S & TW\\
                 & 6$^\dagger$ & 238 & $1.04\pm0.10$ & 0.72 $\times$ 0.55 & 0.56 $\times$ 0.40 & 0.035 & 2019.1.01025.S & S24\\
    \hline
    J0038-1527   & 3 & 92 & $0.03\pm 0.01$ & 0.63 $\times $ 0.99 &0.99 $\times $ 0.75 & 0.006 & 2024.1.01105.S & TW\\
                 & 6$^\dagger$ & 229 & $0.90\pm 0.06$ &0.86 $\times $ 0.68& 0.60 $\times$ 0.54  & - & 2018.1.01188.S & W24\\

    \hline \hline

    \end{tabular}
    \flushleft {\footnotesize {\bf Notes.} Columns: target name, ALMA band, frequency of the continuum emission, derived flux density of the continuum emission, deconvolved size of the source, beam size, r.m.s of the continuum map, project ID of the observation, reference. $^\dagger$: results retrieved from the literature. References: A22 \citep{ansarinejad2022}; V20 \citep{venemans2020}; S25 \citep{spilker2025}; Ak20 \citep{andika2020}; E20 \citep{eilers2020}; S24 \citep{salvestrini2025}.}
    \label{tab:res-cont}
\end{table*}

\section{Figures}

\begin{figure*}
    \centering
    \includegraphics[width=0.33\linewidth]{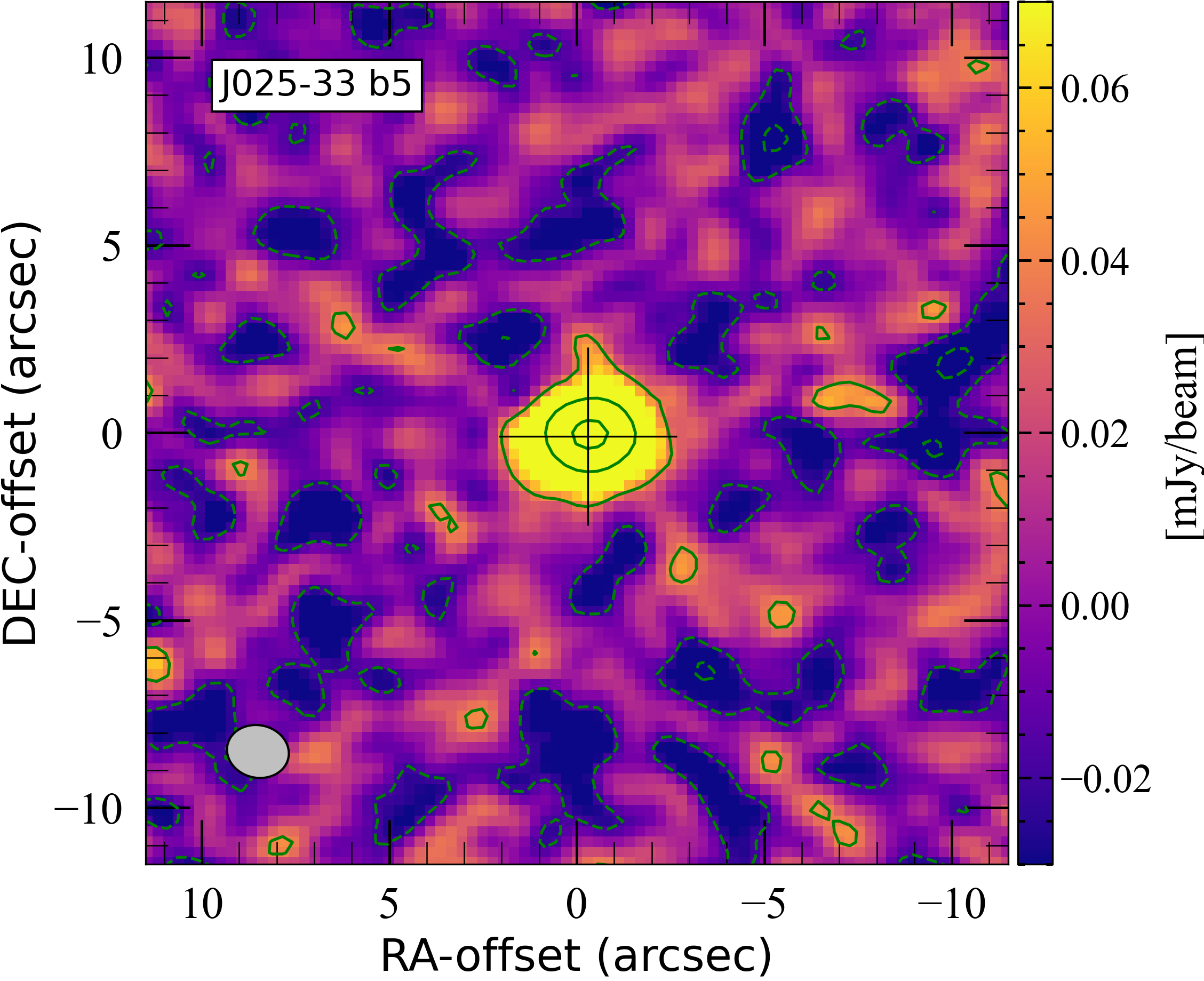}
    \includegraphics[width=0.33\linewidth]{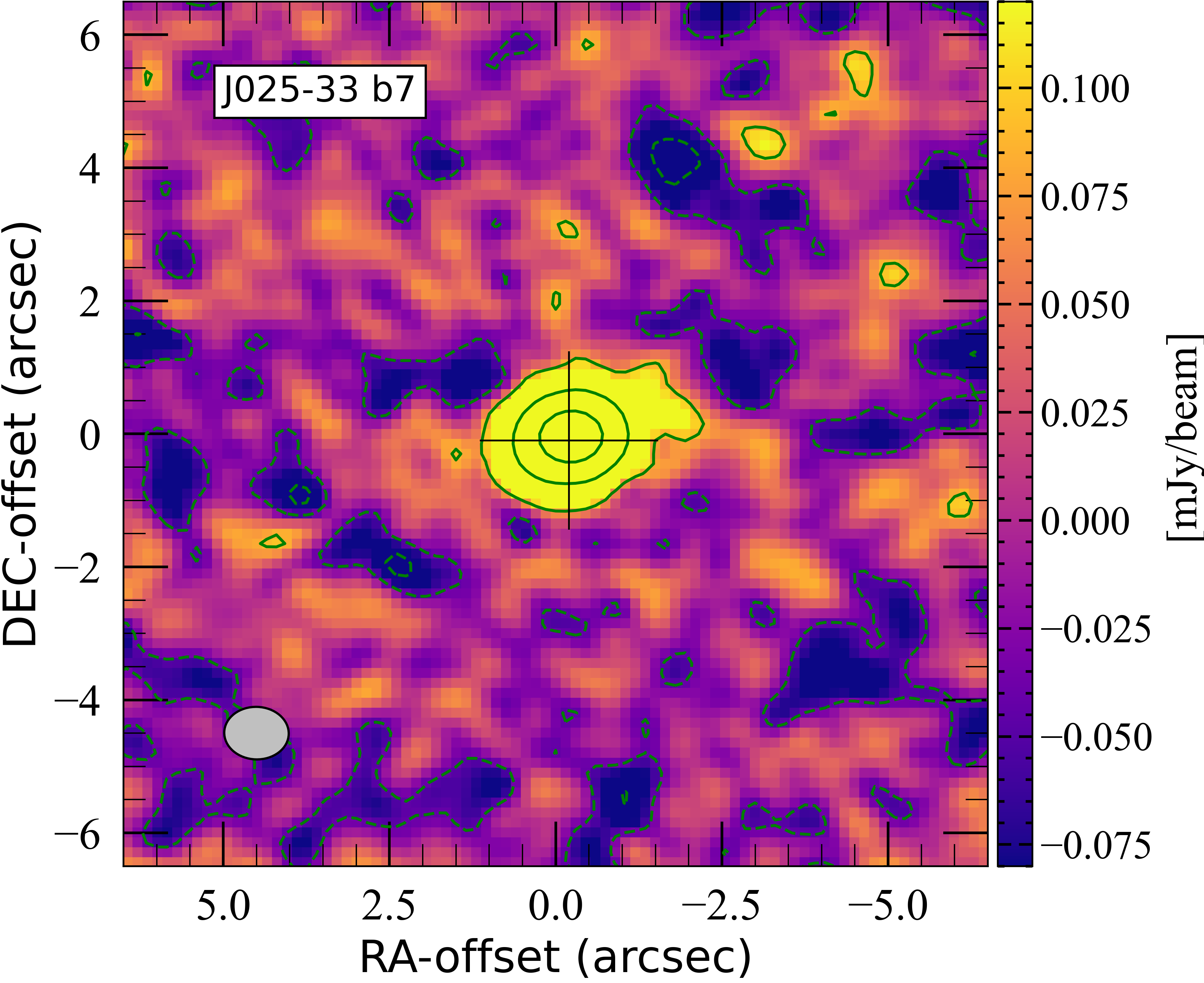}
    \includegraphics[width=0.33\linewidth]{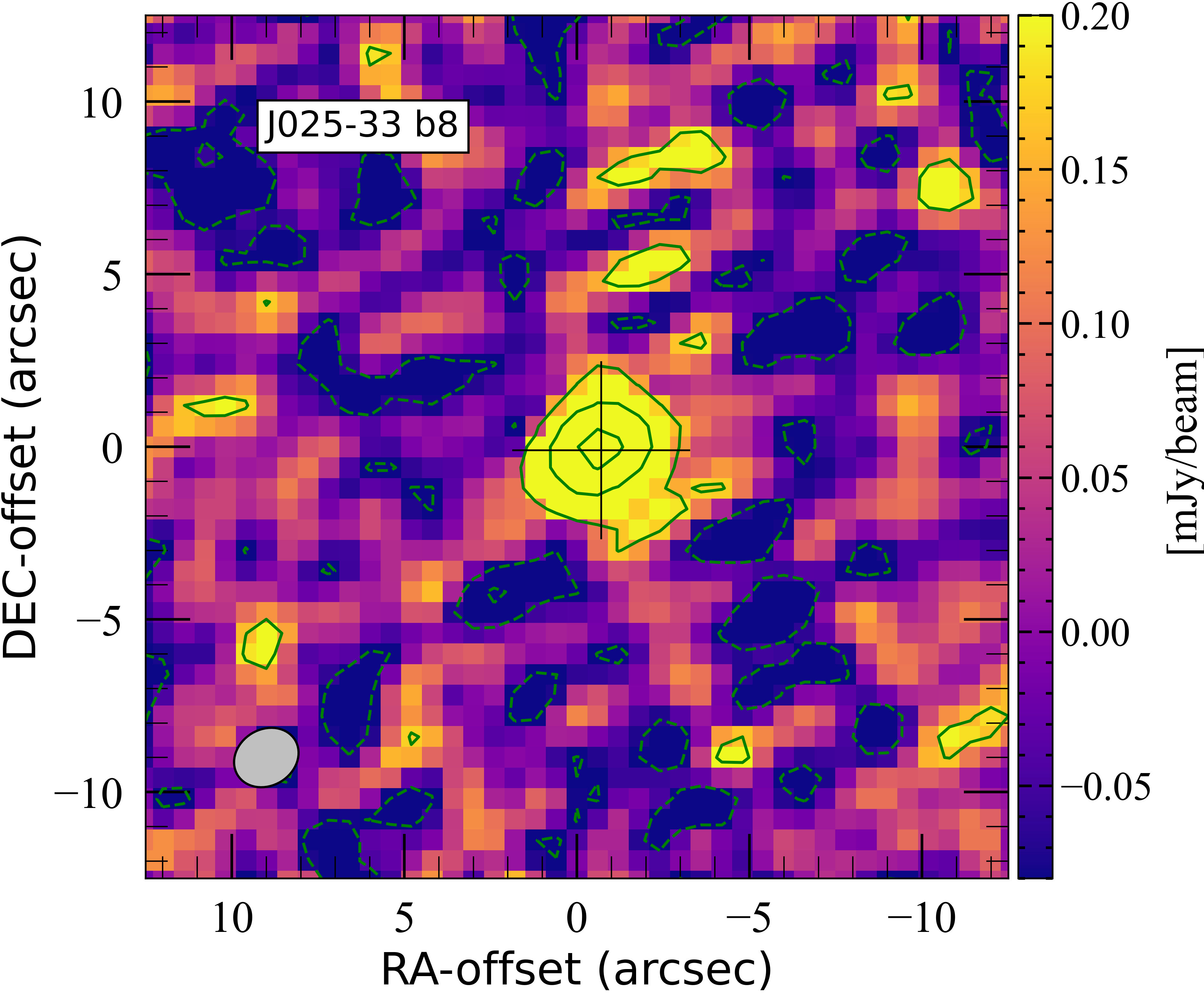}\\
    \includegraphics[width=0.33\linewidth]{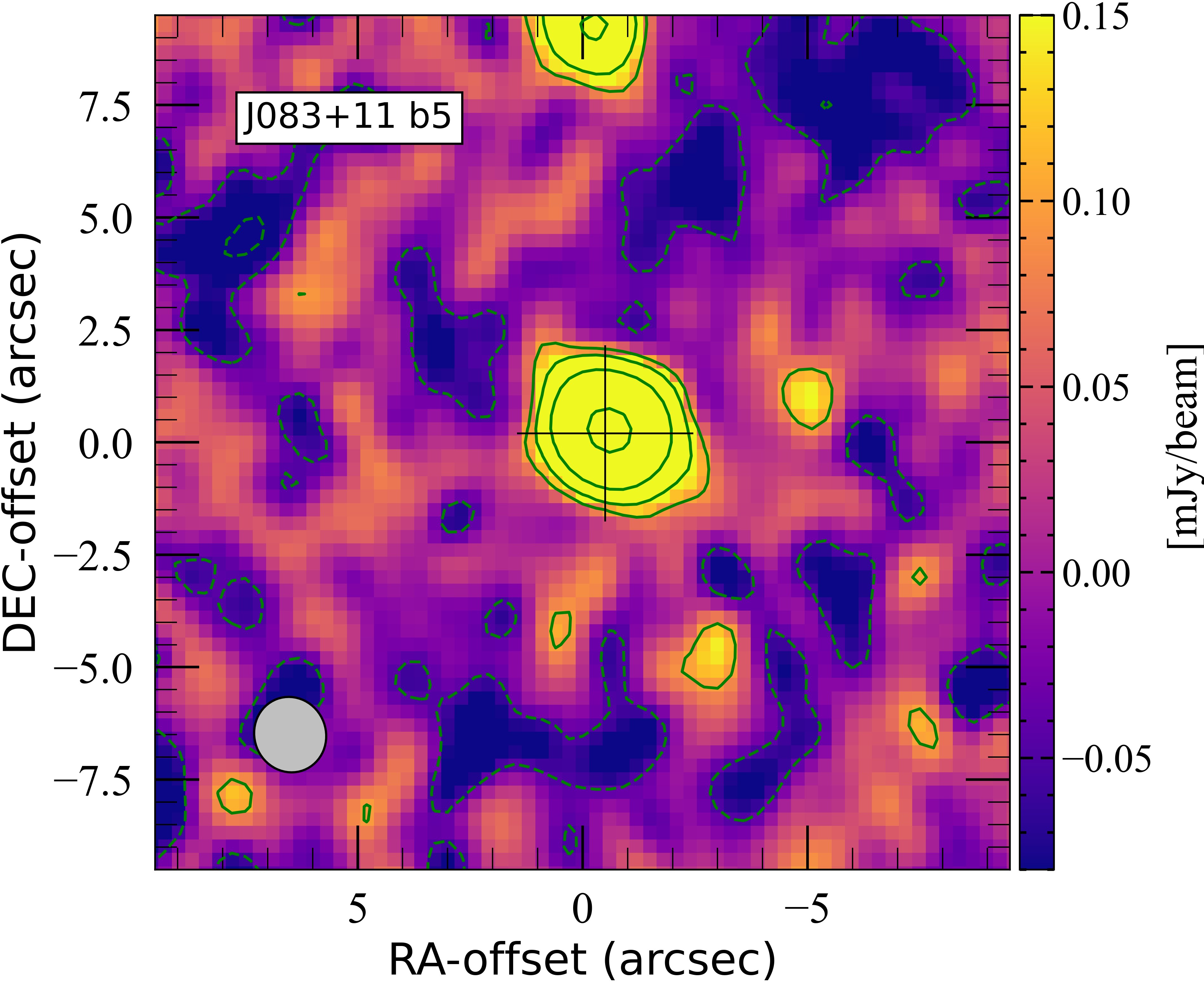}
    \includegraphics[width=0.33\linewidth]{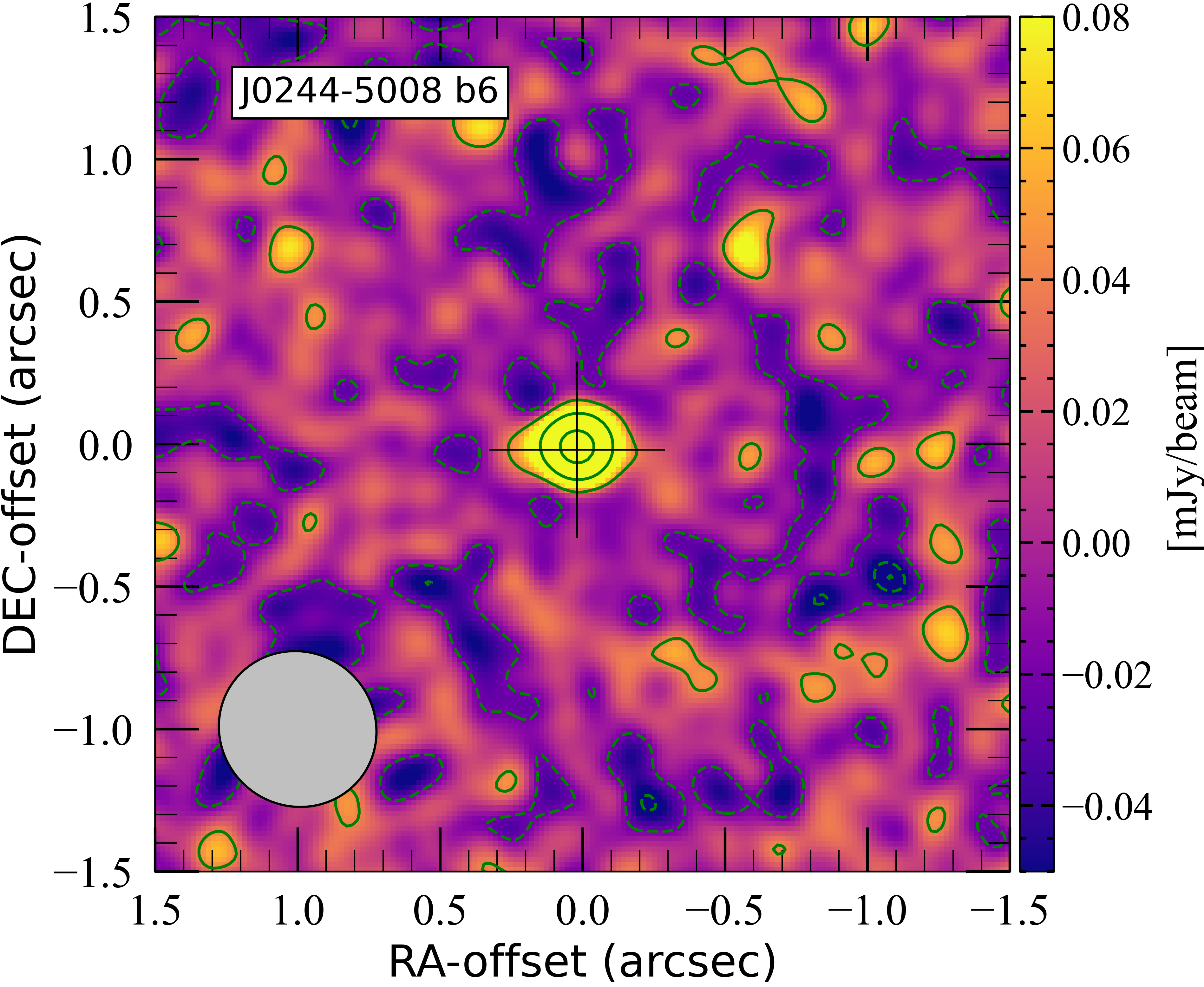}
    \caption{Dust continuum maps. Top panels: from left to right, dust continuum maps in B5, B7 and B8 of QSO J025-33. Bottom left panel: dust continuum maps in B5 of QSO J083+11. Bottom right panel: dust continuum maps in B6 of QSO J0244-5008. The clean beam is shown in gray in the lower left corner of each panel. The size of the beam and the rms of each map is reported in Tab. \ref{tab:res-cont}. Contours start from $2\sigma$ level for each map.}
    \label{fig:cont-otherbands}
\end{figure*}

\begin{figure*}
    \centering
    \includegraphics[width=1\linewidth]{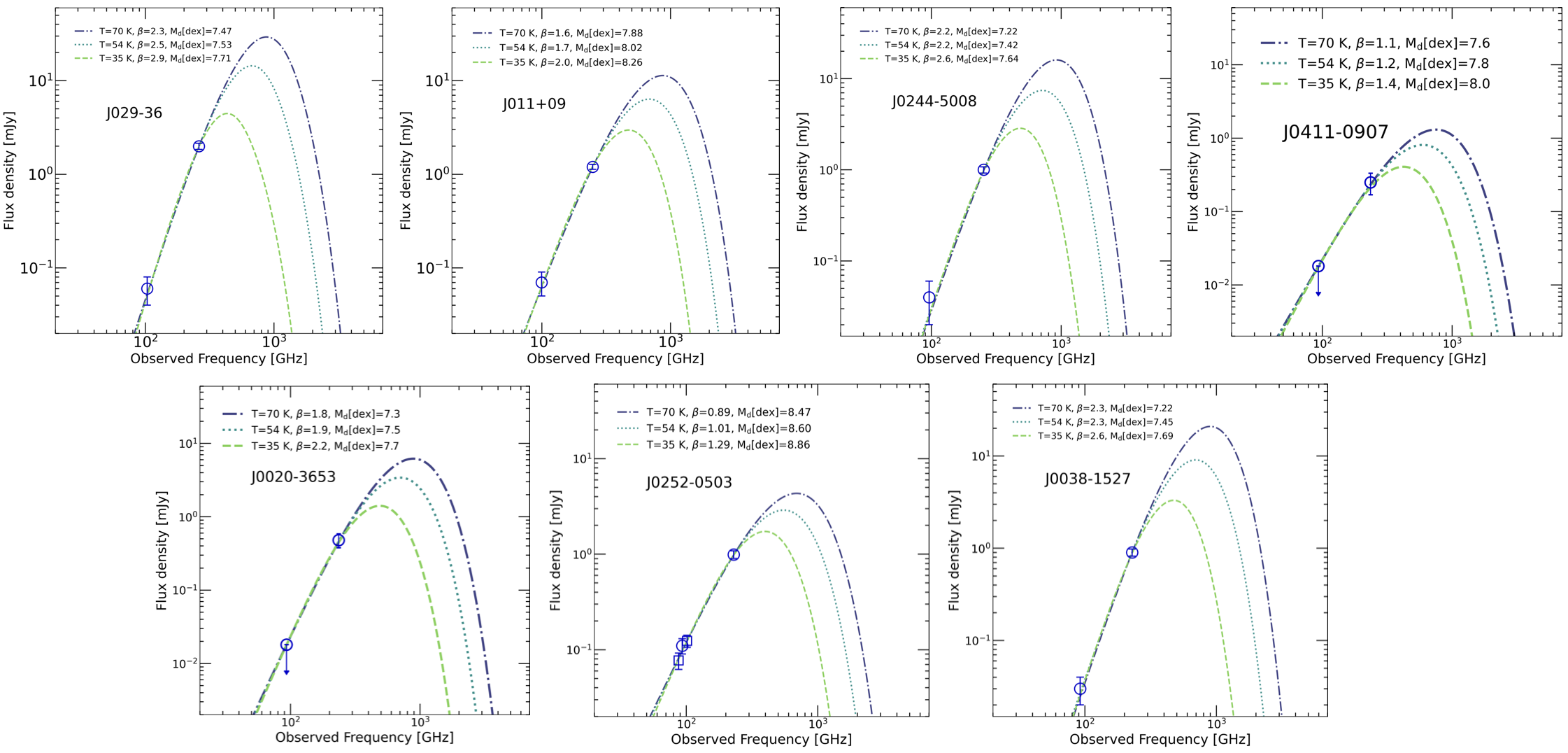}
    \caption{Cold dust SED fitting results for targets with B3 and B6 observations only. ALMA observed photometric data are plotted as blue edged circles with errorbars, and NOEMA data as blue edged squares. Best-fitting models are shown as dashed lines for models at fixed $T_{\rm dust}$ and $\beta$ color coded as in the legend of each panel.}
    \label{fig:sed}
\end{figure*}

\begin{figure}
    \centering
    \includegraphics[width=0.9\linewidth]{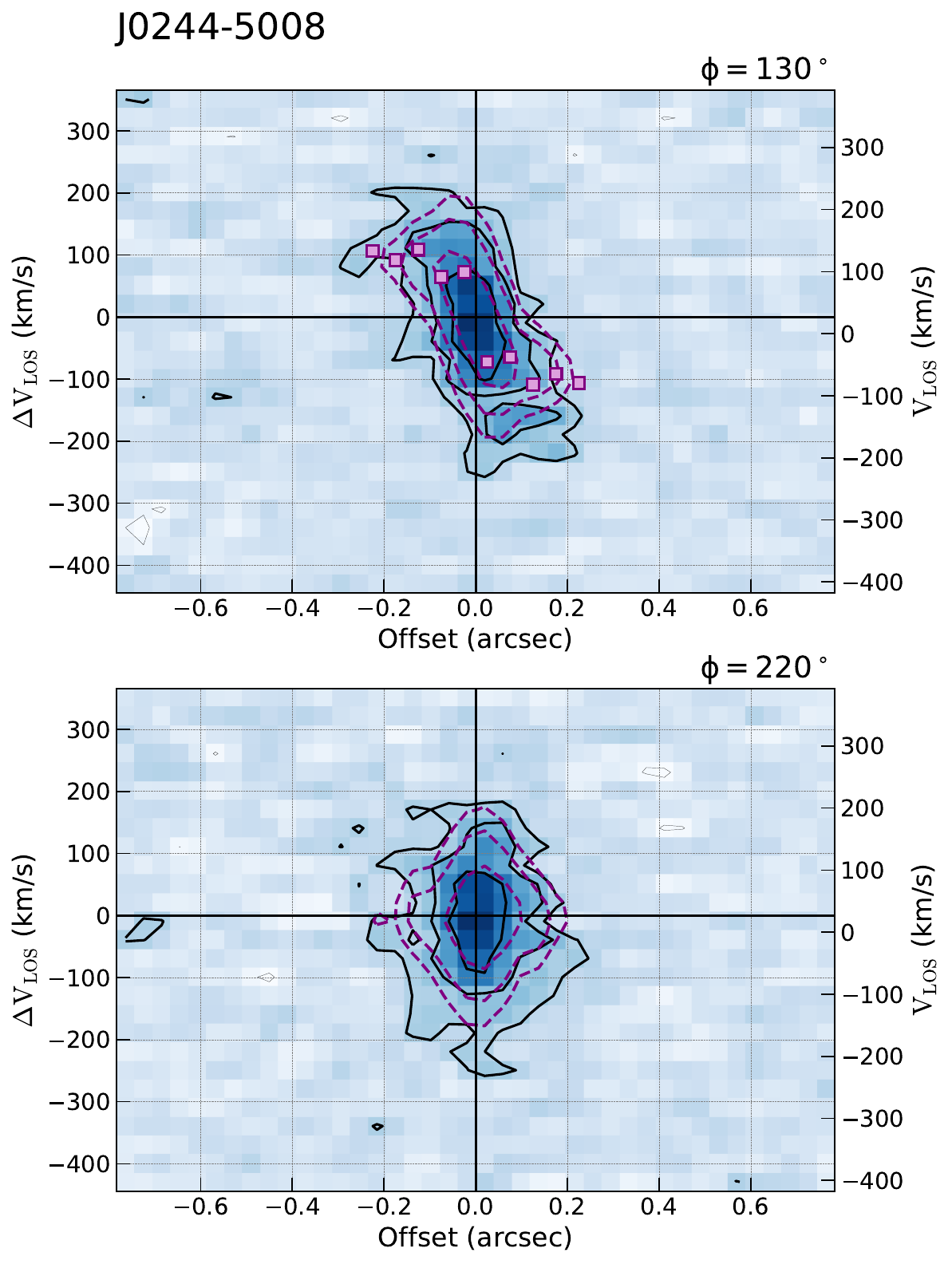}
    \caption{PV diagram of the [CII] emission of J0244-5088 along the major (top) and minor (bottom) axes. The observed data are colored in blue, while the $^{\rm 3D}$Barolo model is marked as purple solid line. Violet squares are the modelled line-of-sight velocities.}
    \label{fig:pv}
\end{figure}

Fig.~\ref{fig:cont-otherbands} shows the maps of the continuum emission for our targets from ancillary ALMA data in bands other than B3 analysed in this work. Fig.~\ref{fig:sed} shows the SED fitting results for the targets that only have B3 and B6 observations available. Fig.~\ref{fig:pv} shows the PV diagram along the major (top) and minor (bottom) axis of the [CII] emission of J0244-5008.

\section{CO SLED model: Corner plot and tables}
\label{app:corner-COSLED}

\begin{figure}
    \centering
    \includegraphics[width=0.8\linewidth]{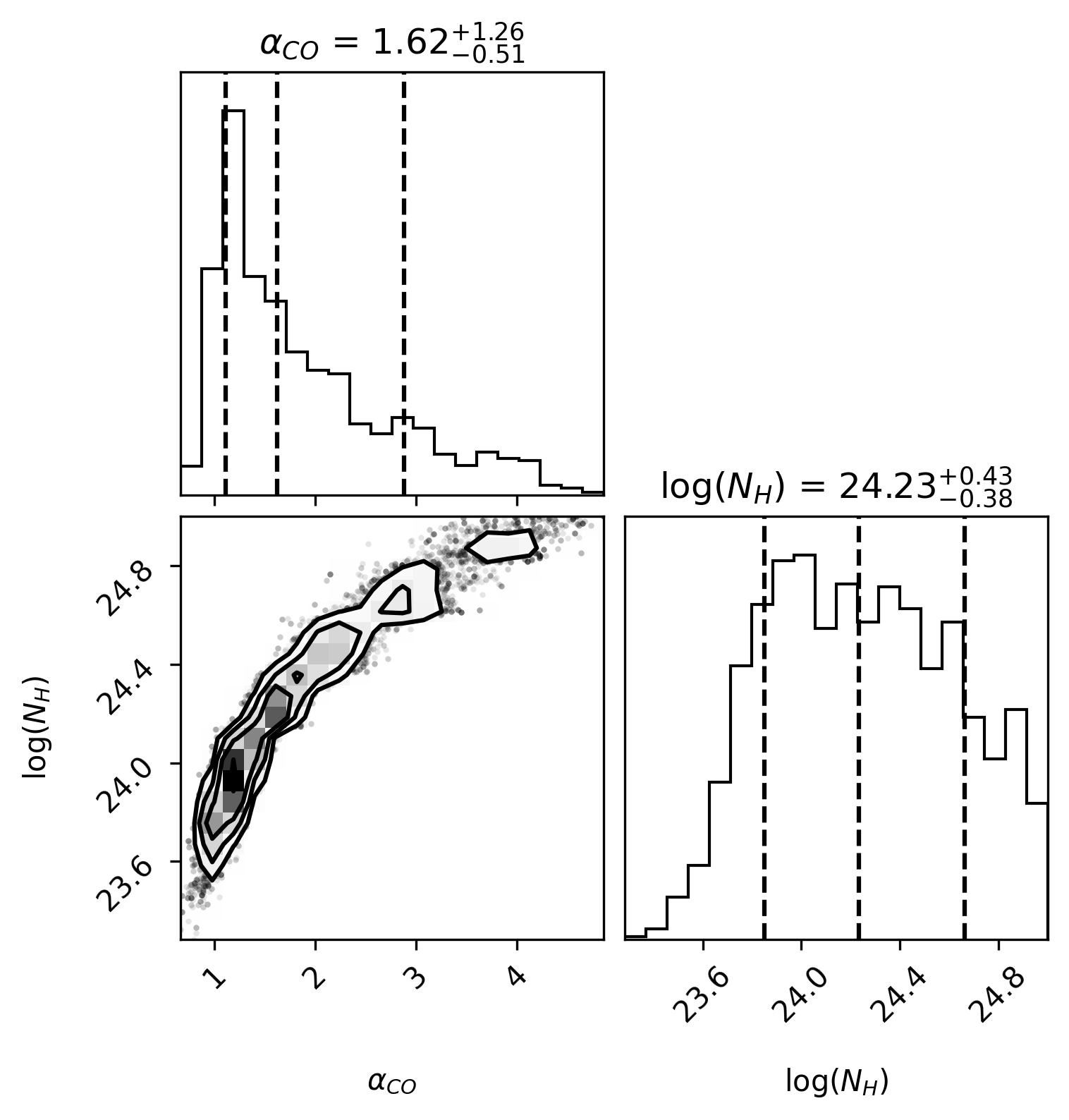}
    \caption{Corner plot showing the posterior distributions of the CO-to-H$2$ conversion factor, $\alpha{\rm CO}$, and the hydrogen column density, $\log(N_{\rm H})$. Dashed lines indicate the median and 16th–84th percentile ranges, while contours show the joint probability distribution.}
    \label{fig:corner-COSLED}
\end{figure}

Fig.~\ref{fig:corner-COSLED} shows the corner plot of the posterior distribution of the $\alpha_{CO}$ and $N_H$ parameters of the best-fit model for the CO SLED of P231-20, described in Section~\ref{sec:disc-cosled}. 
Table~\ref{tab:co_sled} reports the best-fit parameters obtained from the CO SLED modelling. The first column indicates the prescription adopted to derive the input cold gas mass, $M_{\rm H2,input}$. The first three rows correspond to the cases in which a fixed GDR is assumed; K24 refers to the use of the mean CO SLED ladder from \cite{kaasinen2024}, while S25 refers to the calibration based on [CII] emission from \cite{salvestrini2025}. The third column lists the output cold gas mass, $M_{\rm H2,output}$, obtained using the parameters reported in the following columns. The fourth and fifth columns show the 50th percentiles of the posterior distributions of $\alpha_{\rm CO}$ and $N_{\rm H}$, respectively, together with their $1\sigma$ uncertainties. The corresponding $\chi^2_{\rm dof}$ values are reported in the last column.

\begin{table*}[t]
\centering
\caption{Best-fit molecular gas parameters.}
\label{tab:co_sled}
\begin{tabular}{cccccc}
\hline
Prescription
& $M_{\rm H2, input}$ 
& $M_{\rm H2, output}$ 
& $\alpha_{CO}$ 
& $N_{\rm H}$ 
& $\chi^{2}_{dof}$\\
\hline
 
& $[10^{10} M_\odot]$ 
& $10^{10} [M_\odot]$ 
& $[M_\odot\,({\rm K\,km\,s^{-1}\,pc^2})^{-1}]$ 
& $[{\rm cm^{-2}}]$ 
& \\
\hline
$GDR=100$ & 5.3 & 2.75 & $ 0.41 ^{+ 0.33 }_{- 0.14 }$ & $ 24.27 ^{+ 0.45 }_{- 0.39 }$ & 2.05 \\[0.1cm]
$GDR=75$ & 3.98 & 2.91 & $ 0.59 ^{+ 0.43 }_{- 0.21 }$ & $ 24.33 ^{+ 0.4 }_{- 0.43 }$ & 2.18 \\[0.1cm]
$GDR=50$ & 2.65 & 2.7 & $ 0.82 ^{+ 0.62 }_{- 0.27 }$ & $ 24.24 ^{+ 0.44 }_{- 0.39 }$ & 1.97 \\[0.1cm]
K24 & 2.1 & 2.72 & $ 1.04 ^{+ 0.8 }_{- 0.34 }$ & $ 24.25 ^{+ 0.44 }_{- 0.39 }$ & 1.98 \\[0.1cm]
S25 & 1.31 & 2.65 & $ 1.62 ^{+ 1.26 }_{- 0.51 }$ & $ 24.23 ^{+ 0.43 }_{- 0.38 }$ & 1.96 \\[0.1cm]

\hline
\end{tabular}
\caption*{\small Columns: prescription adopted to derive the input cold gas mass; input and best-fit cold gas masses; 50th percentiles of the posterior distributions of $\alpha_{\rm CO}$ and $N_{\rm H}$, with their $1\sigma$ uncertainties; and $\chi^2_{\rm dof}$ of the CO SLED model corresponding to the 50th-percentile solution.} 
\end{table*}

\section{Comparison between CO- and [CII]-based molecular gas masses}
\label{sec:comparison}

\begin{figure}
    \centering
    \includegraphics[width=0.9\linewidth]{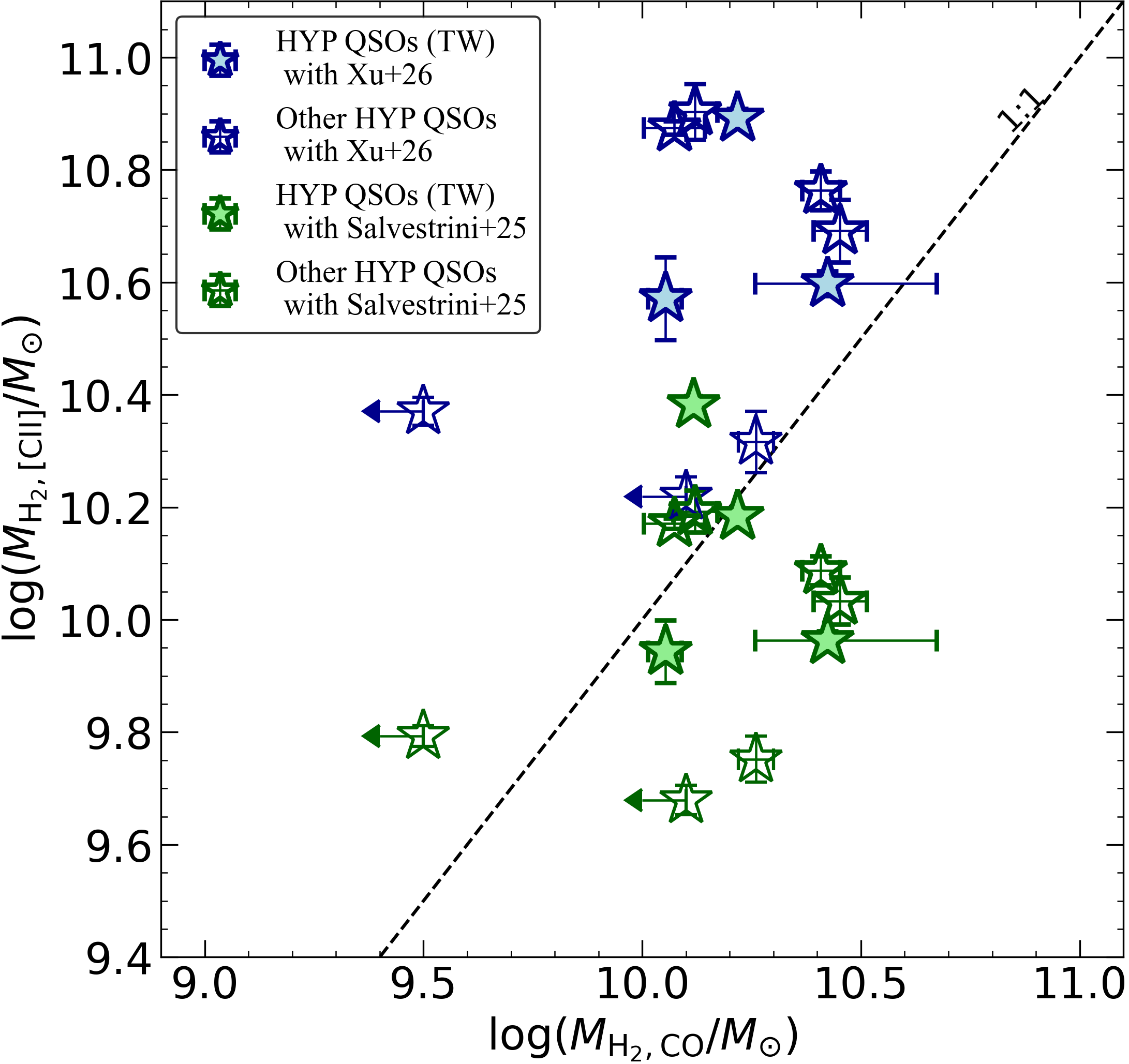}
    \caption{Molecular gas mass estimated via [CII] line vs via CO line. Different colors indicate the different relation used to derive $M_{\rm H_2}$ from the [CII] emission line. Green color is for the relation of \citet{salvestrini2025}, while blue for the one of \citet{xu2026}. Filled stars mark HYPERION QSOs whose CO emission has been analysed in this work.}
    \label{fig:comp-Mh2}
\end{figure}

Before proceeding with the comparison between the [CII] and CO-based estimate of the molecular gas masses, we summarize here the [CII] properties of our target QSOs.

\textbf{J029-36.} \citet{ansarinejad2022} reported the continuum detection in ALMA B6 for J029-36, finding $F_{\rm cont}=1.99\pm 0.09$ mJy at 262 GHz. Due to an inaccurate determination of the redshift of J029-36 in the near-infrared spectral properties, the [CII] emission line was missed by the setup of the B6 observation (see Tab.~\ref{tab:res-cont}). 
 
\textbf{J025-33.} The results for the [CII] emission and its underlying continuum from ALMA B6 data have been already reported in \citet{ansarinejad2022, decarli2018, venemans2020}. The [CII] emission line is detected at $z_{\rm [CII]}=6.3373\pm 0.0002$ with an integrated flux of $5.53\pm0.21\rm ~Jy~km~s^{-1}$, corresponding to $L_{\rm [CII]}=(5.65\pm 0.22)\times 10^9\rm ~L_\odot$ \citep{venemans2020}. \citet{venemans2020} report a [CII] size for J025-33 which is more compact by a factor of 2 than the CO(6-5) sizes measured in our work \citep[see Tab. 3 in][]{venemans2020}. This is mainly due to the higher resolution of their observations ($\sim 5$ times higher than ours), which yields to the loss of the fainter and more extended flux, due to the surface brightness dimming effect \citep[e.g.,][]{carniani2020}. 

\textbf{J083+11.} \citet{andika2020} also reported the detection of the [CII] emission line at $z_{\rm [CII]}=6.3401\pm 0.0004$ with an integrated flux of $F_{\rm [CII]}=10.22\pm 0.35~\rm Jy ~km~s^{-1}$, corresponding to $L_{\rm [CII]}=(1.04\pm0.04) \times 10^{10}\rm ~L_\odot$.  

\textbf{J011+09.}  \citet{eilers2020} reported the detection of the [CII] emission line and its underlying continuum detection in ALMA B6, finding $z_{\rm [CII]}=6.4694\pm0.0002$, an integrated line flux and corresponding luminosity of $F_{\rm [CII]}=0.27\pm0.01\rm ~Jy~km~s^{-1}$, $L_{\rm [CII]}=(2.90\pm0.06 )\times 10^8\rm ~L_\odot$, and a continuum flux of  $F_{\rm cont}=1.20\pm 0.01$ mJy at 254 GHz. 

\textbf{J231-20.} \citet{Pensabene2021} reported the detection of the [CII] emission line in B6, with a luminosity of $L_{\rm [CII]}=(2.87\pm0.15)\times 10^9\rm ~L_\odot$. Similarly to J025-33, \citep{venemans2020} report a [CII] size for J231-20 that is more compact by a factor of 10 than the CO(6-5) sizes measured in our work \citep[see Tab. 3 in][]{venemans2020}, mainly due to the higher resolution of their observations ($20$ times higher than ours). 

\textbf{J0244-5008.}  We also analysed an unpublished observation in B6, targeting and detecting the [CII] line and its underlying continuum emission (see Sect.~\ref{sec:cii-J0244}). 

\textbf{J0411-0907.} We analysed the publicly archival B6 observation, finding a continuum emission of $0.19\pm 0.06$ mJy, which is resolved with a size of $0.72''\times 0.91''$. No [CII] emission has been detected down to $0.32~\rm Jy~km~s^{-1}$ ($3\sigma$ limit, assuming a FWHM$=300~\rm km~s^{-1}$), corresponding to a limit luminosity of $L_{\rm [CII]}<3.7\times 10^8~\rm L_\odot$. 

\textbf{J0020-3653.} We analysed the publicly archival B6 observation, finding a continuum emission of $0.45\pm 0.10$ mJy, which is resolved with a size of $1.5''\times 2.5''$. No [CII] emission has been detected down to $0.52~\rm Jy~km~s^{-1}$ ($3\sigma$ limit, assuming a FWHM$=300~\rm km~s^{-1}$), corresponding to a limit luminosity of $L_{\rm [CII])}<6.1\times 10^8~\rm L_\odot$.

\textbf{J0252-0503.} \citet{wang2024} reported the detection of the [CII] emission line and its underlying continuum detection in ALMA B6, finding $z_{\rm [CII]}=7.0006\pm0.0009$, a [CII] luminosity of  $L_{\rm [CII]}=(2.7\pm0.5 )\times 10^9\rm ~L_\odot$, and a continuum flux of  $F_{\rm cont}=0.99\pm 0.07$ mJy at 230 GHz. Interestingly, the [CII] emission size is consistent with the one found for CO(6-5) in our work.

\textbf{J0038-1527.} \citet{wang2024} reported the detection of the [CII] emission line and its underlying continuum detection in ALMA B6, finding $z_{\rm [CII]}=7.0340\pm0.0003$, a [CII] luminosity of $L_{\rm [CII]}=(3.2\pm0.3 )\times 10^9\rm ~L_\odot$, and a continuum flux of  $F_{\rm cont}=0.90\pm 0.06$ mJy at 229 GHz. Both our estimates of continuum and line emission are fully consistent with previous upper limits placed by \citet{salvestrini2025} with lower sensitivity NOEMA observations. 

To test the robustness of [CII]$\lambda 158\,\mu$m as a tracer of the molecular gas content in HYPERION QSOs, we compare the CO-based molecular gas masses with those inferred from the [CII] luminosity. We consider all HYPERION QSOs for which both [CII] and CO measurements, or meaningful upper limits, are available \citep{neeleman2019, venemans2012, banados2019, yang2020, decarli2018, Pensabene2021, feruglio2018}. The CO-based $M_{\rm H_2}$ values have been recomputed homogeneously for the full comparison sample using the assumptions described in Sect. \ref{sec:cont-line}, namely a common CO excitation correction and CO-to-H$_2$ conversion factor. For the [CII]-based estimates, we adopt two recent calibrations specifically relevant for high-redshift QSOs: Eq.~1 from \citet{salvestrini2025}, calibrated on luminous $z\gtrsim6$ QSO hosts with CO constraints, and the [CII]-based conversion implied by Eq.~14 of \citet{xu2026}, derived within a hierarchical cross-calibration of multiple molecular-gas tracers.

The comparison is shown in Fig.~\ref{fig:comp-Mh2}. The dashed line marks the one-to-one relation between the [CII]- and CO-based molecular gas masses. Overall, the calibration by \citet{salvestrini2025} provides the closest agreement with the CO-based estimates for the HYPERION sample. Most of the green points lie close to, or within a 0.1-0.5 dex from, the one-to-one relation, indicating that this prescription gives a broadly consistent estimate of the molecular gas reservoir in these luminous QSO hosts. In contrast, the masses obtained using the \citet{xu2026} calibration are systematically higher, often overestimating the CO-based $M_{\rm H_2}$ by 0.2-0.8 dex.

This systematic offset highlights the intrinsic difficulty of using [CII] as a molecular-gas tracer in QSO host galaxies. Unlike CO, [CII] does not arise exclusively from the cold molecular phase. Its luminosity can be boosted by emission from ionized or atomic gas, by extended photodissociation regions, by tidally disturbed gas, or by outflowing material, especially in luminous and dynamically complex quasar hosts. Therefore, a single [CII]-to-$M_{\rm H_2}$ conversion may not be universally applicable, even within the high-redshift QSO population. The better agreement obtained with the \citet{salvestrini2025} relation likely reflects the fact that it tied directly to CO-based molecular gas constraints.

We conclude that [CII] remains a valuable proxy for estimating molecular gas masses when CO observations are unavailable, but its calibration must be chosen with caution. For HYPERION-like systems, the \citet{salvestrini2025} relation appears to provide a more consistent match to the CO-based molecular gas scale, while the \citet{xu2026} prescription tends to yield larger gas masses for the same [CII] luminosity. Deeper CO observations and spatially resolved [CII] maps will be essential to determine whether the observed offsets are driven by excitation effects, multiphase [CII] emission, or genuine source-to-source variations in the molecular gas content.
\end{document}

%% file: authors.tex
\author{Claudia~Maria~Pierro
    \inst{1}$^\dagger$
    \and Roberta~Tripodi
    \inst{1,2}$^\dagger$\thanks{\email{roberta.tripodi@inaf.it}}
    \and Laura~Pentericci
    \inst{1}
    \and Francesco~Salvestrini
    \inst{3}
    \and Luca~Zappacosta
    \inst{1}
    \and Alessia~Tortosa
    \inst{4}
    \and Blessing Musiimenta
    \inst{1}
    \and Enrico~Piconcelli
    \inst{1}
    \and Livia~Vallini
    \inst{4}
    \and Fabio~Di~Mascia
    \inst{5}
    \and Federico~Esposito
    \inst{6}
    \and Chiara Feruglio
    \inst{3}
    \and Fabio~Vito
    \inst{4}
    \and Maria~Vittoria~Zanchettin
    \inst{7}
}

 \institute{INAF - Osservatorio Astronomico di Roma, via Frascati 33, I-00078, Monte Porzio Catone, Italy 
         \and
         IFPU - Institute for Fundamental Physics of the Universe, via Beirut 2, I-34151 Trieste, Italy \\ $^\dagger$ These authors equally contributed to this work 
         \and INAF, Osservatorio Astronomico di Trieste, via Tiepolo 11, I-34131, Trieste, Italy
         \and INAF – Osservatorio di Astrofisica e Scienza dello Spazio di
Bologna, Via Gobetti 93/3, I-40129 Bologna, Italy
        \and Scuola Normale Superiore, Piazza dei Cavalieri 7, 56126 Pisa, Italy
        \and Observatorio de Madrid, OAN-IGN, Alfonso XII, 3, E-28014 Madrid, Spain
        \and INAF - Osservatorio Astrofisico di Arcetri, Largo E. Fermi 5, 50125 Firenze, Italy
    }

%% file: main.bbl
\begin{thebibliography}{75}
\expandafter\ifx\csname natexlab\endcsname\relax\def\natexlab#1{#1}\fi

\bibitem[{{Algera} {et~al.}(2026){Algera}, {Rowland}, {Stefanon}, {Palla}, {Sommovigo}, {Inami}, {Bouwens}, {Aravena}, {Bowler}, {Dayal}, {De Looze}, {Ferrara}, {Fisher}, {Graziani}, {Gulis}, {Heintz}, {Hodge}, {Laza-Ramos}, {van Leeuwen}, {Pallottini}, {Phillips}, {Schouws}, {Smit}, {Stark}, \& {van der Werf}}]{algera2026}
{Algera}, H. S.~B., {Rowland}, L., {Stefanon}, M., {et~al.} 2026, \mnras, 545, staf1897

\bibitem[{{Andika} {et~al.}(2020){Andika}, {Jahnke}, {Onoue}, {Ba{\~n}ados}, {Mazzucchelli}, {Novak}, {Eilers}, {Venemans}, {Schindler}, {Walter}, {Neeleman}, {Simcoe}, {Decarli}, {Farina}, {Marian}, {Pensabene}, {Cooper}, \& {Rojas}}]{andika2020}
{Andika}, I.~T., {Jahnke}, K., {Onoue}, M., {et~al.} 2020, \apj, 903, 34

\bibitem[{{Ansarinejad} {et~al.}(2022){Ansarinejad}, {Shanks}, {Bielby}, {Metcalfe}, {Infante}, {Murphy}, {Rosario}, \& {Stach}}]{ansarinejad2022}
{Ansarinejad}, B., {Shanks}, T., {Bielby}, R.~M., {et~al.} 2022, \mnras, 510, 4976

\bibitem[{{Ba{\~n}ados} {et~al.}(2019){Ba{\~n}ados}, {Novak}, {Neeleman}, {Walter}, {Decarli}, {Venemans}, {Mazzucchelli}, {Carilli}, {Wang}, {Fan}, {Farina}, \& {Rix}}]{banados2019}
{Ba{\~n}ados}, E., {Novak}, M., {Neeleman}, M., {et~al.} 2019, \apjl, 881, L23

\bibitem[{{Birkin} {et~al.}(2021){Birkin}, {Weiss}, {Wardlow}, {Smail}, {Swinbank}, {Dudzevi{\v{c}}i{\={u}}t{\.{e}}}, {An}, {Ao}, {Chapman}, {Chen}, {da Cunha}, {Dannerbauer}, {Gullberg}, {Hodge}, {Ikarashi}, {Ivison}, {Matsuda}, {Stach}, {Walter}, {Wang}, \& {van der Werf}}]{birkin2021}
{Birkin}, J.~E., {Weiss}, A., {Wardlow}, J.~L., {et~al.} 2021, \mnras, 501, 3926

\bibitem[{{Bischetti} {et~al.}(2021){Bischetti}, {Feruglio}, {Piconcelli}, {Duras}, {P{\'e}rez-Torres}, {Herrero}, {Venturi}, {Carniani}, {Bruni}, {Gavignaud}, {Testa}, {Bongiorno}, {Brusa}, {Circosta}, {Cresci}, {D'Odorico}, {Maiolino}, {Marconi}, {Mingozzi}, {Pappalardo}, {Perna}, {Traianou}, {Travascio}, {Vietri}, {Zappacosta}, \& {Fiore}}]{bischetti2021}
{Bischetti}, M., {Feruglio}, C., {Piconcelli}, E., {et~al.} 2021, \aap, 645, A33

\bibitem[{{Bolatto} {et~al.}(2013){Bolatto}, {Wolfire}, \& {Leroy}}]{Bolatto2013}
{Bolatto}, A.~D., {Wolfire}, M., \& {Leroy}, A.~K. 2013, \araa, 51, 207

\bibitem[{{Carilli} \& {Walter}(2013)}]{carilli2013}
{Carilli}, C.~L. \& {Walter}, F. 2013, \araa, 51, 105

\bibitem[{{Carniani} {et~al.}(2020){Carniani}, {Ferrara}, {Maiolino}, {Castellano}, {Gallerani}, {Fontana}, {Kohandel}, {Lupi}, {Pallottini}, {Pentericci}, {Vallini}, \& {Vanzella}}]{carniani2020}
{Carniani}, S., {Ferrara}, A., {Maiolino}, R., {et~al.} 2020, \mnras, 499, 5136

\bibitem[{{Carniani} {et~al.}(2019){Carniani}, {Gallerani}, {Vallini}, {Pallottini}, {Tazzari}, {Ferrara}, {Maiolino}, {Cicone}, {Feruglio}, {Neri}, {D'Odorico}, {Wang}, \& {Li}}]{carniani2019}
{Carniani}, S., {Gallerani}, S., {Vallini}, L., {et~al.} 2019, \mnras, 489, 3939

\bibitem[{{Chabrier}(2003)}]{chabrier2003}
{Chabrier}, G. 2003, \pasp, 115, 763

\bibitem[{{Chevance} {et~al.}(2023){Chevance}, {Krumholz}, {McLeod}, {Ostriker}, {Rosolowsky}, \& {Sternberg}}]{Chevance2023}
{Chevance}, M., {Krumholz}, M.~R., {McLeod}, A.~F., {et~al.} 2023, in Astronomical Society of the Pacific Conference Series, Vol. 534, Protostars and Planets VII, ed. S.~{Inutsuka}, Y.~{Aikawa}, T.~{Muto}, K.~{Tomida}, \& M.~{Tamura}, 1

\bibitem[{{Costa} {et~al.}(2026){Costa}, {Decarli}, {Pozzi}, {Cox}, {Meyer}, {Pensabene}, {Venemans}, {Walter}, \& {Xu}}]{costa2026}
{Costa}, M., {Decarli}, R., {Pozzi}, F., {et~al.} 2026, \aap, 706, A285

\bibitem[{{Decarli} {et~al.}(2023){Decarli}, {Pensabene}, {Diaz-Santos}, {Ferkinhoff}, {Strauss}, {Venemans}, {Walter}, {Ba{\~n}ados}, {Bertoldi}, {Fan}, {Farina}, {Riechers}, {Rix}, \& {Wang}}]{decarli2023}
{Decarli}, R., {Pensabene}, A., {Diaz-Santos}, T., {et~al.} 2023, \aap, 673, A157

\bibitem[{{Decarli} {et~al.}(2022){Decarli}, {Pensabene}, {Venemans}, {Walter}, {Ba{\~n}ados}, {Bertoldi}, {Carilli}, {Cox}, {Fan}, {Farina}, {Ferkinhoff}, {Groves}, {Li}, {Mazzucchelli}, {Neri}, {Riechers}, {Uzgil}, {Wang}, {Wang}, {Weiss}, {Winters}, \& {Yang}}]{decarli2022}
{Decarli}, R., {Pensabene}, A., {Venemans}, B., {et~al.} 2022, \aap, 662, A60

\bibitem[{{Decarli} {et~al.}(2017){Decarli}, {Walter}, {Venemans}, {Ba{\~n}ados}, {Bertoldi}, {Carilli}, {Fan}, {Farina}, {Mazzucchelli}, {Riechers}, {Rix}, {Strauss}, {Wang}, \& {Yang}}]{decarli2017}
{Decarli}, R., {Walter}, F., {Venemans}, B.~P., {et~al.} 2017, \nat, 545, 457

\bibitem[{{Decarli} {et~al.}(2018){Decarli}, {Walter}, {Venemans}, {Ba{\~n}ados}, {Bertoldi}, {Carilli}, {Fan}, {Farina}, {Mazzucchelli}, {Riechers}, {Rix}, {Strauss}, {Wang}, \& {Yang}}]{decarli2018}
{Decarli}, R., {Walter}, F., {Venemans}, B.~P., {et~al.} 2018, \apj, 854, 97

\bibitem[{{D{\'\i}az-Santos} {et~al.}(2017){D{\'\i}az-Santos}, {Armus}, {Charmandaris}, {Lu}, {Stierwalt}, {Stacey}, {Malhotra}, {van der Werf}, {Howell}, {Privon}, {Mazzarella}, {Goldsmith}, {Murphy}, {Barcos-Mu{\~n}oz}, {Linden}, {Inami}, {Larson}, {Evans}, {Appleton}, {Iwasawa}, {Lord}, {Sanders}, \& {Surace}}]{diazsantos2017}
{D{\'\i}az-Santos}, T., {Armus}, L., {Charmandaris}, V., {et~al.} 2017, \apj, 846, 32

\bibitem[{{D{\'\i}az-Santos} {et~al.}(2016){D{\'\i}az-Santos}, {Assef}, {Blain}, {Tsai}, {Aravena}, {Eisenhardt}, {Wu}, {Stern}, \& {Bridge}}]{diazsantos2016}
{D{\'\i}az-Santos}, T., {Assef}, R.~J., {Blain}, A.~W., {et~al.} 2016, \apjl, 816, L6

\bibitem[{{Draine} {et~al.}(2007){Draine}, {Dale}, {Bendo}, {Gordon}, {Smith}, {Armus}, {Engelbracht}, {Helou}, {Kennicutt}, {Li}, {Roussel}, {Walter}, {Calzetti}, {Moustakas}, {Murphy}, {Rieke}, {Bot}, {Hollenbach}, {Sheth}, \& {Teplitz}}]{draine2007}
{Draine}, B.~T., {Dale}, D.~A., {Bendo}, G., {et~al.} 2007, \apj, 663, 866

\bibitem[{{Duras} {et~al.}(2017){Duras}, {Bongiorno}, {Piconcelli}, {Bianchi}, {Pappalardo}, {Valiante}, {Bischetti}, {Feruglio}, {Martocchia}, {Schneider}, {Vietri}, {Vignali}, {Zappacosta}, {La Franca}, \& {Fiore}}]{duras2017}
{Duras}, F., {Bongiorno}, A., {Piconcelli}, E., {et~al.} 2017, \aap, 604, A67

\bibitem[{{Eilers} {et~al.}(2020){Eilers}, {Hennawi}, {Decarli}, {Davies}, {Venemans}, {Walter}, {Ba{\~n}ados}, {Fan}, {Farina}, {Mazzucchelli}, {Novak}, {Schindler}, {Simcoe}, {Wang}, \& {Yang}}]{eilers2020}
{Eilers}, A.-C., {Hennawi}, J.~F., {Decarli}, R., {et~al.} 2020, \apj, 900, 37

\bibitem[{{Esposito} {et~al.}(2024){Esposito}, {Vallini}, {Pozzi}, {Casasola}, {Alonso-Herrero}, {Garc{\'\i}a-Burillo}, {Decarli}, {Calura}, {Vignali}, {Mingozzi}, {Gruppioni}, \& {Sengupta}}]{Esposito2024}
{Esposito}, F., {Vallini}, L., {Pozzi}, F., {et~al.} 2024, \mnras, 527, 8727

\bibitem[{{Fern{\'a}ndez Aranda} {et~al.}(2024){Fern{\'a}ndez Aranda}, {D{\'\i}az Santos}, {Hatziminaoglou}, {Assef}, {Aravena}, {Eisenhardt}, {Ferkinhoff}, {Pensabene}, {Nikola}, {Andreani}, {Vishwas}, {Stacey}, {Decarli}, {Blain}, {Brisbin}, {Charmandaris}, {Jun}, {Li}, {Liao}, {Martin}, {Stern}, {Tsai}, {Wu}, \& {Zewdie}}]{fernandez2024}
{Fern{\'a}ndez Aranda}, R., {D{\'\i}az Santos}, T., {Hatziminaoglou}, E., {et~al.} 2024, \aap, 682, A166

\bibitem[{{Feruglio} {et~al.}(2018){Feruglio}, {Fiore}, {Carniani}, {Maiolino}, {D'Odorico}, {Luminari}, {Barai}, {Bischetti}, {Bongiorno}, {Cristiani}, {Ferrara}, {Gallerani}, {Marconi}, {Pallottini}, {Piconcelli}, \& {Zappacosta}}]{feruglio2018}
{Feruglio}, C., {Fiore}, F., {Carniani}, S., {et~al.} 2018, \aap, 619, A39

\bibitem[{{Feruglio} {et~al.}(2023){Feruglio}, {Maio}, {Tripodi}, {Winters}, {Zappacosta}, {Bischetti}, {Civano}, {Carniani}, {D'Odorico}, {Fiore}, {Gallerani}, {Ginolfi}, {Maiolino}, {Piconcelli}, {Valiante}, \& {Zanchettin}}]{feruglio2023}
{Feruglio}, C., {Maio}, U., {Tripodi}, R., {et~al.} 2023, \apjl, 954, L10

\bibitem[{{Foreman-Mackey} {et~al.}(2013){Foreman-Mackey}, {Hogg}, {Lang}, \& {Goodman}}]{foreman2013}
{Foreman-Mackey}, D., {Hogg}, D.~W., {Lang}, D., \& {Goodman}, J. 2013, \pasp, 125, 306

\bibitem[{{Herrera-Camus} {et~al.}(2016){Herrera-Camus}, {Bolatto}, {Smith}, {Draine}, {Pellegrini}, {Wolfire}, {Croxall}, {de Looze}, {Calzetti}, {Kennicutt}, {Crocker}, {Armus}, {van der Werf}, {Sandstrom}, {Galametz}, {Brandl}, {Groves}, {Rigopoulou}, {Walter}, {Leroy}, {Boquien}, {Tabatabaei}, \& {Beirao}}]{herrera-camus2016}
{Herrera-Camus}, R., {Bolatto}, A., {Smith}, J.~D., {et~al.} 2016, \apj, 826, 175

\bibitem[{{Kaasinen} {et~al.}(2024){Kaasinen}, {Venemans}, {Harrington}, {Boogaard}, {Meyer}, {Ba{\~n}ados}, {Decarli}, {Walter}, {Neeleman}, {Calistro Rivera}, \& {da Cunha}}]{kaasinen2024}
{Kaasinen}, M., {Venemans}, B., {Harrington}, K.~C., {et~al.} 2024, \aap, 684, A33

\bibitem[{{Kroupa} \& {Weidner}(2003)}]{kroupa2003}
{Kroupa}, P. \& {Weidner}, C. 2003, \apj, 598, 1076

\bibitem[{{Leroy} {et~al.}(2011){Leroy}, {Bolatto}, {Gordon}, {Sandstrom}, {Gratier}, {Rosolowsky}, {Engelbracht}, {Mizuno}, {Corbelli}, {Fukui}, \& {Kawamura}}]{leroy2011}
{Leroy}, A.~K., {Bolatto}, A., {Gordon}, K., {et~al.} 2011, \apj, 737, 12

\bibitem[{{Li} {et~al.}(2020){Li}, {Wang}, {Riechers}, {Walter}, {Decarli}, {Venamans}, {Neri}, {Shao}, {Fan}, {Gao}, {Carilli}, {Omont}, {Cox}, {Menten}, {Wagg}, {Bertoldi}, \& {Narayanan}}]{li2020}
{Li}, J., {Wang}, R., {Riechers}, D., {et~al.} 2020, \apj, 889, 162

\bibitem[{{Liao} {et~al.}(2024){Liao}, {Chen}, {Wang}, {Smail}, {Ao}, {Chapman}, {Dudzevi{\v{c}}i{\={u}}t{\.{e}}}, {Frias Castillo}, {Lee}, {Serjeant}, {Swinbank}, {Taylor}, {Umehata}, \& {Zhao}}]{liao2023}
{Liao}, C.-L., {Chen}, C.-C., {Wang}, W.-H., {et~al.} 2024, \apj, 961, 226

\bibitem[{{Madden} {et~al.}(2020){Madden}, {Cormier}, {Hony}, {Lebouteiller}, {Abel}, {Galametz}, {De Looze}, {Chevance}, {Polles}, {Lee}, {Galliano}, {Lambert-Huyghe}, {Hu}, \& {Ramambason}}]{madden2020}
{Madden}, S.~C., {Cormier}, D., {Hony}, S., {et~al.} 2020, \aap, 643, A141

\bibitem[{{McMullin} {et~al.}(2007){McMullin}, {Waters}, {Schiebel}, {Young}, \& {Golap}}]{mcmullin2007}
{McMullin}, J.~P., {Waters}, B., {Schiebel}, D., {Young}, W., \& {Golap}, K. 2007, in Astronomical Society of the Pacific Conference Series, Vol. 376, Astronomical Data Analysis Software and Systems XVI, ed. R.~A. {Shaw}, F.~{Hill}, \& D.~J. {Bell}, 127

\bibitem[{{Meyer} {et~al.}(2022){Meyer}, {Walter}, {Cicone}, {Cox}, {Decarli}, {Neri}, {Novak}, {Pensabene}, {Riechers}, \& {Weiss}}]{meyer2022}
{Meyer}, R.~A., {Walter}, F., {Cicone}, C., {et~al.} 2022, \apj, 927, 152

\bibitem[{{Miettinen} {et~al.}(2017){Miettinen}, {Delvecchio}, {Smol{\v{c}}i{\'c}}, {Aravena}, {Brisbin}, {Karim}, {Magnelli}, {Novak}, {Schinnerer}, {Albrecht}, {Aussel}, {Bertoldi}, {Capak}, {Casey}, {Hayward}, {Ilbert}, {Intema}, {Jiang}, {Le F{\`e}vre}, {McCracken}, {Mu{\~n}oz Arancibia}, {Navarrete}, {Padilla}, {Riechers}, {Salvato}, {Scott}, {Sheth}, \& {Tasca}}]{miettinen2017}
{Miettinen}, O., {Delvecchio}, I., {Smol{\v{c}}i{\'c}}, V., {et~al.} 2017, \aap, 606, A17

\bibitem[{{Neeleman} {et~al.}(2019){Neeleman}, {Ba{\~n}ados}, {Walter}, {Decarli}, {Venemans}, {Carilli}, {Fan}, {Farina}, {Mazzucchelli}, {Novak}, {Riechers}, {Rix}, \& {Wang}}]{neeleman2019}
{Neeleman}, M., {Ba{\~n}ados}, E., {Walter}, F., {et~al.} 2019, \apj, 882, 10

\bibitem[{{Neeleman} {et~al.}(2021){Neeleman}, {Novak}, {Venemans}, {Walter}, {Decarli}, {Kaasinen}, {Schindler}, {Ba{\~n}ados}, {Carilli}, {Drake}, {Fan}, \& {Rix}}]{neeleman2021}
{Neeleman}, M., {Novak}, M., {Venemans}, B.~P., {et~al.} 2021, \apj, 911, 141

\bibitem[{{Novak} {et~al.}(2019){Novak}, {Ba{\~n}ados}, {Decarli}, {Walter}, {Venemans}, {Neeleman}, {Farina}, {Mazzucchelli}, {Carilli}, {Fan}, {Rix}, \& {Wang}}]{novak2019}
{Novak}, M., {Ba{\~n}ados}, E., {Decarli}, R., {et~al.} 2019, \apj, 881, 63

\bibitem[{{Parente} {et~al.}(2026){Parente}, {Salvestrini}, {Granato}, {Narayanan}, {Tripodi}, {Bianchi}, {Bischetti}, {Feruglio}, {Fiore}, \& {Silva}}]{parente2026}
{Parente}, M., {Salvestrini}, F., {Granato}, G.~L., {et~al.} 2026, arXiv e-prints, arXiv:2603.04505

\bibitem[{{Pensabene} {et~al.}(2021){Pensabene}, {Decarli}, {Ba{\~n}ados}, {Venemans}, {Walter}, {Bertoldi}, {Fan}, {Farina}, {Li}, {Mazzucchelli}, {Novak}, {Riechers}, {Rix}, {Strauss}, {Wang}, {Wei{\ss}}, {Yang}, \& {Yang}}]{Pensabene2021}
{Pensabene}, A., {Decarli}, R., {Ba{\~n}ados}, E., {et~al.} 2021, \aap, 652, A66

\bibitem[{{R{\'e}my-Ruyer} {et~al.}(2014){R{\'e}my-Ruyer}, {Madden}, {Galliano}, {Galametz}, {Takeuchi}, {Asano}, {Zhukovska}, {Lebouteiller}, {Cormier}, {Jones}, {Bocchio}, {Baes}, {Bendo}, {Boquien}, {Boselli}, {DeLooze}, {Doublier-Pritchard}, {Hughes}, {Karczewski}, \& {Spinoglio}}]{remy2014}
{R{\'e}my-Ruyer}, A., {Madden}, S.~C., {Galliano}, F., {et~al.} 2014, \aap, 563, A31

\bibitem[{{Ronconi} {et~al.}(2024){Ronconi}, {Lapi}, {Torsello}, {Bressan}, {Donevski}, {Pantoni}, {Behiri}, {Boco}, {Cimatti}, {D'Amato}, {Danese}, {Giulietti}, {Perrotta}, {Silva}, {Talia}, \& {Massardi}}]{ronconi2024}
{Ronconi}, T., {Lapi}, A., {Torsello}, M., {et~al.} 2024, \aap, 685, A161

\bibitem[{{Saccheo} {et~al.}(2025){Saccheo}, {Bongiorno}, {Piconcelli}, {Zappacosta}, {Bischetti}, {D'Odorico}, {Done}, {Temple}, {Testa}, {Tortosa}, {Brusa}, {Carniani}, {Civano}, {Comastri}, {Cristiani}, {De Cicco}, {Elvis}, {Fan}, {Feruglio}, {Fiore}, {Gallerani}, {Giallongo}, {Gilli}, {Grazian}, {Guainazzi}, {Haardt}, {Maiolino}, {Menci}, {Miniutti}, {Nicastro}, {Paolillo}, {Puccetti}, {Salvestrini}, {Schneider}, {Tombesi}, {Tripodi}, {Valiante}, {Vallini}, {Vanzella}, {Vietri}, {Vignali}, {Vito}, {Volonteri}, \& {La Franca}}]{saccheo2025}
{Saccheo}, I., {Bongiorno}, A., {Piconcelli}, E., {et~al.} 2025, \aap, 693, A157

\bibitem[{{Saintonge} {et~al.}(2013){Saintonge}, {Lutz}, {Genzel}, {Magnelli}, {Nordon}, {Tacconi}, {Baker}, {Bandara}, {Berta}, {F{\"o}rster Schreiber}, {Poglitsch}, {Sturm}, {Wuyts}, \& {Wuyts}}]{saintonge2013}
{Saintonge}, A., {Lutz}, D., {Genzel}, R., {et~al.} 2013, \apj, 778, 2

\bibitem[{{Salpeter}(1955)}]{salpeter1955}
{Salpeter}, E.~E. 1955, \apj, 121, 161

\bibitem[{{Salvestrini} {et~al.}(2025{\natexlab{a}}){Salvestrini}, {Bianchi}, \& {Corbelli}}]{salvestrini2025b}
{Salvestrini}, F., {Bianchi}, S., \& {Corbelli}, E. 2025{\natexlab{a}}, \aap, 699, A346

\bibitem[{{Salvestrini} {et~al.}(2025{\natexlab{b}}){Salvestrini}, {Feruglio}, {Tripodi}, {Fontanot}, {Bischetti}, {De Lucia}, {Fiore}, {Hirschmann}, {Maio}, {Piconcelli}, {Saccheo}, {Tortosa}, {Valiante}, {Xie}, \& {Zappacosta}}]{salvestrini2025}
{Salvestrini}, F., {Feruglio}, C., {Tripodi}, R., {et~al.} 2025{\natexlab{b}}, \aap, 695, A23

\bibitem[{{Salvestrini} {et~al.}(2022){Salvestrini}, {Gruppioni}, {Hatziminaoglou}, {Pozzi}, {Vignali}, {Casasola}, {Paladino}, {Aalto}, {Andreani}, {Marchesi}, \& {Stanke}}]{salvestrini2022}
{Salvestrini}, F., {Gruppioni}, C., {Hatziminaoglou}, E., {et~al.} 2022, \aap, 663, A28

\bibitem[{{Schneider} {et~al.}(2015){Schneider}, {Bianchi}, {Valiante}, {Risaliti}, \& {Salvadori}}]{schneider2015}
{Schneider}, R., {Bianchi}, S., {Valiante}, R., {Risaliti}, G., \& {Salvadori}, S. 2015, \aap, 579, A60

\bibitem[{{Solomon} \& {Vanden Bout}(2005)}]{solomon2005}
{Solomon}, P.~M. \& {Vanden Bout}, P.~A. 2005, \araa, 43, 677

\bibitem[{{Spilker} {et~al.}(2025){Spilker}, {Champagne}, {Fan}, {Fujimoto}, {van der Werf}, {Yang}, \& {Yue}}]{spilker2025}
{Spilker}, J.~S., {Champagne}, J.~B., {Fan}, X., {et~al.} 2025, \apj, 982, 72

\bibitem[{Stefan {et~al.}(2015)Stefan, Carilli, Wagg, Walter, Riechers, Bertoldi, Green, Fan, Menten, \& Wang}]{stefan2015}
Stefan, I.~I., Carilli, C.~L., Wagg, J., {et~al.} 2015, Monthly Notices of the Royal Astronomical Society, 451, 1713

\bibitem[{{Sun} {et~al.}(2026){Sun}, {Yang}, {Wang}, {Eisenstein}, {Decarli}, {Fan}, {Rieke}, {Ba{\~n}ados}, {Bosman}, {Cai}, {Champagne}, {Colina}, {D'Eugenio}, {Fudamoto}, {Li}, {Lin}, {Liu}, {Lyu}, {Mazzucchelli}, {Jin}, {Jun}, {Tee}, {Wu}, \& {Zhang}}]{sun2026}
{Sun}, F., {Yang}, J., {Wang}, F., {et~al.} 2026, \apj, 1003, 206

\bibitem[{{Tadaki} {et~al.}(2025){Tadaki}, {Esposito}, {Vallini}, {Tsukui}, {Saito}, {Iono}, \& {Michiyama}}]{tadaki2025}
{Tadaki}, K., {Esposito}, F., {Vallini}, L., {et~al.} 2025, Nature Astronomy, 9, 720

\bibitem[{{Tortosa} {et~al.}(2024){Tortosa}, {Zappacosta}, {Piconcelli}, {Bischetti}, {Done}, {Miniutti}, {Saccheo}, {Vietri}, {Bongiorno}, {Brusa}, {Carniani}, {Chilingarian}, {Civano}, {Cristiani}, {D'Odorico}, {Elvis}, {Fan}, {Feruglio}, {Fiore}, {Gallerani}, {Giallongo}, {Gilli}, {Grazian}, {Guainazzi}, {Haardt}, {Luminari}, {Maiolino}, {Menci}, {Nicastro}, {Petrucci}, {Puccetti}, {Salvestrini}, {Schneider}, {Testa}, {Tombesi}, {Tripodi}, {Valiante}, {Vallini}, {Vanzella}, {Vasylenko}, {Vignali}, {Vito}, {Volonteri}, \& {La Franca}}]{tortosa2024}
{Tortosa}, A., {Zappacosta}, L., {Piconcelli}, E., {et~al.} 2024, \aap, 691, A235

\bibitem[{{Tripodi} {et~al.}(2024{\natexlab{a}}){Tripodi}, {Feruglio}, {Fiore}, {Zappacosta}, {Piconcelli}, {Bischetti}, {Bongiorno}, {Carniani}, {Civano}, {Chen}, {Cristiani}, {Cupani}, {Di Mascia}, {D'Odorico}, {Fan}, {Ferrara}, {Gallerani}, {Ginolfi}, {Maiolino}, {Mainieri}, {Marconi}, {Saccheo}, {Salvestrini}, {Tortosa}, \& {Valiante}}]{tripodi2024}
{Tripodi}, R., {Feruglio}, C., {Fiore}, F., {et~al.} 2024{\natexlab{a}}, \aap, 689, A220

\bibitem[{{Tripodi} {et~al.}(2022){Tripodi}, {Feruglio, C.}, {Fiore, F.}, {Bischetti, M.}, {D\'{}Odorico, V.}, {Carniani, S.}, {Cristiani, S.}, {Gallerani, S.}, {Maiolino, R.}, {Marconi, A.}, {Pallottini, A.}, {Piconcelli, E.}, {Vallini, L.}, \& {Zana, T.}}]{Tripodi2022}
{Tripodi}, R., {Feruglio, C.}, {Fiore, F.}, {et~al.} 2022, A\&A, 665, A107

\bibitem[{{Tripodi} {et~al.}(2023){Tripodi}, {Lelli}, {Feruglio}, {Fiore}, {Fontanot}, {Bischetti}, \& {Maiolino}}]{tripodi2023a}
{Tripodi}, R., {Lelli}, F., {Feruglio}, C., {et~al.} 2023, \aap, 671, A44

\bibitem[{{Tripodi} {et~al.}(2024{\natexlab{b}}){Tripodi}, {Scholtz}, {Maiolino}, {Fujimoto}, {Carniani}, {Silverman}, {Feruglio}, {Ginolfi}, {Zappacosta}, {Costa}, {Jones}, {Piconcelli}, {Bischetti}, \& {Fiore}}]{tripodi2023c}
{Tripodi}, R., {Scholtz}, J., {Maiolino}, R., {et~al.} 2024{\natexlab{b}}, \aap, 682, A54

\bibitem[{{Vallini} {et~al.}(2018){Vallini}, {Pallottini}, {Ferrara}, {Gallerani}, {Sobacchi}, \& {Behrens}}]{vallini2018}
{Vallini}, L., {Pallottini}, A., {Ferrara}, A., {et~al.} 2018, \mnras, 473, 271

\bibitem[{{Venemans} {et~al.}(2012){Venemans}, {McMahon}, {Walter}, {Decarli}, {Cox}, {Neri}, {Hewett}, {Mortlock}, {Simpson}, \& {Warren}}]{venemans2012}
{Venemans}, B.~P., {McMahon}, R.~G., {Walter}, F., {et~al.} 2012, \apjl, 751, L25

\bibitem[{{Venemans} {et~al.}(2020){Venemans}, {Walter}, {Neeleman}, {Novak}, {Otter}, {Decarli}, {Ba{\~n}ados}, {Drake}, {Farina}, {Kaasinen}, {Mazzucchelli}, {Carilli}, {Fan}, {Rix}, \& {Wang}}]{venemans2020}
{Venemans}, B.~P., {Walter}, F., {Neeleman}, M., {et~al.} 2020, \apj, 904, 130

\bibitem[{{Walter} {et~al.}(2003){Walter}, {Bertoldi}, {Carilli}, {Cox}, {Lo}, {Neri}, {Fan}, {Omont}, {Strauss}, \& {Menten}}]{walter2003}
{Walter}, F., {Bertoldi}, F., {Carilli}, C., {et~al.} 2003, \nat, 424, 406

\bibitem[{{Walter} {et~al.}(2004){Walter}, {Carilli}, {Bertoldi}, {Menten}, {Cox}, {Lo}, {Fan}, \& {Strauss}}]{walter2004}
{Walter}, F., {Carilli}, C., {Bertoldi}, F., {et~al.} 2004, \apjl, 615, L17

\bibitem[{{Wang} {et~al.}(2024){Wang}, {Yang}, {Fan}, {Venemans}, {Decarli}, {Ba{\~n}ados}, {Walter}, {Barth}, {Bian}, {Davies}, {Eilers}, {Farina}, {Hennawi}, {Li}, {Mazzucchelli}, {Wang}, {Wu}, \& {Yue}}]{wang2024}
{Wang}, F., {Yang}, J., {Fan}, X., {et~al.} 2024, \apj, 968, 9

\bibitem[{{Wang} {et~al.}(2019){Wang}, {Yang}, {Fan}, {Wu}, {Yue}, {Li}, {Bian}, {Jiang}, {Ba{\~n}ados}, {Schindler}, {Findlay}, {Davies}, {Decarli}, {Farina}, {Green}, {Hennawi}, {Huang}, {Mazzuccheli}, {McGreer}, {Venemans}, {Walter}, {Dye}, {Lyke}, {Myers}, \& {Nunez}}]{wang2019b}
{Wang}, F., {Yang}, J., {Fan}, X., {et~al.} 2019, \apj, 884, 30

\bibitem[{{Witstok} {et~al.}(2023){Witstok}, {Jones}, {Maiolino}, {Smit}, \& {Schneider}}]{witstok2023}
{Witstok}, J., {Jones}, G.~C., {Maiolino}, R., {Smit}, R., \& {Schneider}, R. 2023, \mnras, 523, 3119

\bibitem[{{Witstok} {et~al.}(2022){Witstok}, {Smit}, {Maiolino}, {Kumari}, {Aravena}, {Boogaard}, {Bouwens}, {Carniani}, {Hodge}, {Jones}, {Stefanon}, {van der Werf}, \& {Schouws}}]{witstok2022}
{Witstok}, J., {Smit}, R., {Maiolino}, R., {et~al.} 2022, \mnras, 515, 1751

\bibitem[{{Wu} {et~al.}(2025){Wu}, {Wang}, {Liu}, {Tan}, {Ho}, {Zhang}, {Shi}, {Xu}, {Kohno}, {Wang}, {Izumi}, \& {Li}}]{wu2025}
{Wu}, Y., {Wang}, T., {Liu}, D., {et~al.} 2025, arXiv e-prints, arXiv:2506.14896

\bibitem[{{Xu} {et~al.}(2026){Xu}, {Decarli}, {Wang}, {Borea}, {Pensabene}, {Fan}, {Riechers}, {Ba{\~n}ados}, {Wei{\ss}}, {Costa}, {Walter}, {Wang}, {Yang}, {Venemans}, {Li}, \& {Farina}}]{xu2026}
{Xu}, F., {Decarli}, R., {Wang}, R., {et~al.} 2026, arXiv e-prints, arXiv:2605.20698

\bibitem[{{Yang} {et~al.}(2020){Yang}, {Wang}, {Fan}, {Hennawi}, {Davies}, {Yue}, {Banados}, {Wu}, {Venemans}, {Barth}, {Bian}, {Boutsia}, {Decarli}, {Farina}, {Green}, {Jiang}, {Li}, {Mazzucchelli}, \& {Walter}}]{yang2020}
{Yang}, J., {Wang}, F., {Fan}, X., {et~al.} 2020, \apjl, 897, L14

\bibitem[{{Zappacosta} {et~al.}(2023){Zappacosta}, {Piconcelli}, {Fiore}, {Saccheo}, {Valiante}, {Vignali}, {Vito}, {Volonteri}, {Bischetti}, {Comastri}, {Done}, {Elvis}, {Giallongo}, {La Franca}, {Lanzuisi}, {Laurenti}, {Miniutti}, {Bongiorno}, {Brusa}, {Civano}, {Carniani}, {D'Odorico}, {Feruglio}, {Gallerani}, {Gilli}, {Grazian}, {Guainazzi}, {Marinucci}, {Menci}, {Middei}, {Nicastro}, {Puccetti}, {Tombesi}, {Tortosa}, {Testa}, {Vietri}, {Cristiani}, {Haardt}, {Maiolino}, {Schneider}, {Tripodi}, {Vallini}, \& {Vanzella}}]{zappacosta2023}
{Zappacosta}, L., {Piconcelli}, E., {Fiore}, F., {et~al.} 2023, \aap, 678, A201

\bibitem[{{Zhao} {et~al.}(2013){Zhao}, {Lu}, {Xu}, {Gao}, {Lord}, {Howell}, {Isaak}, {Charmandaris}, {Diaz-Santos}, {Appleton}, {Evans}, {Iwasawa}, {Leech}, {Mazzarella}, {Petric}, {Sanders}, {Schulz}, {Surace}, \& {van der Werf}}]{zhao2013}
{Zhao}, Y., {Lu}, N., {Xu}, C.~K., {et~al.} 2013, \apjl, 765, L13

\end{thebibliography}
